\documentclass[mnsc, nonblindrev]{informs1}
\OneAndAHalfSpacedXI 
\usepackage{rotating}
\usepackage{bbding}
\usepackage[left=1in,top=1in,right=1in,bottom=1in]{geometry}
\usepackage{amssymb,amsmath,amsxtra,amstext,xfrac,mathtools}
\usepackage{graphicx,booktabs,enumerate,paralist,mdwlist,tabularx}
\usepackage{dsfont,verbatim,latexsym}
\usepackage{mathrsfs}
\usepackage{bbm}
\usepackage{multirow}
\usepackage[table]{xcolor}
\newcolumntype{C}[1]{>{\centering\let\newline\\\arraybackslash\hspace{0pt}}m{#1}}
\usepackage{longtable,multirow,threeparttable}
\usepackage{float}
\usepackage{tikz,qtree}
\usetikzlibrary{arrows,positioning,decorations.pathreplacing}
\usepackage{ulem}
\usepackage{mwe} 
\usepackage{makecell}
\usepackage{eurosym}
\usepackage[english]{babel}
\usepackage[utf8]{inputenc}
\usepackage{fancyhdr}
\usepackage{color}
\usepackage{array}
\usepackage{subfig}
\usepackage{url}
\usepackage{diagbox}
\usepackage[colorlinks=TRUE,citecolor=MyDarkBlue,urlcolor=MyDarkBlue,linkcolor=MyDarkBlue]{hyperref}
\definecolor{MyDarkBlue}{RGB}{158,0,0}

\usepackage{natbib}
\bibpunct[, ]{(}{)}{,}{a}{}{,}%
\def\bibfont{\small}%
\TheoremsNumberedThrough
\EquationsNumberedThrough

\begin{document}
\RUNAUTHOR{Xu et al.}
\RUNTITLE{Organizational Adoption of GenAI}
\TITLE{Generative AI and Organizational Structure in the Knowledge Economy}
\ARTICLEAUTHORS{%
\AUTHOR{Fasheng Xu$^1$ ~~~ Jing Hou$^2$ ~~~ Wei Chen$^1$ ~~~ Karen Xie$^1$\\[-2mm]}
\AFF{$^1$School of Business, University of Connecticut} 
\AFF{$^2$School of Management, Fudan University} 
} 
\ABSTRACT{%
Generative AI (GenAI) is rapidly transforming knowledge work, yet its implications for organizational hierarchies remain poorly understood. Unlike earlier automation technologies, GenAI can both perform tasks autonomously and assist human workers, while its intrinsic fallibility, the tendency to produce confident but incorrect outputs, demands continuous human oversight. We develop a theoretical model to study how GenAI reshapes workforce composition and organizational structure in knowledge-based hierarchies. Our analysis highlights two deployment dimensions, namely mode (automation vs.\ augmentation) and location (worker vs.\ expert layer), which generate a $2 \times 2$ design space whose organizational implications are not predicted by traditional technology adoption theories. We obtain three main findings. First, GenAI's effect on entry-level skill requirements is critically mode-dependent. Worker-level automation leads firms to hire fewer but more skilled workers who validate AI outputs and limit costly escalation to experts. Worker-level augmentation, by contrast, expands workers' effective capability, allowing firms to relax entry-level knowledge requirements while sustaining performance. The decline in junior employment documented in recent studies therefore reflects deployment choices favoring automation over augmentation, not an inevitable consequence of GenAI itself. Second, expert-level deployment uniformly lowers entry-level skill requirements, regardless of whether GenAI automates or augments. By expanding experts' capacity to support downstream workers, it enables organizations to employ a broader base of less specialized workers, thereby broadening entry-level access to knowledge work. Third, organizational structure evolves non-monotonically as GenAI improves: across all four deployment architectures, the span of control initially contracts before eventually expanding. Demand for senior expertise may therefore remain stable or even increase during early-to-intermediate stages before hierarchies ultimately flatten. Together, these results demonstrate that GenAI's organizational consequences depend on deployment design rather than adoption intensity alone, and that the same technology can upskill or deskill the workforce depending on how and where it is deployed.

}%
\HISTORY{Current version: March, 2026}
\maketitle
\vspace{-8mm}
\section{Introduction}
The knowledge economy, in which expertise serves as a key economic resource, has long relied on organizational hierarchies to efficiently allocate specialized knowledge \citep{powell2004knowledge}. A canonical form is the ``knowledge-based hierarchy,'' in which lower-level workers handle routine problems while escalating more complex cases to higher layers of expertise, such as those in analyst--manager, attorney--partner, and software engineer--tech lead structures. When workers encounter problems beyond their capacity, they turn to experts at higher levels for guidance \citep{garicano2000hierarchies}. The advent of Generative AI (GenAI), however, is fundamentally challenging this established order by enabling machines to perform complex knowledge-based tasks through natural language interactions, potentially transforming how organizations structure their hierarchies and deploy their workforce.

Emerging deployment patterns across industries reveal systematic variation in how GenAI is integrated into daily workflows. In software development, AI coding agents such as Claude Code and Cursor autonomously generate code, shifting the developer's role from execution toward validation and integration. In creative fields, tools like Adobe Firefly and Canva's AI features expand users' capabilities, allowing non-designers to produce content that previously required specialized expertise. GenAI is also transforming expert-level roles. In professional services, AI platforms like McKinsey's Lilli function as virtual knowledge repositories, providing direct answers to junior employees' queries and reducing their routine reliance on senior staff. 
Meanwhile, organizations are deploying AI-powered synthesis tools to assist senior consultants and partners directly, enabling them to rapidly retrieve firm expertise and respond to complex escalations more efficiently. This heterogeneity reveals that GenAI is not a uniform technology; rather, it is being integrated in fundamentally different ways that vary both in how it interacts with human ability and where it is positioned within the organizational hierarchy.

Despite widespread adoption and intense public interest, we lack rigorous theoretical frameworks for understanding how GenAI reshapes organizational hierarchies and the distribution of skills within firms. Recent empirical evidence reveals striking workforce composition effects: early-career employment has declined in the most AI-exposed occupations, while senior employment has remained stable or even risen \citep{brynjolfsson2025canaries, hosseini2025generative}. Furthermore, these entry-level declines are concentrated in settings where GenAI primarily automates tasks and are much more muted where GenAI augments human work \citep{brynjolfsson2025canaries}. Yet existing organizational theories cannot explain why the same technology produces simultaneous junior displacement and senior stability, nor why these effects vary systematically across deployment contexts. Without clear conceptual foundations, it remains impossible to predict whether a given GenAI deployment will raise or lower skill requirements, steepen or flatten hierarchies, or displace or preserve senior roles.

Two features distinguish GenAI from prior workplace technologies and drive the organizational dynamics we characterize. The first is \textit{intrinsic fallibility}: regardless of how GenAI is deployed, it produces probabilistically incorrect outputs that appear authoritative---a phenomenon known as \textit{hallucination}---thereby creating a novel organizational burden of continuous human validation \citep{bubeck2023sparks, zhou2024larger}. Unlike rule-based automation, which either produces a correct output or fails transparently, or earlier AI systems that attach explicit uncertainty estimates to their recommendations, GenAI's errors are indistinguishable from correct outputs without domain expertise, making validation central to organizational design. Critically, validation demands domain knowledge: workers can verify outputs only within their own area of expertise. GenAI therefore does not simply substitute for human knowledge; it transforms the nature of knowledge work from execution to oversight, requiring organizations to retain the very expertise that GenAI ostensibly replaces, not to perform tasks, but to verify them.

The second distinctive feature is \textit{deployment flexibility}. GenAI can be integrated into knowledge work along two dimensions: deployment mode and deployment location. Deployment mode determines how GenAI effectiveness relates to human ability, specifically through automation or augmentation. Under automation, GenAI autonomously executes tasks within a defined scope, and its effectiveness is independent of users' baseline skill. Under augmentation, GenAI extends the user's capability boundary beyond their intrinsic expertise, with effectiveness inherently coupled to baseline ability. Deployment location specifies which role within the hierarchy GenAI reshapes. In knowledge-based hierarchies, workers focus on task execution, whereas experts provide guidance on exceptions rather than routine execution \citep{garicano2000hierarchies}. The organizational consequences of GenAI adoption are therefore inherently layer-dependent. At the worker level, GenAI either executes tasks that workers would otherwise perform (automation) or operates alongside workers as a capability amplifier (augmentation). At the expert level, GenAI either substitutes for experts as a first-line solution provider (automation) or accelerates experts' consultation efficiency (augmentation). Together, this two-dimensional flexibility generates a $2 \times 2$ design space (Table~\ref{tab:design_space}), each cell carrying distinct implications for workforce composition.

\begin{table}[htbp]
\centering
\small
\caption{GenAI Deployment Design Space}
\label{tab:design_space}
\linespread{1.5}
\begin{tabular}{lcc}
\toprule
& \textbf{Automation} & \textbf{Augmentation} \\
\midrule
\textbf{Worker Level~~} & Task Executor & Capability Amplifier \\[0.5em]
\textbf{Expert Level} & Virtual Supervisor & ~~Communication Accelerator \\
\bottomrule
\end{tabular}
\end{table}

The interaction between intrinsic fallibility and deployment flexibility generates organizational dynamics that traditional technology-adoption theories do not predict. These two features play complementary but distinct roles. Intrinsic fallibility establishes \textit{why} organizational design matters under GenAI: human oversight remains essential regardless of deployment configuration. Deployment flexibility determines \textit{how} organizational consequences vary: the choice of mode and location shapes who validates, what is validated, and how validation reshapes skill requirements and workforce composition. Traditional automation executed rule-based tasks without probabilistic errors requiring cognitive oversight; earlier AI systems provided predictions or recommendations but left execution to human workers; and information and communication technologies reduced coordination frictions without performing cognitive work. GenAI uniquely combines autonomous cognitive execution with intrinsic fallibility and unprecedented deployment flexibility, creating organizational design challenges and opportunities that existing frameworks cannot address.

This paper asks: \textit{How does GenAI adoption reshape organizational structure in knowledge-based hierarchies?} To answer this question, we develop a formal analytical model building on \citet{garicano2000hierarchies}, where hierarchies emerge endogenously to economize on knowledge acquisition costs. We incorporate three GenAI-specific elements. First, a capability parameter determines the scope of tasks that GenAI can address. Second, intrinsic fallibility is captured by a hallucination rate that is not mechanically tied to capability, consistent with evidence that more advanced models may exhibit comparable or higher error rates across tasks \citep{zhou2024larger}. Third, we model explicit validation and correction time requirements that draw on scarce worker or expert labor. We analyze four deployment architectures spanning the $2 \times 2$ design space and characterize their effects on optimal workforce composition, skill requirements, and managerial spans of control (i.e., the ratio of low-level workers to high-level experts).\footnote{We emphasize that this $2\times 2$ framework is an analytical decomposition of distinct organizational mechanisms rather than a rigid product taxonomy. In practice, a single technology deployment may blend elements across cells. For example, GitHub Copilot operates primarily as worker-level automation when generating code blocks autonomously, but shifts toward worker-level augmentation when providing in-line suggestions during active editing. Our analytical separation is precisely what enables the key insight: these blended deployments produce qualitatively different, and sometimes opposing, organizational consequences depending on which mechanism dominates.} Our analysis yields findings that challenge conventional wisdom about technology-driven organizational change.

First, the effect of GenAI on entry-level skill requirements is highly mode-dependent. Under worker-level automation, GenAI takes over routine execution, yet the firm still relies on workers to validate AI outputs and resolve residual tasks. The firm optimally raises entry-level skill requirements, hiring fewer but more knowledgeable juniors to curb costly escalation to senior experts and to ensure reliable oversight of AI outputs. Under worker-level augmentation, in contrast, GenAI enlarges workers' effective problem-solving range, allowing the firm to lower the knowledge requirement for junior roles while sustaining performance through AI-assisted workflows; deskilling is optimal. This distinction suggests that the decline in junior employment is not an inevitable consequence of GenAI per se, but is more consistent with deployment choices that tilt toward automation rather than augmentation, in line with emerging evidence on early-career impacts \citep{brynjolfsson2025canaries}.

Second, deployment location can be as consequential as deployment mode. Under expert-level automation, AI-based substitutes for expert input lower the cost of escalation, so the firm optimally reduces entry-level skill requirements while preserving problem-solving capacity through AI plus expert oversight. A distinctive implication is what we term \textit{post-adoption capability insensitivity}: once the firm adopts expert-level automation, further improvements in GenAI capability shift queries between GenAI and human experts but leave the optimal entry-level skill threshold unchanged within the AI-covered region. Under expert-level augmentation, GenAI accelerates experts' handling of consultations and lowers the effective cost of accessing high-level knowledge, with deskilling strengthening as the technology advances. Taken together, expert-level adoption yields a clean first-order prediction: it uniformly reduces entry-level skill requirements, broadening access to knowledge work regardless of deployment mode.

Third, departing from the conventional view that technological progress monotonically flattens hierarchies, we uncover a robust non-monotonic evolution of the span of control across all deployment modes: the span contracts initially and expands later as GenAI advances. This implies that even if organizations ultimately flatten, the demand for senior expertise can remain stable or even rise during early-to-intermediate deployment stages, consistent with large-scale evidence \citep{brynjolfsson2025canaries, hosseini2025generative}. The mechanisms, however, differ by deployment mode and location. Under worker-level automation, early automation gains reduce frontline labor demand while leaving expert demand largely unchanged, narrowing the span of control. As GenAI matures, induced upskilling curbs upward referrals, compresses expert demand, and widens the span. In the other deployment modes, deskilling at the entry level takes precedence in the early stage, increasing the volume of tasks that require expert intervention and thereby tightening the span of control. At later stages, GenAI's direct efficiency gains substantially reduce reliance on expert input, and the span of control eventually widens.

Our analysis yields actionable insights for stakeholders navigating the GenAI transformation. For firms, technology procurement cannot be decoupled from hiring strategy. Under worker-level automation, the role of entry-level employees shifts from primary task execution to oversight and validation of AI outputs. In this setting, organizations benefit from hiring workers with sufficient domain expertise and judgment to detect subtle errors and escalate problematic cases appropriately. Training should therefore strengthen workers' ability to critically evaluate AI-generated outputs, recognize potential failure modes, and apply domain knowledge when validating or correcting automated results. Conversely, under augmentation, firms can broaden recruitment toward candidates with strong foundational traits but without deep technical mastery, pivoting HR criteria from peak technical execution toward adaptability and human--AI collaborative aptitude.

Deployment decisions at the expert layer carry different implications. By expanding experts' capacity to supervise and support entry-level workers, expert-level GenAI enables organizations to sustain performance while lowering entry-level knowledge requirements. Critically, firms should anticipate that demand for senior expertise may remain stable or even increase during early-to-intermediate stages of GenAI deployment. Preemptive hierarchy flattening is therefore counterproductive, as it risks creating bottlenecks in exception handling and quality control precisely when expert oversight becomes more critical.

For policymakers and educators, our findings suggest that the labor-market consequences of GenAI depend less on the overall rate of adoption than on how the technology is deployed within organizational hierarchies. Policies that encourage augmentation-oriented applications can help broaden participation in knowledge work, while deployments that expand expert capacity may similarly enable firms to support a larger base of less specialized workers. 
Furthermore, workforce development programs should  move beyond generic AI literacy toward mode-specific curricula that emphasize deep domain expertise for validation in automation settings versus adaptive problem-solving in augmentation contexts.

The rest of the paper is organized as follows. Section \ref{section-literature} compares and contrasts our paper with the related literature. Section \ref{section-model} outlines the model setup for pre-GenAI organizations. Sections \ref{section-w-auto} and \ref{section-w-aug} analyze worker-level automation and augmentation, respectively. Section \ref{section-expert-adoption} examines GenAI adoption at the expert level. We conclude in Section \ref{section-conclusion}. All proofs and supplemental materials are presented in the appendices.

\section{Literature Review}\label{section-literature}
Our work contributes to three research streams: the economics of organizational design and knowledge-based hierarchies, the literature on AI's labor market and economic impacts, and related work on human--AI collaboration.

Our work builds most directly on the foundational literature examining how organizations structure knowledge work and allocate problem-solving across hierarchical layers.\footnote{The relationship between technology adoption, skill demand, and organizational structure is a foundational inquiry in economics. Traditional frameworks---the skill-biased technical change hypothesis \citep{katz1992changes, autor1998computing}, the routine-biased technical change refinement \citep{autor2003skill, goos2007lousy}, and the task-based framework \citep{acemoglu2011skills}---generally predict uniform directional effects for a given technology, typically upskilling or the displacement of routine tasks.} \citet{garicano2000hierarchies} establishes the seminal framework in which hierarchy emerges endogenously to economize on knowledge acquisition costs: workers handle routine problems independently while escalating exceptional cases to more knowledgeable experts, and the optimal span of control balances the benefits of knowledge specialization against communication costs. This framework has been extended to incorporate product positioning \citep{wu2015organizational}, heterogeneous communication technologies \citep{gumpert2018organization}, and variations in decision complexity \citep{bloom2014distinct}. \citet{garicano2006organization} and \citet{garicano2015knowledge} provide comprehensive surveys. 

A key insight from this literature is that technologies altering communication costs can shift the balance between centralized and decentralized problem-solving, thereby reshaping spans of control \citep{bloom2014distinct}. However, these earlier information and communication technologies (ICTs) primarily reduced coordination and communication frictions without substituting for cognitive labor or introducing systematic unreliability. GenAI differs on both dimensions: it can autonomously generate substantive task outputs, yet does so with non-trivial error risk that requires human validation.

The paper most closely related to ours is \citet{ide2025artificial}, which embeds AI in a competitive economy of knowledge hierarchies, modeling AI as agents with an exogenously fixed knowledge level and showing that equilibrium outcomes hinge on whether agents act autonomously as ``co-workers'' or only as advisory ``co-pilots.'' Our paper shares this knowledge-hierarchy foundation but introduces a mechanism absent from perfect-substitute agent models: GenAI's intrinsic fallibility creates an endogenous validation-and-escalation workload that reshapes workforce predictions. This mechanism yields workforce predictions---specifically, how entry-level skill thresholds and spans of control respond to deployment mode and location---that do not emerge from frameworks treating AI as an error-free additional layer of knowledge.

Our theoretical predictions connect to a rapidly growing empirical literature documenting AI's heterogeneous effects on workforce composition and job quality. Early research on pre-GenAI automation technologies established that AI adoption can simultaneously displace workers in routine cognitive tasks while increasing demand for workers in complementary roles \citep{acemoglu2019automation, acemoglu2023rebalancing, autor2024applying}. Recent empirical studies of GenAI adoption document substantial productivity gains but reveal considerable heterogeneity across skill levels and contexts. \citet{noy2023experimental} and \citet{brynjolfsson2025generative} find that GenAI tools substantially accelerate task completion, with particularly pronounced benefits for initially lower-performing workers. \citet{peng2023impact} and \citet{cui2024effects} document that GitHub Copilot enables developers to code significantly faster, though with quality concerns that require heightened code review. \citet{dell2023navigating} show that consultants using GenAI exhibit significant productivity improvements but face what they term a ``jagged frontier'': some tasks benefit dramatically while others see minimal gains or quality degradation.

Two recent large-scale studies provide convergent evidence of GenAI's seniority-biased labor market effects that are particularly relevant to our paper. \citet{brynjolfsson2025canaries} report that early-career employment declined by approximately 13\% in the most AI-exposed occupations after 2022, while senior employment in those occupations remained stable or rose. 
Complementing this reduced-form evidence with a firm-level adoption design, \citet{hosseini2025generative} use direct measures of GenAI adoption and show that adopting firms reduce junior employment by roughly 9--10\% within six quarters while senior employment remains largely unchanged, with effects concentrated in high-exposure occupations and driven primarily by reduced hiring rather than separations. Taken together, these papers establish a robust empirical regularity---entry-level opportunities can contract even as senior roles remain stable---but they are not designed to pin down why the effects differ across contexts. Our paper provides the missing organizational mechanism by endogenizing deployment architecture within the firm: not only whether GenAI is used for automation versus augmentation, but also where it is deployed in the hierarchy and who bears the validation and rework burden created by intrinsic fallibility. This framework offers a lens for interpreting the seniority-biased pattern by tracing how deployment architecture shapes entry-level skill requirements and spans of control.

Lastly, our analysis relates to a growing literature examining how human workers and AI systems interact. \citet{fugener2021will} study ``humans-in-the-loop'' configurations where workers oversee algorithmic decision-making, documenting both benefits and potential cognitive pitfalls. \citet{wang2024friend} find that less experienced workers benefit more from AI assistance, a pattern consistent with our augmentation mode. \citet{bastani2021improving} and \citet{ibrahim2021eliciting} examine optimal designs for human--algorithm collaboration in prediction and decision tasks. More broadly, recent work has begun to investigate how the modality and structure of human--AI collaboration shape creative and knowledge-intensive outcomes \citep[e.g.,][]{chen2024large, hou2025double, lu2025}. \citet{bastani2025human} show that as AI errors become rarer, motivating human vigilance can become paradoxically more expensive. \cite{zhang2025incentivizing} study a platform setting in which human creators can delegate tasks to AI and show that hidden effort distortions can induce excessive delegation. \citet{dai2026designing} further show that enterprise GenAI design choices (e.g., temperature) interact with hidden human effort under moral hazard to endogenously shape the optimal hallucination--creativity trade-off. \cite{siderius2026use} study a dynamic principal-agent model in which reliance on organizational AI can erode worker expertise over time, showing that incentive frictions can cause firms to underinvest in oversight and tolerate skill atrophy. Together, these studies reinforce that organizations' skill and oversight demands depend as much on how GenAI is deployed and governed as on its raw capability. Our work complements this literature by embedding deployment flexibility within a knowledge-based hierarchy where intrinsic fallibility makes validation a binding organizational constraint, showing that the same GenAI technology can produce opposite workforce effects depending on where and how it is deployed.

\section{Benchmark: Pre-GenAI Organizational Structure}\label{section-model}
We construct a parsimonious model of knowledge-based hierarchy and establish the pre-GenAI organizational structure as a benchmark. We consider a two-layer firm consisting of low-level junior workers and high-level senior experts. While our model formally depicts a two-layer hierarchy, the ``senior expert" role structurally proxies for middle-management functions centered on exception handling, quality assurance, and oversight. The firm faces one problem per unit of time, a normalization that captures a steady-state workflow without loss of generality. The difficulty of each problem $d$ is ex-ante unknown and uniformly distributed over the interval $[0, 1]$. If the problem is solved, the firm can produce one unit of output; otherwise, no output is produced. For simplicity, we normalize the value of this output to 1. We assume that each worker can process one problem per unit of time. Therefore, the firm employs one worker and $n$ experts, which gives a \textit{span of control} (the ratio of workers to experts) of $s=\frac{1}{n}$.

The labor market consists of individuals with varying levels of expertise, quantified by their \textit{knowledge level} $x \in [0, 1]$, which represents an individual's ability to solve problems up to difficulty $x$. The firm perfectly observes each candidate's knowledge level and, in an efficient labor market, selects the precise knowledge levels needed to maximize profitability. 

The worker is employed to independently handle the easier problems and seeks assistance for more complex ones. Specifically, a worker with knowledge level $x$ independently solves problems with difficulty $d\in[0, x]$, where $0\le x\le 1$. If the problem's difficulty exceeds the worker's knowledge level, i.e., $d\in(x, 1]$, the worker requests help from experts, who then provide solutions, allowing the worker to handle the problem and produce one unit of output. Importantly, the worker remains the sole producer of output. For problems exceeding their knowledge boundary, the worker relies on expert consultation to obtain the solution rather than delegating the task itself to the expert for execution.  

Following \cite{bloom2014distinct}, we assume that experts possess complete knowledge over the entire difficulty spectrum $[0, 1]$. In the context of a two-layer hierarchy, this assumption simply implies that all problems under consideration fall within the firm's collective problem-solving capacity, so that escalations represent transfers of work upward rather than unresolved uncertainty. Importantly, it ensures that changes in organizational structure are driven entirely by endogenous reorganization rather than by limits to top-layer competence, cleanly isolating our focal mechanism: how GenAI's fallibility reallocates verification and escalation burdens across layers. We relax this assumption in Appendix \ref{app-expert} and show that our insights remain robust.

Each time a worker consults an expert, the communication process consumes $t_c$ units of the expert's time, where $t_c>0$ is the \textit{communication cost}. Since the worker escalates problems with probability $1-x$ and each consultation takes $t_c$ units of expert time, the firm requires $n_0=(1-x)t_c$ experts. We assume that the communication cost $t_c$ lies within the range $(0, \overline{t}_c)$, where $\overline{t}_c=\frac{2k}{k+2w}$. If $t_c$ reaches or exceeds $\overline{t}_c$, the communication cost becomes high enough that the firm would find it more profitable to employ a worker with complete knowledge, eliminating the need for expert support. 

Drawing on the literature on hierarchical organization in the knowledge economy \citep[e.g.,][]{garicano2000hierarchies, antras2006offshoring}, we assume that the firm pays an individual with knowledge level $x$ a wage of $w + \frac{1}{2}kx^2$, where $w$ is the market price of a unit of work time and $\frac{1}{2}kx^2$ represents the compensation for the individual's knowledge, with $k$ as the knowledge premium.\footnote{This wage structure reflects the convex cost of acquiring knowledge at increasing levels. To satisfy participation constraints, the firm aims to set wages that align with the costs employees incur to attain their knowledge levels.}  
The firm's profit function is thus:
\begin{equation}\label{equ-profit-ben}
    \Pi_0(x)=1-\left(w+\frac{1}{2}kx^2\right)-\left(w+\frac{1}{2}k\right)(1-x)t_c.
\end{equation}
The presence of experts with comprehensive knowledge ensures that any problems the firm encounters can be resolved, allowing the firm to maintain one unit of output per unit of time. The term $\left(w+\frac{1}{2}kx^2\right)$ represents the wage the firm must pay to hire one worker with knowledge level $x$, while $\left(w+\frac{1}{2}k\right)(1-x)t_c$ reflects the total wage required to employ $(1-x)t_c$ experts. Here, since the expert's knowledge level is 1, the firm pays each expert a wage of $w+\frac{1}{2}k$. The firm's objective is to determine the optimal knowledge level $x^*$ for its workers to maximize profit per unit of time.

The following Lemma \ref{lemma-noGenAI} reveals the firm's optimal organizational structure in the pre-GenAI environment. Throughout this paper, ``organizational structure'' refers specifically to two endogenous outcomes: the optimal knowledge level of entry-level workers ($x^*$), which determines workforce composition, and the span of control ($s^* = 1/n$), which captures hierarchical shape.
\begin{lemma}\label{lemma-noGenAI}
 In the pre-GenAI organization, the optimal worker knowledge level is $x_0^*=\frac{(k+2w)t_c}{2k}$ and the span of control is $s_0^*=\frac{1}{(1-x_0^*)t_c}$.
\end{lemma}

Organizations face a fundamental trade-off when determining the optimal knowledge level of workers. While lower knowledge requirements reduce worker wage costs, they simultaneously increase the need for worker-expert communications, necessitating more senior positions. As shown in Lemma \ref{lemma-noGenAI}, the optimal worker knowledge level $x_0^*$ responds differently to changes in knowledge premium $k$ and communication costs $t_c$. A lower knowledge premium reduces the cost of hiring more knowledgeable workers, increasing worker autonomy. Conversely, lower communication costs make it more efficient to rely on managerial consultation. This finding aligns with previous research by \cite{garicano2000hierarchies} and \cite{bloom2014distinct} on how information and communication technology (ICT) development influences organizational hierarchies. 

\section{Worker-Level Automation}\label{section-w-auto}
Having established the pre-GenAI benchmark, we now examine how the four GenAI deployment modes reshape this organizational structure. In this section, we begin with a common deployment mode in which GenAI autonomously executes low-level knowledge tasks that would otherwise be performed by junior workers. We refer to this mode as \textit{worker-level automation} (denoted by subscript $t$).
Specifically, GenAI automatically completes tasks in the interval $[0, r_t]$, where $0<r_t<1$ represents GenAI's \textit{automation capability}, while the remaining tasks in $(r_t, 1]$ are handled by workers (see Figure \ref{illustration-worker-auto}). 

\begin{figure}[ht]
\centering \footnotesize
\quad \quad
\begin{tikzpicture}[scale=1.2]
\linespread{0.8} 
\foreach \x in {0,3,6,9}{
    \draw (\x cm,5pt) -- (\x cm,-5pt);
}
\node[align=center] at (3,0.5) {$r_t$};
\node[align=center] at (6,0.5) {$x$};
\node[align=center] at (9,0.5) {1};
\node[align=center] at (0,0.5) {0};

\fill[yellow!50] (0.01,-0.16) rectangle (2.99,0.16);
\draw [thick,->] (0,0) -- (9.3,0)node[right]{\text{Tasks}};
\draw [decorate,decoration={brace,amplitude=5pt,mirror,raise=2ex}] (0,0) -- (2.95,0) node[midway,yshift=-2.2em]{GenAI Automation};
\node[align=center] at (1.5,-1) {\textcolor{gray}{(Worker Validation)}};
\draw [decorate,decoration={brace,amplitude=5pt,mirror,raise=2ex}] (3.05,0) -- (5.95,0) node[midway,yshift=-2.3em]{Worker};
\draw [decorate,decoration={brace,amplitude=5pt,mirror,raise=2ex}] (6.05,0) -- (8.95,0) node[midway,yshift=-2.3em]{Worker+Expert};
\end{tikzpicture}
\caption{Delegation of tasks under worker-level automation}
\label{illustration-worker-auto}
\end{figure}

To formalize the presence of hallucinations, we assume that each GenAI-automated task is solved incorrectly with probability $h$, where $0<h<1$ denotes the \textit{hallucination rate}. Because erroneous AI outputs may trigger losses far exceeding the value of the task itself, such as reputational damage or strained business relationships, we assume that the firm mandates human validation for all AI-processed tasks to identify and resolve potential hallucinations. We treat the hallucination rate $h$ as an exogenous primitive independent of GenAI's capability $r_t$. This reflects growing evidence that improvements in model scale or capability do not necessarily translate into lower error rates, and that more advanced language models may exhibit equal or even higher hallucination rates across tasks \citep{zhou2024larger}. In Appendix \ref{app-extension-rh}, we show that our main results remain robust when $h$ is inversely correlated with $r_t$.

Since GenAI automates tasks in the interval $[0, r_t]$, the firm requires only $(1-r_t)$ workers per unit of time to process the remaining tasks in $(r_t, 1]$. To mitigate the potentially severe losses caused by hallucinations, workers are also responsible for validating all AI-completed outputs and reworking tasks whenever hallucinations are detected. Consequently, the firm must hire workers whose knowledge level satisfies $x\ge r_t$. We assume that each validation consumes $t_v$ units of worker time, while reworking a task following a detected hallucination consumes $t_r$ units of worker time. We impose $t_v<1$ and $t_r<1$, implying that either validation or rework alone is less time-consuming than manual task execution (normalized to one). However, we assume $t_v+t_r>1$, so that when a hallucination occurs, the combined effort of validation and rework strictly exceeds the time required for manual processing. This condition captures the realistic scenario where catching and correcting an AI error is more costly than handling the task manually from the outset, which is precisely what makes the hallucination risk organizationally consequential. Absent this condition (i.e., if $t_v+t_r\le1$), GenAI adoption would always be optimal regardless of the hallucination magnitude. Therefore, after GenAI adoption, the firm's profit function is given by 
\begin{equation}\label{equ-profit-w-auto}
    \Pi_t(x)=1-\left(w+\frac{1}{2}kx^2\right)[(1-r_t)+(t_v+ht_r)r_t]-\left(w+\frac{1}{2}k\right)(1-x)t_c.
\end{equation}

Since worker validation eliminates the impact of hallucinations on the final output, the firm's output per unit of time remains normalized to one. The total worker time requirement $[(1-r_t)+(t_v+ht_r)r_t]$ corresponds to the number of workers the firm must employ per unit of time. The term $[(1-r_t)+(t_v+ht_r)r_t]$ consists of three components: $(1-r_t)$ units of time devoted to non-automated tasks; $t_v r_t$ units allocated to validating AI-completed tasks; and an expected rework time of $h t_r r_t$ when hallucinations occur. By contrast, the demand for senior experts remains at $(1-x)t_c$, as the probability that a worker encounters a problem beyond their knowledge level is unchanged at $1-x$, with each consultation requiring $t_c$ units of expert time.\footnote{Our baseline treats capability $r_t$ and hallucination rate $h$ as exogenous and abstracts from direct monetary adoption costs (e.g., licensing, integration) to isolate the organizational mechanism. The organizational costs of GenAI integration, including validation, rework, and consultation burdens, are modeled explicitly. Appendix \ref{app-cost} endogenizes capability and reliability via convex investment costs and confirms that our main insights are robust.} 
\subsection{Impact of GenAI Advancements on Worker Knowledge}
We first examine when firms should implement worker-level automation and how GenAI technology advancements reshape organizational knowledge hierarchies.
\begin{lemma}\label{lemma-adoption-wauto}
The firm should adopt worker-level automation if and only if the hallucination rate satisfies $h<\overline{h}_t$, where $\overline{h}_t=\frac{1-t_v}{t_r}$ for $r_t\le x_0^*$ and $\overline{h}_t<\frac{1-t_v}{t_r}$ for $r_t>x_0^*$.    
\end{lemma}

Lemma \ref{lemma-adoption-wauto} establishes that the firm's adoption of worker-level automation is contingent upon a critical hallucination threshold, $\overline{h}_t$. When GenAI capability is relatively low ($r_t\le x_0^*$), the threshold is given by $\overline{h}_t=\frac{1-t_v}{t_r}$. 
This condition ensures that automating tasks in $[0, r_t]$ yields a net reduction in worker demand: the labor saved exceeds the validation and rework burden, i.e., $r_t > (t_v + h t_r) r_t$. However, as GenAI capability expands beyond the firm's pre-adoption worker knowledge level ($r_t>x_0^*$), the adoption decision becomes more restrictive. While higher $r_t$ enables the automation of increasingly complex tasks, it simultaneously requires that the firm employ workers with a knowledge level at least equal to $r_t$ to effectively validate AI-generated outputs. This requirement inflates worker wage costs, making adoption more sensitive to inefficiencies caused by hallucinations. Consequently, the hallucination threshold for adoption tightens as GenAI becomes more capable ($\overline{h}_t<\frac{1-t_v}{t_r}$). This result highlights that GenAI adoption is governed by the strategic interaction between the automation capability and the resulting organizational burden of human oversight.

\begin{proposition}\label{prop-workerauto-x}
The optimal worker knowledge level under worker-level automation is $x_t^*=\max\{r_t,\min\{1,x_t\}\}$, where $x_t=\frac{x_0^*}{(1-r_t)+(t_v+ht_r)r_t}$. Moreover, $x_t^*$ increases in GenAI's automation capability $r_t$ and decreases in hallucination rate $h$, and $x_t^*>x_0^*$.
\end{proposition}

Proposition \ref{prop-workerauto-x} shows that worker-level automation drives an \textit{upskilling} effect: the firm optimally raises entry-level skill requirements after GenAI adoption ($x^*_t > x^*_0$).\footnote{Throughout the paper, ``upskilling" and ``deskilling" refer specifically to changes in the entry-level knowledge threshold $x^*$: upskilling means the firm raises the worker knowledge requirement, deskilling means it lowers it. These terms do not characterize overall skill demand in the firm, as changes in span of control can simultaneously move demand for senior expert roles in the opposite direction.} The mechanism operates through a cost-ratio channel distinct from traditional skill-biased technical change. Automation reduces the firm's total worker-time requirement from $1$ to $(1-r_t) + (t_v + ht_r)r_t$, lowering the effective cost of worker knowledge per unit of output. The marginal cost of raising $x$, captured by the convex knowledge premium $\frac{1}{2}kx^2$, is now borne by fewer worker units, while the marginal benefit from reducing expert consultations remains unchanged at $(w+\frac{1}{2}k)t_c$ per unit of escalation probability. As a result, the firm's optimal worker knowledge level tilts upward. Intuitively, when automation handles routine execution, each unit of worker knowledge is leveraged over a smaller labor base, making it cost-effective for the firm to invest in higher-quality frontline labor that reduces costly upward referrals. 

This upskilling intensifies as GenAI's automation capability $r_t$ increases, which further reduces the worker-time base over which knowledge costs are amortized. Furthermore, a decline in the hallucination rate $h$ strengthens the upskilling effect: a lower $h$ reduces expected rework time per automated task, reducing total worker demand and lowering the marginal cost of knowledge investment. 
This comparative static may seem paradoxical: lower error rates lead to higher, not lower, skill requirements. The driver is the cost channel: more reliable automation requires fewer worker-hours, making the convex knowledge premium less expensive to bear, so the firm responds by raising the skill threshold.

\begin{figure}[h]
    \centering
    \subfloat[$h=0.4$]{
    \includegraphics[width=0.4\textwidth]{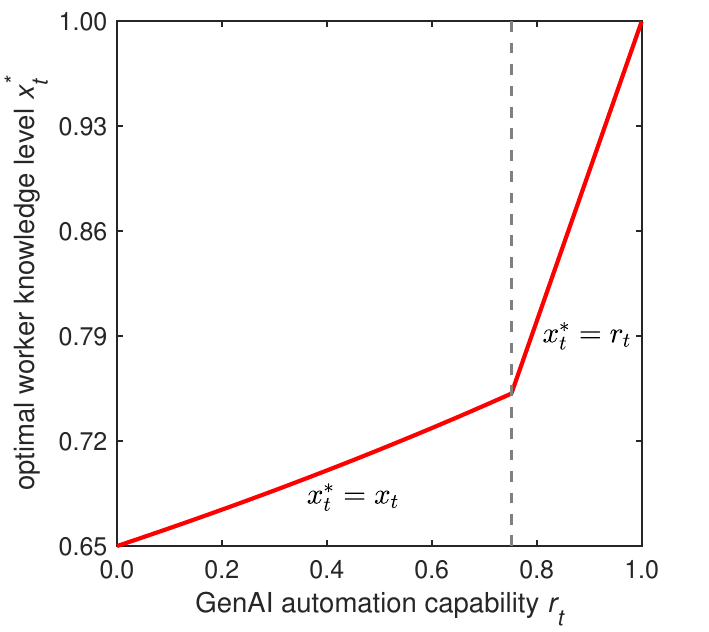}{\label{fig-w-auto-largeh}}
    }
    \quad\quad
    \subfloat[$h=0.1$]{
    \includegraphics[width=0.4\textwidth]{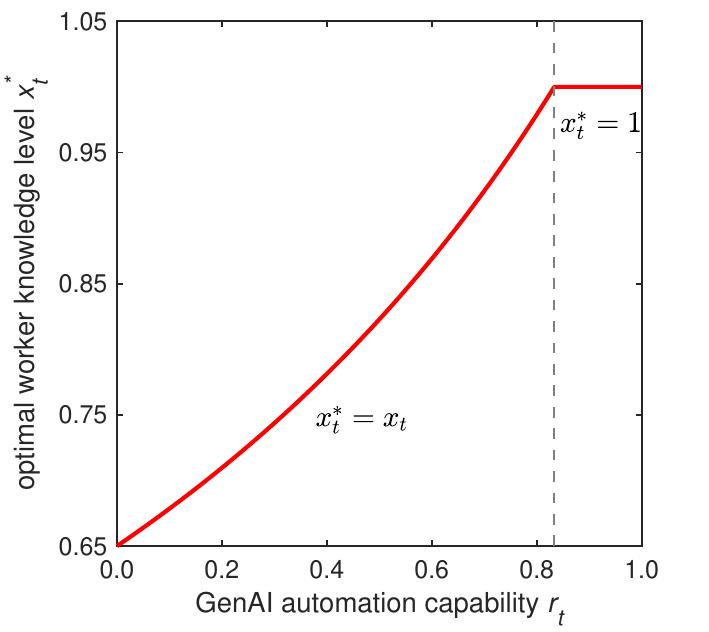}{\label{fig-w-auto-smallh}}
    }
    \caption{Impact of GenAI capability enhancement on worker knowledge level under worker-level automation \\
    ($k=0.8$, $w=0.25$, $t_c=0.8$, $t_v=0.5$, $t_r=0.8$)}
    \label{fig-w-auto-x}
\end{figure}

Beyond this cost-ratio channel, the validation mandate introduces a direct skill-floor effect: because workers must be capable of detecting errors in all AI-automated tasks, the firm is constrained to hire workers with $x \geq r_t$. Figure \ref{fig-w-auto-x} illustrates how these two channels interact to produce distinct upskilling trajectories depending on GenAI's reliability. Under moderate hallucination rates (Figure \ref{fig-w-auto-largeh}), as GenAI capability increases, the optimal worker knowledge level shifts from the unconstrained optimum to exactly matching GenAI's automation capability ($x_t^*=r_t$). The skill-floor constraint becomes binding: the labor effort required to validate and rework AI-generated outputs is substantial enough that hiring workers with knowledge beyond $r_t$ would incur prohibitively high wage costs relative to the marginal benefit. The firm therefore sets the worker knowledge level at the minimum threshold required for validation. In this sense, the validation requirement amplifies automation-induced upskilling: expanding the automation scope enlarges the set of tasks that workers must be able to verify and correct. Conversely, in a high-reliability environment (Figure \ref{fig-w-auto-smallh}), reliable GenAI automation drives the optimal worker knowledge level to its upper bound of one. The traditional hierarchy collapses into a single-layer organization of fully knowledgeable workers.


Our theoretical finding that worker-level automation induces an upskilling effect aligns closely with the empirical evidence discussed earlier \citep{brynjolfsson2025canaries, hosseini2025generative}: the documented decline in junior employment among GenAI-adopting firms is consistent with our mechanism: automating routine execution reduces demand for low-skilled executors, prompting firms to hire fewer but more knowledgeable workers who oversee AI outputs and reduce costly escalations. 

This mechanism differs sharply from traditional automation-driven upskilling \citep[e.g.,][]{acemoglu2019automation}. In those frameworks, automation eliminates routine tasks from workers' portfolios, and upskilling arises because the remaining tasks are inherently more complex. Under GenAI, by contrast, automated tasks do not leave the worker's domain of responsibility: they return as a validation tax that the worker must pay on every AI-executed task. The upskilling imperative arises not from the removal of routine work but from the oversight of AI-executed work, driven by intrinsic fallibility. Furthermore, as we show in Section \ref{section-w-aug}, this upskilling outcome is specific to the automation mode; the same technology deployed for augmentation produces the opposite effect. This mode-dependence has no counterpart in traditional automation theories, which predict a uniform directional skill effect for a given technology.

\subsection{Impact of GenAI Advancements on Organizational Structure}\label{section-wauto-s}
We next examine how advancements in GenAI technology reshape organizational structure under worker-level automation. In particular, we study how improvements along two fundamental dimensions of GenAI (its capability to automate a broader range of tasks and its reliability in generating correct outputs) affect the firm's managerial span of control, defined as the ratio of junior workers to senior experts.
\begin{proposition}\label{prop-workerauto-s}
After the adoption of worker-level automation, 
\begin{enumerate}[$(a)$]
    \item When $h\ge h_0$ or $r_t<r_0$, the optimal span of control is given by $s_t^*=\frac{1-r_t+(t_v+ht_r)r_t}{(1-x_t^*)t_c}$. Furthermore, the span of control $s_t^*$ 
\begin{enumerate}[$(1)$]
    \item decreases in automation capability $r_t$ on $(0,r_1)$ and increases in $r_t$ thereafter;

    \item decreases in hallucination rate $h$ on $(0,h_1)$ and increases in $h$ thereafter.
\end{enumerate}

 \item Otherwise, when $h<h_0$ and $r_t\ge r_0$, the traditional two-layer organizational hierarchy collapses into a single-layer structure composed solely of workers with complete knowledge, i.e., $x_t^*=1$.
\end{enumerate}
\end{proposition}

\begin{figure}[h]
    \centering
    \subfloat[$h=0.1$]{
    \includegraphics[width=0.4\textwidth]{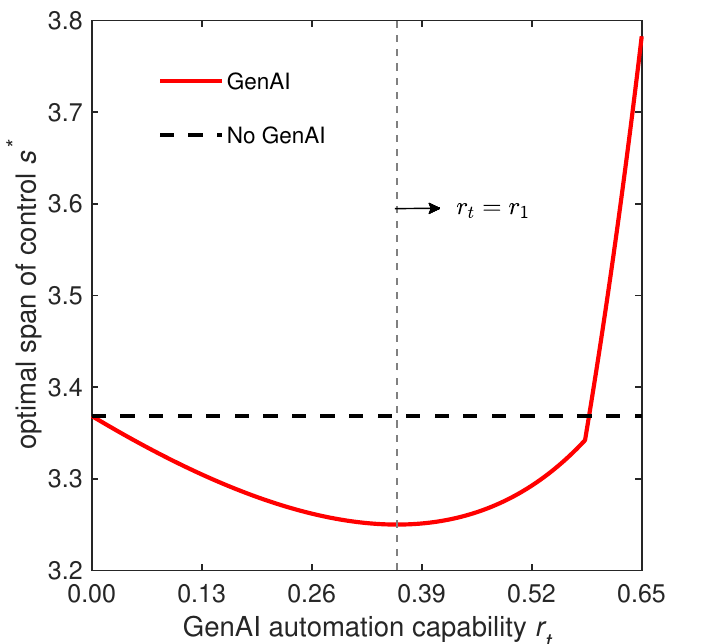}{\label{fig-w-auto-spanr}}
    }
    \quad\quad
    \subfloat[$r_t=0.4$]{
    \includegraphics[width=0.4\textwidth]{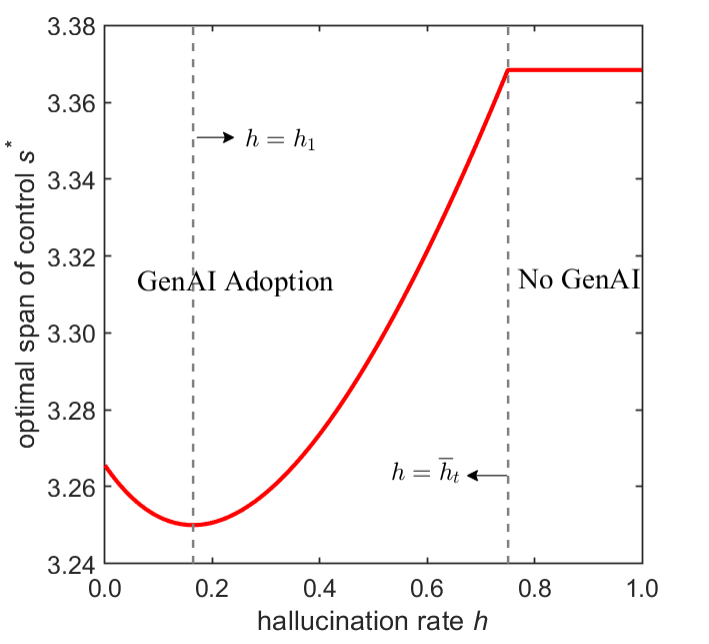}{\label{fig-w-auto-spanh}}
    }
    \caption{Impact of GenAI advancements on span of control under worker-level automation \\
    ($k=0.8$, $w=0.25$, $t_c=0.5$, $t_v=0.4$, $t_r=0.8$)}
    \label{fig-w-auto-s}
\end{figure}

Proposition \ref{prop-workerauto-s} demonstrates a non-monotonic evolution of organizational structure in response to GenAI advancements. Specifically, the optimal span of control exhibits a distinct ``U-shaped" trajectory as the automation capability $r_t$ increases (see Figure \ref{fig-w-auto-spanr}).  
This dynamic stems from the shifting dominance of two competing forces: a direct \textit{task-automating effect} and an indirect \textit{organization-upskilling effect}. When automation capability is limited (i.e., at low $r_t$), the task-automating effect prevails: since the modest reduction in worker demand provides little incentive for upskilling, increases in $r_t$ primarily contract the worker headcount while expert demand remains stable, narrowing the span of control. Conversely, at high $r_t$, the \textit{organization-upskilling effect} takes precedence. As the marginal labor savings from further automation diminish within a significantly contracted workforce, the firm shifts its adjustment margin toward worker knowledge. To minimize costly referrals, the firm substantially elevates the worker knowledge level. As a result, the reduction in expert demand outpaces the decline in worker headcount, thereby widening the span of control. Notably, as illustrated in Figure \ref{fig-w-auto-spanr}, the span of control surges once $r_t$ reaches a sufficiently high level. This is because the firm raises the worker knowledge threshold to $x_t^*=r_t$ to ensure validation and correction of increasingly complex AI outputs. Such intense upskilling effectively internalizes the vast majority of problem-solving within the worker layer, causing the demand for senior experts to collapse and leading to a rapid flattening of the organization.

A similar non-monotonic pattern can arise as GenAI becomes more reliable (Figure \ref{fig-w-auto-spanh}), although the underlying mechanism is subtly different from that for capability improvements. When hallucination rates are relatively high ($h>h_1$), a marginal reduction in $h$ mainly alleviates workers' correction burden on AI-automated tasks. We refer to this channel as the direct \textit{correction-reduction effect}. This effect reduces worker labor demand more than expert demand, thereby narrowing the span of control. Once hallucinations become sufficiently rare ($h<h_1$), further reliability gains operate primarily through the organization-upskilling channel: workers can internalize more problem solving, reduce referrals, and drive a faster contraction of expert demand, which widens the span of control. Importantly, the initial contraction at high hallucination rates hinges on the presence of worker oversight. Without these frictions, improved reliability would chiefly amplify the upskilling channel that substitutes away from experts, so decreases in $h$ would increase the span of control monotonically rather than producing an initial decline.

Finally, Proposition \ref{prop-workerauto-s}(b) identifies an extreme organizational regime. When GenAI is both highly capable and sufficiently reliable, the upskilling effect becomes so strong that the firm optimally hires workers with complete knowledge ($x_t^*=1$). In this case, workers can independently handle all non-automated tasks while validating GenAI outputs, eliminating the need for expert assistance. The traditional two-layer hierarchy therefore collapses into a single-layer organizational structure composed solely of fully knowledgeable workers. 

Overall, worker-level automation induces a staged reconfiguration of organizational structure as automation expands. At low levels of automation, the hierarchy becomes steeper as routine tasks are automated while expert roles remain intact. As automation deepens, firms respond by upgrading worker knowledge, compressing the expert layer and flattening the organization. In the limit, sufficiently capable and reliable GenAI renders expert assistance unnecessary, collapsing the hierarchy into a single-layer organization of fully knowledgeable task executors.

\section{Worker-Level Augmentation}\label{section-w-aug}
While worker-level automation centers on substituting human effort in routine execution, GenAI can also reshape organizations through a fundamentally different mechanism: it is embedded directly into workers' workflows to expand their effective problem-solving capabilities beyond their intrinsic skill endowments. In this deployment mode, GenAI operates alongside workers during task execution, providing real-time suggestions and refinements that workers assess and incorporate, so that outcomes are jointly produced rather than autonomously generated. Conceptually, this pattern resembles a copilot-style use of GenAI, in which the technology augments workers' capabilities without displacing their role in the production process. We refer to this capability-amplifying deployment as our second deployment mode, \textit{worker-level augmentation} (denoted by subscript $g$).

Under worker-level augmentation, workers integrate GenAI into their task execution process and operate as a tightly coupled human--AI collaborative unit. This in-workflow integration expands a worker's effective capability boundary beyond their intrinsic knowledge level $x$, but the magnitude of this expansion depends jointly on the worker's own skill and the augmentation capability of GenAI.\footnote{Beyond expanding the capability boundary, worker--AI collaboration may also improve execution efficiency for tasks that already fall within the worker's intrinsic knowledge range $[0, x]$. We formally incorporate such a productivity improvement in Appendix \ref{app-productivity}.} Specifically, we assume that an augmented worker can resolve problems within the difficulty range $[0, x+r_g(1-x)]$, where $r_g\in(0,1)$ denotes the GenAI's augmentation capability. Problems with difficulty exceeding this augmented boundary, i.e., $d\in (x+r_g(1-x), 1]$, still require upward consultation with senior experts (see Figure \ref{illustration-w-aug}).

\begin{figure}[ht]
\centering \footnotesize
\quad \quad
\begin{tikzpicture}[scale=1.2]
\linespread{0.8} 
\foreach \x in {0,3,6,9}{
    \draw (\x cm,5pt) -- (\x cm,-5pt);
}
\node[align=center] at (3,0.5) {$x$};
\node[align=center] at (6,0.5) {$x+r_g(1-x)$};
\node[align=center] at (9,0.5) {1};
\node[align=center] at (0,0.5) {0};

\fill[yellow!50] (3.01,-0.16) rectangle (5.99,0.16);
\draw [thick,->] (0,0) -- (9.3,0)node[right]{\text{Tasks}};
\draw [decorate,decoration={brace,amplitude=5pt,raise=6.5ex}] (0,0) -- (5.95,0) node[midway,yshift=4em]{Worker $+$ GenAI};
\draw [decorate,decoration={brace,amplitude=5pt,mirror,raise=2ex}] (3.05,0) -- (5.95,0) node[midway,yshift=-2.3em]{GenAI Augmentation};
\node[align=center] at (4.5,-1) {\textcolor{gray}{(Expert Validation)}};
\draw [decorate,decoration={brace,amplitude=5pt,mirror,raise=2ex}] (6.05,0) -- (8.95,0) node[midway,yshift=-2.3em]{Worker+Expert};
\end{tikzpicture}
\caption{Delegation of tasks under worker-level augmentation}
\label{illustration-w-aug}
\end{figure}

This formulation captures a key feature of the copilot-style augmentation: the benefits of GenAI are inherently heterogeneous across workers. As a worker's intrinsic knowledge level $x$ increases, the incremental value of GenAI assistance diminishes, because highly skilled workers already possess much of the relevant problem-solving capability. Conversely, less-skilled workers experience a disproportionately larger expansion of their effective capability boundary from the same augmentation technology. In this sense, worker-level augmentation exhibits an equalizing effect, consistent with empirical evidence that GenAI disproportionately enhances the performance of lower-skilled workers \citep{noy2023experimental, brynjolfsson2025generative, caplin2025abcs}.

However, this expansion of capability boundaries introduces a critical validation challenge. For tasks in the augmented interval $(x, x+r_g(1-x)]$, problem difficulty exceeds the worker's intrinsic knowledge level, preventing the worker from independently verifying the accuracy of the human--AI collaborative output. As in the automation setting, we assume that solutions in this interval are subject to hallucinations, with probability $h$ of being incorrect. To eliminate such hallucinations, senior experts must validate AI-assisted outputs. When errors are detected, experts intervene to assist workers in correcting them. Accordingly, after adopting worker-level augmentation, the firm's profit function is given by
\begin{equation}\label{profit-GenAI}
    \Pi_g(x)=1-\left(w+\frac{1}{2}kx^2\right)-\left(w+\frac{1}{2}k\right)\big[r_g(1-x)(t_v+ht_c)+(1-r_g)(1-x)t_c\big].
\end{equation}

In contrast to worker-level automation, workers remain responsible for executing all tasks within each unit of time. Therefore, the firm continues to employ one worker and incurs a total worker-hiring cost of $w + \frac{1}{2}kx^2$. The total expert time requirement, $[r_g(1-x)(t_v + ht_c) + (1-r_g)(1-x)t_c]$, determines the number of experts needed and reflects multiple managerial responsibilities. The first component, $r_g(1-x)(t_v + ht_c)$, captures the oversight burden associated with human--AI collaboration within the augmented task range: experts devote $r_g(1-x)t_v$ units of time to validating AI-assisted outputs and incur an additional expected time cost of $r_g(1-x)ht_c$ to assist workers in correcting hallucination-related errors. The second component, $(1-r_g)(1-x)t_c$, represents conventional expert intervention for tasks that exceed the combined capabilities of the worker and GenAI, i.e., those in the interval $(x+r_g(1-x), 1]$. We further assume that the worker's hallucination-induced rework cost is negligible under augmentation, since the worker is already engaged in execution and the primary rework friction is the expert-worker communication time.

\subsection{Impact of GenAI Advancements}\label{subsection-waug}
\begin{lemma}\label{lemma-w-aug}
    The firm should adopt worker-level augmentation if and only if the hallucination rate satisfies $h<\overline{h}_g=1-\frac{t_v}{t_c}$.
\end{lemma}

Lemma \ref{lemma-w-aug} shows that firms should adopt worker-level  augmentation only when the hallucination rate $h$ lies below a critical threshold $\overline{h}_g$. Notably, this threshold is determined solely by validation and communication costs. The condition $h < \overline{h}_g$ can be rewritten as $r_g(1-x)(t_v + ht_c) < r_g(1-x)t_c$. The right-hand side of this inequality can be interpreted as the expert time when the worker directly consults them about problems within the difficulty range $(x, x+r_g(1-x)]$. The left-hand side represents experts' working time for the same problems under GenAI augmentation, specifically the time spent validating AI-assisted outputs and correcting hallucination-induced errors. Accordingly, worker-level augmentation is profitable to firms if and only if the expert time devoted to mitigating hallucinations is less than the time required for direct assistance in problem-solving. Unlike worker-level automation, where adoption hinges mainly on the net change in workers' workload, augmentation is pinned down by experts' validation and communication efficiency. The reason is that augmentation shifts the oversight burden from workers to experts.

\begin{proposition}\label{prop-w-aug-equ}
The optimal worker knowledge level under worker-level augmentation is $ x_g^*=\frac{(k+2w)[r_gt_v+(1+r_gh-r_g)t_c]}{2k}$. Moreover, $x_g^*$ decreases in GenAI's augmentation capability $r_g$ and increases in hallucination rate $h$, and $x_g^*<x_0^*$.

\end{proposition} 

Proposition \ref{prop-w-aug-equ} characterizes how worker-level augmentation reshapes the firm's optimal skill demand. The central result is a \textit{deskilling effect}: as firms deploy GenAI to augment frontline workers, they optimally reduce the intrinsic knowledge required of entry-level employees ($x_g^*<x_0^*$). 
Under augmentation, GenAI operates as a capability amplifier that extends a worker's effective problem-solving boundary beyond their own knowledge level, thereby substituting for human expertise. By enabling less specialized workers to handle problems that previously required higher skill levels, the firm can sustain operational performance while economizing on labor costs. As a result, the firm's optimal hiring strategy shifts toward workers with lower intrinsic knowledge, with GenAI filling the resulting capability gap.

The proposition further shows that this deskilling effect strengthens as GenAI technology advances. As augmentation capability increases, workers can rely more on AI support during execution, enabling them to handle a broader range of problems relative to their intrinsic knowledge and reducing the marginal value of higher human expertise. Meanwhile, a lower hallucination rate reduces the expected cost of AI-assisted work by decreasing the frequency of errors that require expert intervention. Together, stronger augmentation capability and higher reliability reinforce the deskilling effect, pushing the optimal worker knowledge level downward.

Under the skill-biased complementarity view, new technologies raise the productivity of highly skilled workers and increase firms' demand for expertise \citep{bartel2007does}. 
In contrast, our research suggests that GenAI augmentation enables a different dynamic. Rather than driving upskilling, GenAI allows for what recent research terms ``\textit{deskilling},'' where specialized human expertise is increasingly substituted by AI capabilities \citep{kinder2024generative, liu2024does}. \citet{agrawal2023turing} formalize this logic as the ``Turing Transformation'': AI equips workers to perform tasks that previously required extensive education and experience, shifting the competitive margin from human expertise to human--AI collaboration.

\begin{proposition}\label{prop-waug-span}
After the adoption of worker-level augmentation, the optimal span of control is $s_g^*=\frac{1}{(1-x_g^*)[r_g(t_v+ht_c)+(1-r_g)t_c]}$. 
\begin{enumerate}[$(a)$]
\item $s_g^*$ decreases in GenAI's augmentation capability $r_g$ on $(0,r_2)$ and increases in $r_g$ on $(r_2,1)$.

\item $s_g^*$ decreases in hallucination rate $h$ on $(0,h_2)$ and increases in $h$ on $(h_2, \overline{h}_g)$.
\end{enumerate}
 
\end{proposition}

Proposition \ref{prop-waug-span}(a) characterizes how worker-level augmentation reshapes the firm's managerial span of control as augmentation capability increases. 
Classic theories and recent empirical evidence suggest that stronger augmentation should monotonically widen the span of control, as AI-assisted workers need less expert guidance \citep{garicano2006organization, bloom2014distinct, babina2023firm, hoffmann2024generative}. Proposition~\ref{prop-waug-span} overturns this prediction: the relationship is non-monotonic. When augmentation capability is limited (i.e., $r_g<r_2$), improvements in GenAI primarily affect organizations through indirect worker skill adjustment rather than through immediate gains in worker autonomy. In this regime, even modest augmentation allows firms to lower worker knowledge requirements to reduce labor costs. As workers become less skilled, their dependence on expert guidance increases, which raises the volume of upward referrals and narrows the span of control. We refer to this channel as the indirect \textit{organization-deskilling effect}.

As augmentation capability becomes sufficiently strong (i.e., $r_g>r_2$), the dominant adjustment margin shifts. At high levels of augmentation, AI-assisted workflows substantially enhance the effective problem-solving capacity of workers, especially when intrinsic skill requirements are low. In this regime, further reductions in worker skill no longer generate large increases in expert dependence. Instead, the ability of augmented workers to resolve problems independently improves, reducing the need for expert involvement. This direct \textit{knowledge-augmenting effect} dominates, leading to a widening of the span of control.

A non-monotonic pattern also arises with respect to improvements in GenAI reliability, as described in Proposition \ref{prop-waug-span}(b). Reductions in the hallucination rate affect organizational structure through two countervailing forces. First, higher reliability lowers the expected cost of AI-assisted workflows, encouraging firms to further relax worker skill requirements to economize on wages; this amplifies the indirect \textit{organization-deskilling effect} and increases workers' reliance on expert support. Second, because experts must intervene to correct AI-induced errors, a lower hallucination rate directly reduces the time experts spend on corrections; this direct \textit{correction-reduction effect} frees managerial capacity and enables experts to supervise more workers. When hallucination rates are relatively high ($h>h_2$), reductions in $h$ mainly operate through the deskilling margin. Firms aggressively lower worker skill requirements, which increases the frequency of expert consultations and narrows the span of control. By contrast, when GenAI becomes sufficiently reliable ($h<h_2$), the correction-reduction effect dominates: expert time savings from fewer errors outweigh the incremental consultation burden from lower-skilled workers, resulting in a broader span of control. Notably, under worker-level augmentation, validation and correction are borne by experts, so the correction-reduction effect reduces expert demand, thereby widening the span of control. This contrasts sharply with worker-level automation, where this burden is borne by workers, so greater correction efficiency reduces worker demand and narrows the span of control.

\subsection{Worker-Level Comparison: Automation vs. Augmentation}
Having characterized worker-level automation and augmentation, we now compare the two deployment modes and summarize their implications for entry-level skill requirements and hierarchy. 
The central difference lies in how the two deployment modes change the knowledge required at the entry level. Under automation, GenAI reduces the amount of entry-level labor needed for task execution, which raises the marginal value of worker expertise: firms respond by hiring fewer but more knowledgeable junior workers who can validate AI outputs and limit costly upward referrals. Under augmentation, by contrast, GenAI partially substitutes for workers' problem-solving capability during execution, allowing firms to relax entry-level knowledge requirements and keep less-skilled junior workers productive with AI augmentation. This mode-dependent logic offers a parsimonious interpretation of emerging evidence: for example, \citet{brynjolfsson2025canaries} document that early-career employment declines are concentrated in occupations where GenAI is used in more automation-oriented ways, whereas entry-level employment changes are muted (and can even turn positive) in more augmentative settings. Together, the theory and evidence suggest that the observed decline in junior jobs is not an inevitable consequence of GenAI per se, but is more consistent with deployment choices that push GenAI toward automation rather than augmentation.

Despite their opposite implications for entry-level skill requirements, the two deployment modes share a common and counterintuitive prediction for hierarchy: as GenAI advances, the optimal span of control can exhibit a U-shaped trajectory along both the capability and reliability dimensions. Under automation, the span of control initially narrows because expert time remains necessary for complex exceptions, while entry-level labor demand falls as the automated scope expands or as higher reliability reduces rework frictions. As GenAI matures, stronger indirect upskilling curbs upward referrals and compresses expert demand, widening the span of control. Under augmentation, early-stage progress encourages deskilling that increases reliance on experts, contracting the span of control; once GenAI becomes sufficiently capable and reliable, firms rely less on experts for consultation and oversight, and the span of control expands. 
Importantly, this early-stage contraction under either deployment mode is consistent with empirical evidence that junior employment can contract while demand for senior expertise remains stable or even rises \citep{brynjolfsson2025canaries, hosseini2025generative}.

Finally, the two deployment modes imply different long-run endpoints. Under sufficiently capable and reliable automation, extreme upskilling can render expert oversight redundant, potentially collapsing the organization toward a single-layer structure. Under augmentation, by contrast, hierarchy persists: AI-augmented workers still require expert validation and correction when errors arise, so augmentation yields flatter but enduring hierarchies.


\section{Expert-Level GenAI Adoption}\label{section-expert-adoption}
Having examined how GenAI reshapes the organizational foundation through worker-level adoption, we now turn to deployment at the expert layer. 
In traditional knowledge-based hierarchies, senior experts primarily function as exception handlers rather than frontline executors, serving as repositories of advanced knowledge that support junior workers in resolving problems beyond their expertise. Because experts' core role is to provide support rather than to perform routine tasks directly, our analysis focuses on how GenAI alters this support mechanism.

We consider two distinct adoption pathways. Under \textit{expert-level automation}, GenAI operates as an autonomous knowledge source that substitutes for human experts in responding to workers' queries. Under \textit{expert-level augmentation}, GenAI acts as a capability multiplier that enhances the efficiency of human experts in assisting junior workers.

\subsection{Expert-Level Automation}\label{section-expert-auto}
We model expert-level automation (denoted by subscript $e$) by treating GenAI as an autonomous knowledge repository with knowledge range $[0,r_e]$, where $0<r_e<1$ captures GenAI's automation capability. When a worker with intrinsic knowledge level $x$ faces a problem of difficulty $d>x$, the worker turns to the expert layer. For problems with $d\in(x, r_e]$, GenAI provides guidance; for problems with $d\in(r_e, 1]$, GenAI fails and human experts supply the solutions through the traditional communication channel.
A key distinction from worker-level augmentation is that the difficulty range that can be resolved without human-expert involvement is determined entirely by GenAI capability $r_e$, independent of the worker's intrinsic knowledge level $x$ (see Figure \ref{illustration-eauto}).

Because enterprise GenAI systems remain probabilistic, we incorporate hallucination risk through a hallucination rate $h$. To mitigate potential losses from authoritative but incorrect outputs, we assume the firm mandates human-in-the-loop validation for all AI-guided work. Specifically, whenever a worker completes a task with GenAI guidance, a human expert reviews the output and intervenes if an error is detected. Accordingly, experts spend $t_v$ per AI-guided task for validation and, with probability $h$, incur an additional $t_c$ to communicate with the worker to correct hallucination-induced errors (detailed profit functions are provided in Appendix \ref{app-tech-eauto}).

\begin{lemma}\label{lemma-adoption-eauto}
The firm should adopt expert-level automation if and only if the hallucination rate $h<\overline{h}_e$ and GenAI's automation capability $r_e>\overline{r}_e$, where $\overline{h}_e=1-\frac{t_v}{t_c}$ and $\overline{r}_e=\frac{(k+2w)(ht_c+t_c+t_v)}{4k}$.    
\end{lemma}

Lemma \ref{lemma-adoption-eauto} characterizes the adoption conditions for expert-level automation. Adoption is contingent upon two thresholds: a reliability threshold $h<\overline{h}_e$ and a capability threshold $r_e>\overline{r}_e$. The reliability condition reflects an efficiency requirement: AI-mediated support necessarily triggers expert validation and potential hallucination correction; hence, the expected per-task overhead $t_v+ht_c$ must be strictly lower than the cost of direct expert consultation $t_c$. Otherwise, routing tasks through GenAI fails to generate time savings and the firm optimally refrains from adoption. The capability condition reflects a scale requirement: if GenAI's knowledge coverage is limited, a substantial mass of tasks still lies beyond its capability range and continues to require direct expert intervention. In that case, GenAI adds an intermediate step without materially reducing expert workload. Only when GenAI covers a sufficiently broad portion of the task space does the reduction in direct referrals render adoption profitable.

\begin{proposition}\label{prop-eauto-x}
The optimal worker knowledge level under expert-level automation is $x_e^*=\frac{(k+2w)(t_v+ht_c)}{2k}$. Furthermore, $x_e^*$ is independent of GenAI's automation capability $r_e$ but increases in hallucination rate $h$, and $x_e^*<x_0^*$.

\end{proposition}

Proposition \ref{prop-eauto-x} characterizes the firm's optimal choice of worker knowledge under expert-level automation. The firm reduces the knowledge requirement for frontline workers, implying a deskilling effect ($x_e^*<x_0^*$). This contrasts sharply with worker-level automation, which drives upskilling (Proposition \ref{prop-workerauto-x}). 
Intuitively, once expert knowledge can be accessed through GenAI, the firm can economize on worker knowledge (and thus wages) while maintaining problem-solving capacity via AI-mediated support.

A striking feature is that the optimal worker knowledge level is independent of GenAI's automation capability $r_e$. This capability insensitivity is unique to expert-level automation and has no parallel in worker-level adoption, where capability enhancements continuously reshape skill requirements (Propositions~\ref{prop-workerauto-x} and \ref{prop-w-aug-equ}). To see why, note that lowering $x$ by one unit shifts one unit of tasks from worker-independent execution to AI-guided execution requiring expert validation. The wage savings from this deskilling depend on the current skill level but not on $r_e$; the added validation cost per shifted task, $t_v + h t_c$, is likewise independent of $r_e$ because each AI-guided task incurs the same oversight cost regardless of how broad the AI-covered region is. Since neither side of the marginal trade-off involves $r_e$, the optimal $x_e^*$ is invariant to capability. By contrast, reliability directly enters the trade-off: a lower hallucination rate $h$ reduces the per-task validation cost $t_v + h t_c$, lowers the marginal cost of deskilling, and therefore decreases $x_e^*$.

\begin{proposition}\label{prop-eauto-span}
After the adoption of expert-level automation, the optimal span of control is $s_e^*=\frac{1}{(1-r_e)t_c+(r_e-x_e^*)(t_v+ht_c)}$.
\begin{enumerate}[$(a)$]
    \item $s_e^*$ exhibits a discontinuous decline at $r_e=\overline{r}_e$ and strictly increases in GenAI's automation capability $r_e$ on $(\overline{r}_e, 1)$.

    \item $s_e^*$ decreases in hallucination rate $h$ on $(0, h_3)$ and increases in $h$ on $(h_3, \overline{h}_e)$.
\end{enumerate}    
\end{proposition}

\begin{figure}[h]
    \centering
    \subfloat[$h=0.1$]{
    \includegraphics[width=0.4\textwidth]{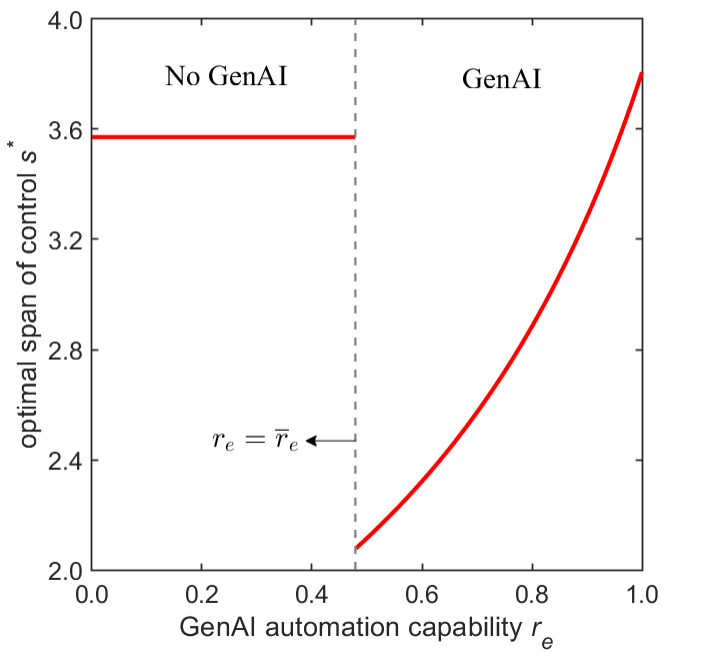}{\label{fig-e-auto-sr}}
    }
    \quad\quad
    \subfloat[$r_e=0.8$]{
    \includegraphics[width=0.4\textwidth]{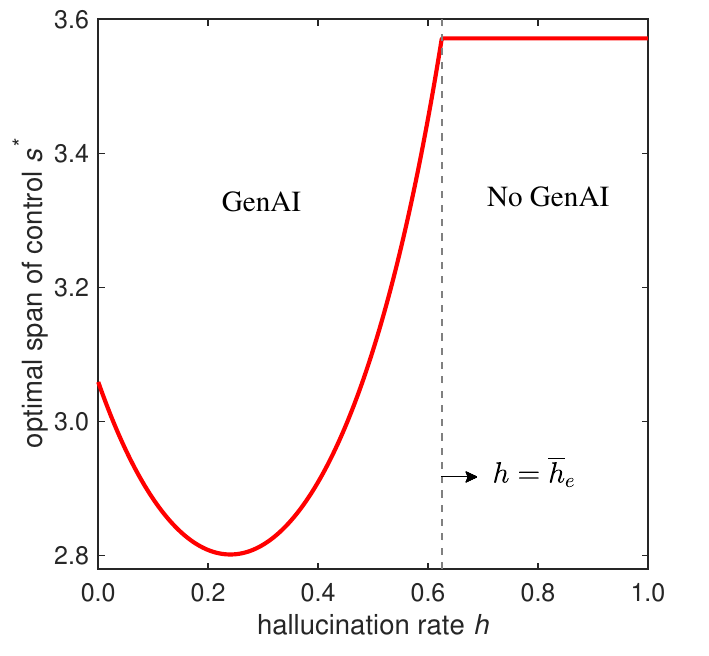}{\label{fig-e-auto-sh}}
    }
    \caption{Impact of GenAI advancements on span of control under expert-level automation \\
    ($k=0.8$, $w=0.25$, $t_c=0.8$, $t_v=0.3$)}
    \label{fig-e-auto-span}
\end{figure}

Proposition \ref{prop-eauto-span} characterizes the organizational implications of expert-level automation. Along the capability dimension, the key feature is a discontinuity at adoption: adopting GenAI induces a discrete drop in worker knowledge level, increasing the mass of AI-guided tasks that require expert oversight and then causing a sudden decline in the span of control (Figure \ref{fig-e-auto-sr}). Conditional on adoption, since $x_e^*$ is insensitive to $r_e$, higher GenAI capability mainly shifts tasks from direct worker-expert consultation to AI guidance. Under the low hallucination rate ($h<\overline{h}_e$), the resulting reduction in direct consultation time dominates the added oversight burden, lowering expert demand and making $s_e^*$ strictly increase in $r_e$.

Along the reliability dimension, Proposition \ref{prop-eauto-span}(b) implies a non-monotonic response driven by two forces. The direct correction-reduction effect lowers experts' expected correction time as $h$ falls, which tends to expand the span of control. Offsetting this, the indirect organization-deskilling effect strengthens as reliability improves, inducing lower $x_e^*$ and increasing the mass of AI-guided tasks that require expert validation. When hallucinations are high, the deskilling channel dominates and $s_e^*$ decreases as $h$ falls; when hallucinations are low, the correction-reduction channel dominates and $s_e^*$ increases with further reductions in $h$. This mechanism parallels the reliability comparative statics under worker-level augmentation.

\subsection{Expert-Level Augmentation}\label{section-e-aug}
Finally, we examine expert-level augmentation (denoted by subscript $u$). In a knowledge-based hierarchy, senior experts primarily create value by supplying solutions when frontline workers face problems beyond their expertise, making worker-expert consultation time the key bottleneck. Expert-level augmentation captures the use of GenAI to streamline this consultation process rather than replacing experts as the solution provider. For example, GenAI can retrieve relevant information, draft proposed solutions, or structure recommendations, enabling experts to respond more quickly.

To formalize this setting, we model expert-level augmentation as a reduction in consultation time from $t_c$ to $(1-r_u)t_c$, where $r_u\in(0, 1)$ measures GenAI's augmentation capability. This proportional reduction implies an equalizing effect, where GenAI delivers the greatest time savings for consultations that are initially the most costly.
However, this efficiency gain is subject to hallucination risk $h$. When hallucination occurs, the expert must revert to the manual process, incurring the original time cost $t_c$. Therefore, the expected expert time per consultation becomes $(1-r_u)t_c+ht_c$ (the detailed profit function is provided in Appendix \ref{app-tech-eaug}). 

\begin{lemma}\label{lemma-adoption-eaug}
The firm should adopt expert-level augmentation if and only if the
hallucination rate $h < r_u$.    
\end{lemma}

The firm adopts expert-level augmentation if and only if the expected consultation time under GenAI is lower than the pre-GenAI benchmark, yielding the threshold $h<r_u$. The threshold highlights a capability-accuracy trade-off: greater augmentation capability expands the set of hallucination rates for which adoption is profitable; equivalently, a lower hallucination rate renders adoption viable even when augmentation capability is modest. 
The adoption criteria differ qualitatively across the four deployment architectures: capability can relax the reliability requirement (expert-level augmentation), tighten it (worker-level automation), or leave it unchanged (worker-level augmentation; expert-level automation), and may itself need to clear a minimum threshold (expert-level automation). This heterogeneity implies that there is no one-size-fits-all adoption rule; firms must assess adoption in a deployment-specific manner.

\begin{proposition}\label{prop-e-aug}
After the adoption of expert-level augmentation:
\begin{enumerate}[$(a)$]
    \item The optimal worker knowledge level is 
    $x_u^*=\frac{(k+2w)(1-r_u+h)t_c}{2k}$, which decreases in GenAI's augmentation capability $r_u$, increases in hallucination rate $h$, and satisfies $x_u^*<x_0^*$.
    

    \item The optimal span of control is $s_u^*=\frac{1}{(1-x_u^*)[(1-r_u)t_c+ht_c]}$:
    \begin{enumerate}[$(1)$]
    \item $s_u^*$ decreases in GenAI's augmentation capability $r_u$ on $(0,r_4)$ and increases in $r_u$ on $(r_4,1)$.

    \item $s_u^*$ decreases in hallucination rate $h$ on $(0,h_4)$ and increases in $h$ on $(h_4, \overline{h}_u)$.

    \end{enumerate}
    
\end{enumerate}
\end{proposition}

Conditional on adoption, expert-level augmentation induces a deskilling effect: the firm optimally lowers the knowledge requirement for frontline workers, because augmentation reduces the marginal cost of relying on experts for exception handling. Moreover, as higher augmentation capability and lower hallucination rates further reduce the expected consultation time, the incentive to deskill becomes stronger (Proposition \ref{prop-e-aug}(a)). Proposition \ref{prop-e-aug}(b) implies a non-monotonic effect of GenAI advancements on organizational structure. Although augmentation ultimately allows each expert to serve more workers by lowering the effective consultation time, early-stage improvements in GenAI can initially narrow the span of control. The reason is that deskilling increases the volume of referrals to experts, which may dominate the per-referral efficiency gain when GenAI's augmentation capability is still moderate or the hallucination rate remains high. As GenAI becomes sufficiently capable and reliable, the direct reduction in consultation time dominates, reducing total expert workload and expanding the span of control, producing a U-shaped trajectory. 


\subsection{Comparison and Cross-Layer Implications}\label{section-expert-com}
The organizational consequences of expert-level GenAI adoption highlight that deployment location can be as consequential as deployment mode. Although worker-level and expert-level implementations rely on the same underlying technology, their interactions with knowledge-based hierarchies generate distinct organizational signatures.

\textbf{Automation Asymmetry:} Expert-level automation exhibits a clear cross-layer reversal relative to worker-level automation. At the worker level, automation drives upskilling by shifting junior roles toward validation and escalation control. At the expert level, however, automation induces deskilling. By providing an autonomous substitute for human-expert consultation, expert-level automation lowers the marginal value of worker knowledge and allows firms to economize on entry-level wages while preserving problem-solving capacity through AI-mediated support. 

A further distinctive feature is post-adoption capability insensitivity. Under worker-level automation, capability improvements expand the scope that workers must validate and therefore continuously push skill requirements upward. Under expert-level automation, by contrast, higher capability mainly reallocates queries between GenAI and human experts and does not change the firm's optimal worker-knowledge choice within the AI-covered region. As a result, once adoption occurs, the optimal worker knowledge level is invariant to further capability gains.

\textbf{Augmentation Symmetry:} In contrast to the asymmetries of automation, augmentation produces qualitatively symmetric effects across both layers of the hierarchy. Whether it expands workers' effective capability boundary or accelerates experts' handling of consultations, it consistently lowers the effective cost of accessing high-level expertise. This leads to a uniform relaxation of frontline knowledge requirements, generating a deskilling effect that intensifies as the technology advances.

Taken together, expert-level adoption delivers a clean first-order skill prediction: both automation and augmentation deskill the frontline, lowering the entry-level skill threshold. The adjustment paths differ: automation produces a discrete drop in skill requirements followed by capability insensitivity, while augmentation generates a more gradual, capability-sensitive response. Both modes also imply similar hierarchical dynamics: direct efficiency gains and endogenous deskilling jointly drive a non-monotonic span of control, contracting initially before expanding as GenAI advances.

\section{Conclusion and Managerial Implications}\label{section-conclusion}
This paper develops an analytical framework showing that GenAI's organizational consequences depend critically on deployment design: not just whether the technology is adopted, but how and where. Building on knowledge-based hierarchy theory, the model incorporates two defining features of GenAI: intrinsic fallibility, which necessitates continuous human validation, and deployment flexibility, which allows GenAI to be integrated across both deployment modes and hierarchical layers. The resulting predictions challenge conventional wisdom about technology-induced organizational change while offering practical guidance for stakeholders navigating the GenAI transformation.

\subsection{Summary of Key Findings}
Our analysis generates three main findings, summarized in Table \ref{tab:mode-industry-summary}. First, GenAI's impact on entry-level skill requirements is mode-dependent. Under worker-level automation, GenAI takes over routine execution but introduces a validation burden due to intrinsic fallibility; the firm responds by hiring fewer but more knowledgeable workers who can verify AI outputs and reduce costly upward referrals to senior experts. Under worker-level augmentation, GenAI instead operates as a capability amplifier that extends workers' effective problem-solving boundaries; the firm responds by relaxing frontline knowledge requirements and sustaining performance through AI-assisted workflows. The observed decline in entry-level employment is therefore not an inevitable consequence of GenAI per se, but reflects deployment choices that tilt GenAI toward automation rather than augmentation. 

Second, deployment location can be as consequential as deployment mode. Expert-level GenAI adoption uniformly lowers entry-level skill requirements regardless of whether GenAI serves as an autonomous knowledge source (automation) or a consultation accelerator (augmentation). In both cases, expert-level deployment reduces the marginal value of frontline knowledge by providing an alternative channel for resolving problems beyond workers' expertise, allowing firms to economize on entry-level wages while preserving problem-solving capacity through AI-mediated or AI-enhanced expert support. This uniform deskilling effect stands in sharp contrast to the mode-dependent outcomes at the worker level. 

Third, organizational structure evolves non-monotonically as GenAI advances. Across all four deployment architectures, spans of control first decrease and then increase, driven by the shifting dominance between direct efficiency gains and indirect workforce restructuring. This pattern implies that demand for senior expertise may remain stable or even increase during early-to-intermediate stages of deployment, even before hierarchies ultimately flatten, a prediction consistent with recent large-scale evidence on GenAI's seniority-biased labor market effects.

\begin{table}[h]
\centering
\small
\renewcommand{\arraystretch}{1.2}
\caption{Predicted organizational impacts across GenAI deployment architectures}
\label{tab:mode-industry-summary}
\begin{tabular}{p{2.9cm} p{5.5cm} p{5.5cm}}
\toprule
 & \textbf{Automation Mode} 
 & \textbf{Augmentation Mode} \\
\midrule
\textbf{Worker Level}
&
Task executor 

\textbf{Upskilling} ($x_t^* > x_0^*$)

Capability-sensitive

U-shaped span of control 
&
Capability amplifier 

\textbf{Deskilling} ($x_g^* < x_0^*$)

Capability-sensitive

U-shaped span of control 
\\
\midrule
\textbf{Expert Level}
&
Virtual supervisor 

\textbf{Deskilling} ($x_e^* < x_0^*$)

Capability-insensitive

Initial drop, then increase
&
Communication accelerator 

\textbf{Deskilling} ($x_u^* < x_0^*$)

Capability-sensitive

U-shaped span of control
\\
\bottomrule
\end{tabular}
\end{table}

Notably, we do not assume that firms can frictionlessly choose a single optimal GenAI architecture in the abstract. In practice, deployment location and mode are shaped by constraints such as compliance regimes, liability and risk tolerance, legacy systems, and the existing distribution of expertise. Our goal is therefore to provide a conditional design guide: given a firm's choice of where GenAI is embedded in the hierarchy and whether it is used primarily for automation or augmentation, the model characterizes how skill thresholds and spans of control adjust as AI capability and fallibility change. This structure yields transparent ``if--then'' prescriptions that firms can use to anticipate second-order organizational consequences of a chosen deployment strategy.

\subsection{Managerial and Policy Implications}
For firms, our analysis offers a strategic framework for aligning technology procurement with human capital planning. The decision to adopt GenAI cannot be decoupled from hiring strategy.
First, firms deploying GenAI for \textit{automation} (e.g., autonomous coding agents or customer service bots) face a counterintuitive skill imperative: the optimal response is to raise, not lower, entry-level knowledge requirements. 
Under automation, the worker's primary role shifts from execution to validation of AI outputs. Firms should therefore raise technical qualification requirements for entry-level roles, ensuring that workers possess the domain expertise to detect subtle AI hallucinations. Failing to upskill the intake pipeline risks a proliferation of undetected errors or an overwhelming escalation load on senior experts. 

Conversely, firms prioritizing \textit{augmentation} (e.g., co-pilots for writing or research) can broaden their recruitment funnels. By deploying GenAI as a capability amplifier, firms can hire candidates who possess strong foundational traits but lack specialized technical mastery, as the AI tool bridges the execution gap. This suggests a strategic pivot for HR: under augmentation strategies, hiring criteria should shift from testing for peak technical execution to assessing adaptability and human-AI collaborative aptitude.

Finally, firms should anticipate that middle-management demand will not shrink in the near term and may even increase during early-to-intermediate stages of GenAI deployment. Organizations that preemptively flatten their hierarchies in anticipation of future efficiencies risk creating significant bottlenecks in exception handling and quality control during the transition period.

For junior workers, our model speaks directly to career strategy. Under worker-level \textit{automation}, the returns to validation skills (the ability to detect and correct AI errors) rise sharply, because the hallucination risk makes human verification the binding constraint on AI-automated production. Entry-level workers in automation-heavy environments should invest in domain expertise sufficient to evaluate AI outputs, not merely to execute tasks that AI can perform. This contrasts with the conventional advice to develop AI-complementary soft skills: our model suggests that deep technical knowledge becomes more valuable, not less, precisely because someone must verify what the machine produces. Under worker-level \textit{augmentation}, the strategic calculus reverses. Because augmentation extends workers' effective capability boundary, the marginal return to intrinsic knowledge declines. Workers in augmentation-heavy environments face a different risk: skill atrophy, as reduced demand for deep expertise may erode the human capital that currently differentiates them from less experienced peers \citep{siderius2026use}. 

For senior experts, our model delivers a nuanced career implication. The near-term outlook is more stable than commonly feared: across all deployment modes, exception handling, validation, and oversight demands sustain or even expand expert workloads during early-to-intermediate stages of GenAI deployment, even as frontline execution becomes more efficient. As GenAI matures and direct efficiency gains accumulate, the demand for senior positions will eventually moderate, but the remaining roles do not vanish. They recenter on validation, error-correction, and judgment-intensive oversight---tasks that remain resistant to full automation. In sectors where automation reaches sufficient maturity to flatten the hierarchy substantially, senior experts can sustain their relevance by evolving beyond pure exception handling toward a hybrid role that combines high-skill execution with AI supervision, effectively serving as elite task performers who also bear responsibility for monitoring and correcting AI-generated outputs.

For policymakers, our findings refine the policy debate beyond the binary framing of ``AI displaces workers'' versus ``AI augments workers.'' Two policy-relevant insights emerge. First, the distributional consequences of GenAI depend critically on deployment architecture, not just adoption intensity. The observed concentration of entry-level employment declines in automation-intensive settings \citep{brynjolfsson2025canaries} is not an inevitable consequence of GenAI but reflects specific, and potentially redirectable, deployment choices. Policies that incentivize augmentation-oriented development \citep{acemoglu2023proworker} are well-targeted from our model's perspective, but our results add the nuance that expert-level deployment of any mode similarly lowers the entry-level skill threshold. Second, training programs should be mode-specific. Our model implies that generic ``AI literacy'' curricula may be insufficient: workers entering automation-heavy roles need deep domain expertise for validation, while those entering augmentation-heavy roles need adaptive problem-solving skills to leverage AI-expanded capability boundaries. 


\subsection{Limitations and Future Research}
While our model provides a general framework for GenAI adoption in knowledge-based hierarchies, its predictions are sharpest in industries characterized by verifiable correctness and high error costs. Our mechanism hinges on the concept of hallucination, an objectively incorrect output that requires remediation. This dynamic maps closely to domains such as software engineering, legal services, and financial analysis, where outputs are functionally binary (correct vs. buggy; legally sound vs. invalid) and the cost of deploying an error is non-trivial. In these settings, the trade-off between execution efficiency and validation rigor is the binding constraint on organizational structure. Conversely, in creative industries (e.g., graphic design or marketing copy generation), where hallucination may be interpreted as creativity and validation is subjective, the strict upskilling/deskilling dynamics we predict may be less pronounced.

Several directions for future research emerge naturally from our framework. First, extending the analysis to multi-tier hierarchies could reveal how GenAI affects intermediate layers, particularly middle management roles that combine execution and supervision. Second, incorporating dynamics, such as learning curves, path dependence, and transition costs, would characterize whether early adoption patterns (e.g., entry-level hollowing under automation) are self-correcting or self-reinforcing as GenAI matures. Third, embedding our single-firm analysis in a competitive labor-market equilibrium would characterize industry-wide wage and employment effects, addressing the general-equilibrium feedback that our partial-equilibrium framework deliberately abstracts from. Finally, integrating worker agency, including strategic skill investment, technology resistance, and behavioral responses such as automation complacency, would enrich the model's predictions about the co-evolution of human capital and AI capability.

\makeatletter
\renewcommand{\@biblabel}[1]{} 
\let\OLDthebibliography\thebibliography
\renewcommand{\thebibliography}[1]{%
  \OLDthebibliography{#1}%
  \setlength{\itemsep}{1pt}%
  \setlength{\parskip}{1pt}%
  \setlength{\baselineskip}{14pt}%
}
\makeatother
\bibliographystyle{ormsv080}
\bibliography{2026_GenAI_Workforce.bib}
\newpage
\ECSwitch
\vspace{-20pt}
\begin{APPENDICES}
\section{Supplementary Materials}
In this appendix, we provide several supplemental materials to the main text.
\subsection{Technical Details in \S\ref{section-w-auto}}
The expressions of thresholds in Proposition \ref{prop-workerauto-x}:
\begin{equation*}
     \overline{h}_t
    =\left \{ 
    \begin{aligned}
    &\; \frac{1-t_v}{t_r}, \quad\quad\quad && \text{if \;} r_t\le x_0^*; \\
    &\; \frac{-(k+2w)^2t_c^2+4kr_t(k+2w)t_c-4kr_t[kr_t-(1-t_v)(kr_t^2+2w)]}{4kr_tt_r(kr_t^2+2w)}, \quad\quad\quad && \text{otherwise. \;}      \quad\quad\quad  \\
    \end{aligned}
    \right.
\end{equation*}
~\\
The expressions of thresholds in Proposition \ref{prop-workerauto-s}: 
$$h_0=\frac{(k+2w)t_c-2kt_v}{2kt_r},$$ $$r_0=\frac{2k-(k+2w)t_c}{2k(1-t_v-ht_r)}.$$
\begin{equation*}
     r_1
    =\left \{ 
    \begin{aligned}
    &\; \frac{k-(k+2w)t_c}{k(1-t_v-ht_r)}, \quad && \text{if \;} h\le \frac{2t_c(1-t_v)(k+2w)-k}{2t_ct_r(k+2w)}; \\
    &\; \min\left\{\frac{k-(k+2w)t_c}{k(1-t_v-ht_r)}, \frac{1-\sqrt{1-4(1-t_v-ht_r)x_0^*}}{2(1-t_v-ht_r)}\right\}, \quad && \text{otherwise. \;}      \quad\quad\quad  \\
    \end{aligned}
    \right.
\end{equation*}
\begin{equation*}
     h_1
    =\left \{ 
    \begin{aligned}
    &\; \frac{(k+2w)t_c-k+r_tk(1-t_v)}{kt_rr_t}, \quad && \text{if \;} r_t\le x_0^*; \\
    &\; \min\left\{\frac{(k+2w)t_c-k+r_tk(1-t_v)}{kt_rr_t}, \frac{2kr_t^2(1-t_v)+(k+2w)t_c-2kr_t}{2kt_rr_t^2}\right\}, \quad && \text{otherwise. \;}      \quad\quad\quad  \\
    \end{aligned}
    \right.
\end{equation*}
\vspace{0.4pt}
\subsection{Technical Details in \S\ref{section-w-aug}}
The expressions of thresholds in Proposition \ref{prop-waug-span}: 
$$r_2=\frac{(k+2w)t_c-k}{(k+2w)(t_c-t_v-ht_c)},$$ $$h_2=\frac{r_g(k+2w)(t_c-t_v)-(k+2w)t_c+k}{r_g(k+2w)t_c}.$$

The following Figure \ref{fig-w-aug-span} illustratively demonstrates the impact of GenAI advancements on span of control under worker-level augmentation.

\begin{figure}[H]
    \centering
    \subfloat[$h=0.2$]{
    \includegraphics[width=0.4\textwidth]{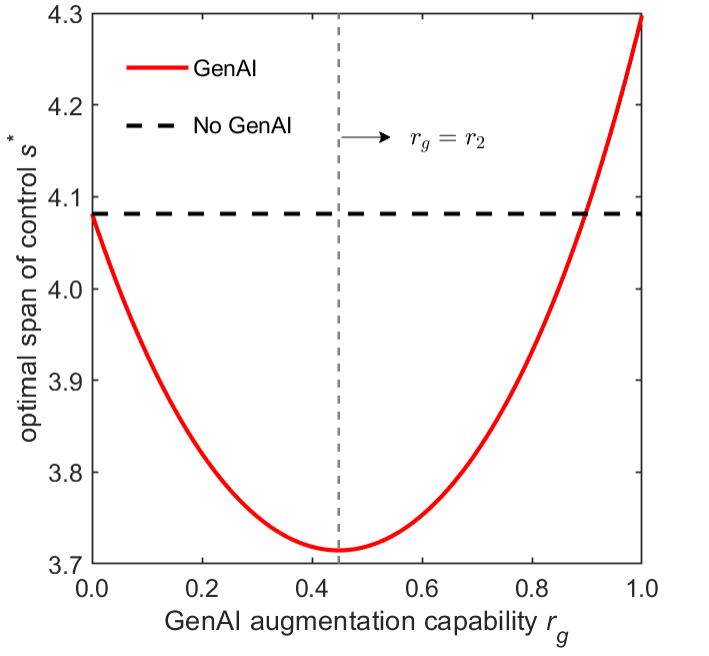}{\label{fig-waug-s-r}}
    }
    \quad\quad
    \subfloat[$r_g=0.5$]{
    \includegraphics[width=0.4\textwidth]{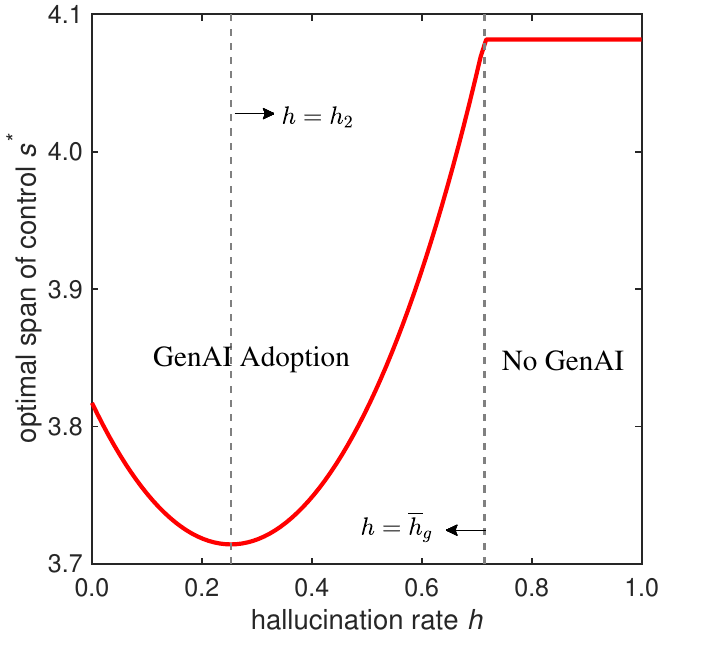}{\label{fig-waug-s-h}}
    }
    \caption{Impact of GenAI advancements on span of control under worker-level augmentation \\
    ($k=0.7$, $w=0.3$, $t_c=0.7$, $t_v=0.2$)}
    \label{fig-w-aug-span}
\end{figure}
\subsection{Technical Details in \S\ref{section-expert-auto}}\label{app-tech-eauto}
Under expert-level automation, the firm's profit function is given by:
\begin{equation*}
     \Pi_e(x) 
    =\left \{ 
    \begin{aligned}
    &1-\left(w+\frac{1}{2}kx^2\right)-\left(w+\frac{1}{2}k\right)[(1-r_e)t_c+(r_e-x)(t_v+ht_c)], && \text{if $x\le r_e$}; \\
    &1-\left(w+\frac{1}{2}kx^2\right)-\left(w+\frac{1}{2}k\right)(1-x)t_c, && \text{if $x\ge r_e$}.\\
    \end{aligned}
    \right.
\end{equation*}

\begin{figure}[H]
\centering \footnotesize
\quad \quad
\begin{tikzpicture}[scale=1.2]
\linespread{0.8} 
\foreach \x in {0,3,6,9}{
    \draw (\x cm,5pt) -- (\x cm,-5pt);
}
\node[align=center] at (3,0.5) {$x$};
\node[align=center] at (6,0.5) {$r_e$};
\node[align=center] at (9,0.5) {1};
\node[align=center] at (0,0.5) {0};

\fill[yellow!50] (3.01,-0.16) rectangle (5.99,0.16);
\draw [thick,->] (0,0) -- (9.3,0)node[right]{\text{Tasks}};
\draw [decorate,decoration={brace,amplitude=5pt,mirror,raise=2ex}] (0,0) -- (2.95,0) node[midway,yshift=-2.2em]{Worker};
\draw [decorate,decoration={brace,amplitude=5pt,mirror,raise=2ex}] (3.05,0) -- (5.95,0) node[midway,yshift=-2.3em]{Worker+GenAI};
\node[align=center] at (4.5,-1) {\textcolor{gray}{(Expert Validation)}};
\draw [decorate,decoration={brace,amplitude=5pt,mirror,raise=2ex}] (6.05,0) -- (8.95,0) node[midway,yshift=-2.3em]{Worker+Expert};
\end{tikzpicture}
\caption{Delegation of tasks under expert-level automation}
\label{illustration-eauto}
\end{figure}

The expression of the threshold in Proposition \ref{prop-eauto-span}:
$$h_3=\frac{kr_e-(k+2w)t_v}{(k+2w)t_c}.$$
\subsection{Technical Details in \S\ref{section-e-aug}}\label{app-tech-eaug}
Under expert-level augmentation, the firm's profit function is given by
$$\Pi_u(x)=1-\left(w+\frac{1}{2}kx^2\right)-\left(w+\frac{1}{2}k\right)(1-x)[(1-r_u)t_c+ht_c].$$

The expressions of thresholds in Proposition \ref{prop-e-aug}: \\
$$r_4=\frac{t_c(k+2w)(1+h)-k}{t_c(k+2w)},$$ $$ h_4=\frac{k-t_c(k+2w)(1-r_u)}{t_c(k+2w)}.$$

The following Figure \ref{fig-e-aug-span} illustratively demonstrates the impact of GenAI advancements on span of control under expert-level augmentation.

\begin{figure}[H]
    \centering
    \subfloat[$h=0.2$]{
    \includegraphics[width=0.4\textwidth]{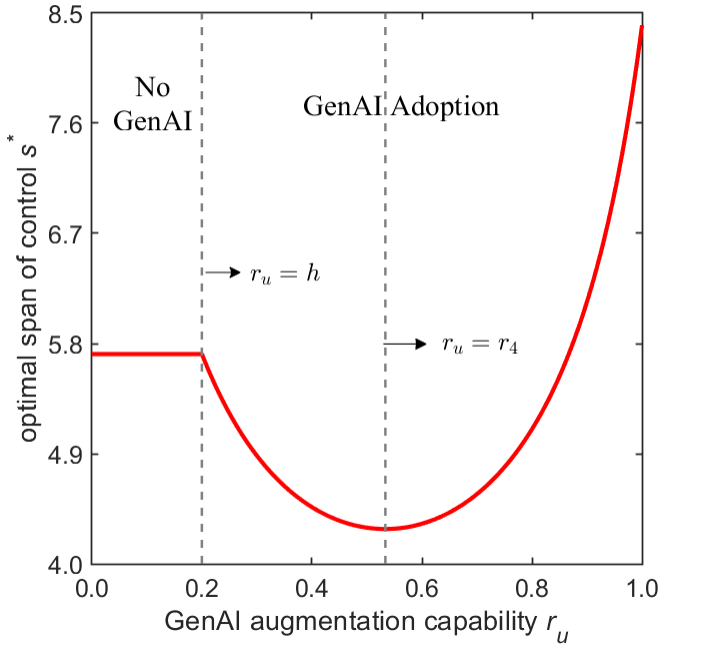}{\label{fig-eaug-s-r}}
    }
    \quad\quad
    \subfloat[$r_g=0.5$]{
    \includegraphics[width=0.4\textwidth]{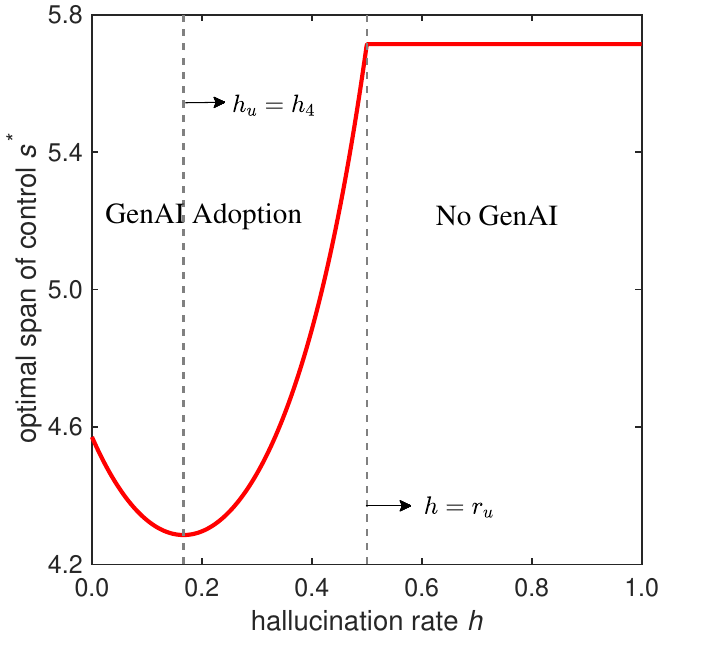}{\label{fig-eaug-s-h}}
    }
    \caption{Impact of GenAI advancements on span of control under expert-level augmentation \\
    ($k=0.7$, $w=0.4$, $t_c=0.7$)}
    \label{fig-e-aug-span}
\end{figure}

\clearpage
\section{Productivity Effect under Worker-Level Augmentation}\label{app-productivity}
In the main text, we emphasize the distinctive role of GenAI under worker-level augmentation as a capability amplifier that enables workers to address problems beyond their intrinsic knowledge. In practice, GenAI can also operate as a co-pilot for tasks that already fall within the worker's expertise, thereby improving execution efficiency. For example, evidence from consulting settings shows that AI-assisted workers completed 12.2\% more tasks and worked 25.1\% faster \citep{dell2023navigating}. Motivated by this dual role, this appendix extends the worker-level augmentation model by incorporating direct productivity improvement brought by GenAI augmentation in addition to capability expansion.

To formalize this productivity gain, we assume that for tasks falling within the worker's intrinsic knowledge range $[0, x]$, the human-AI collaborative unit operates at an accelerated rate. Specifically, a worker who previously handled one unit of tasks per unit of time now achieves a productivity level of $1 + r_g A$, where $r_g \in (0, 1)$ denotes GenAI's augmentation capability and $A \ge 0$ is a productivity gain intensity parameter. A higher $A$ captures settings in which copilot-style assistance yields larger throughput improvements, for example due to better tool integration or more structured workflows. When $A=0$, this extension reduces to the baseline setting in \S\ref{section-w-aug}. Notably, this productivity improvement scales with GenAI's augmentation capability $r_g$, reflecting the intuition that more advanced models are more effective at streamlining routine workflows, such as drafting, coding, or information retrieval, thereby further reducing execution time.

Under this extension, the firm's profit function becomes:
\begin{equation*}
    \Pi_g(x)=1-\left(w+\frac{1}{2}kx^2\right)\left(\frac{x}{1+r_g A}+1-x\right)-\left(w+\frac{1}{2}k\right)\big[r_g(1-x)(t_v+ht_c)+(1-r_g)(1-x)t_c\big].
\end{equation*}
This formulation departs from the baseline only by introducing productivity gains on the ``easy" tasks. Since accelerating task execution over the interval $[0, x]$ does not affect the worker's augmented capability boundary, the set of complex tasks requiring expert intervention remains unchanged. Consequently, the expert-side workload and the associated validation burdens coincide with those in the baseline augmentation mode: the total demand for experts remains $r_g(1-x)(t_v+ht_c)+(1-r_g)(1-x)t_c$. The structural change occurs in the frontline labor demand. Because higher productivity shortens the expected processing time for tasks within the worker's knowledge domain, fewer workers are needed to deliver a unit mass of tasks. Specifically, for tasks in $(x, 1]$, worker productivity is unchanged and thus the firm still requires $(1 - x)$ units of labor. For tasks in $[0, x]$, productivity increases to $1 + r_g A$, reducing the labor requirement from $x$ to $\frac{x}{1+r_g A}$. Hence, total worker demand becomes $\frac{x}{1+r_g A} + 1 - x$. Furthermore, since the worker possesses the requisite expertise for tasks in $[0, x]$, the final output accuracy for this range is decoupled from GenAI hallucinations: the worker serves as a real-time monitor, identifying and correcting any AI errors during execution. Profits are then given by the resulting surplus net of wage payments, where each worker and expert incurs costs $w+\frac{1}{2}kx^2$ and $w+\frac{1}{2}k$, respectively.

Given the analytical intractability of characterizing the optimal worker knowledge level, we employ numerical experiments to investigate the impacts of productivity improvements, with a particular focus on how such productivity improvement interacts with GenAI's unique features---augmentation capability and hallucination rates.  

\begin{figure}[h]
    \centering
    \subfloat[$A=0.2$]{
    \includegraphics[width=0.32\textwidth]{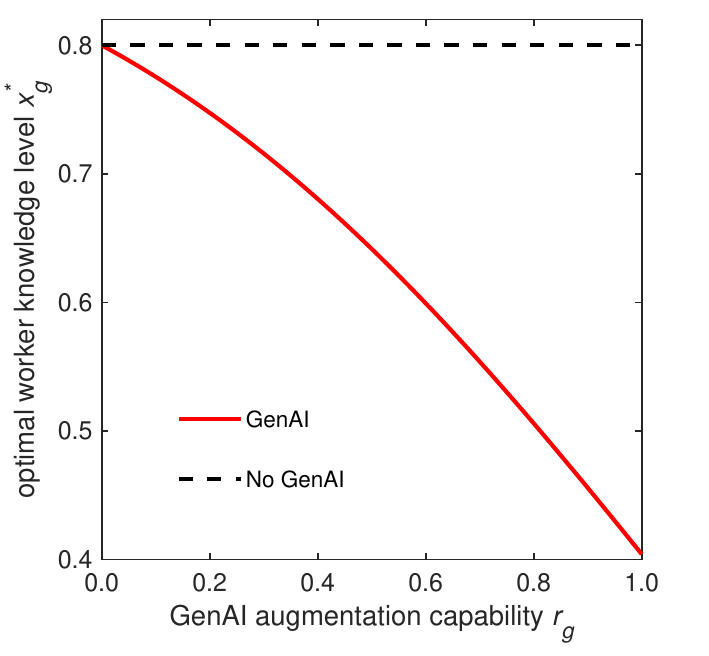}\label{fig-apppro-x-r1}
    }
    \subfloat[$A=0.5$]{
    \includegraphics[width=0.32\textwidth]{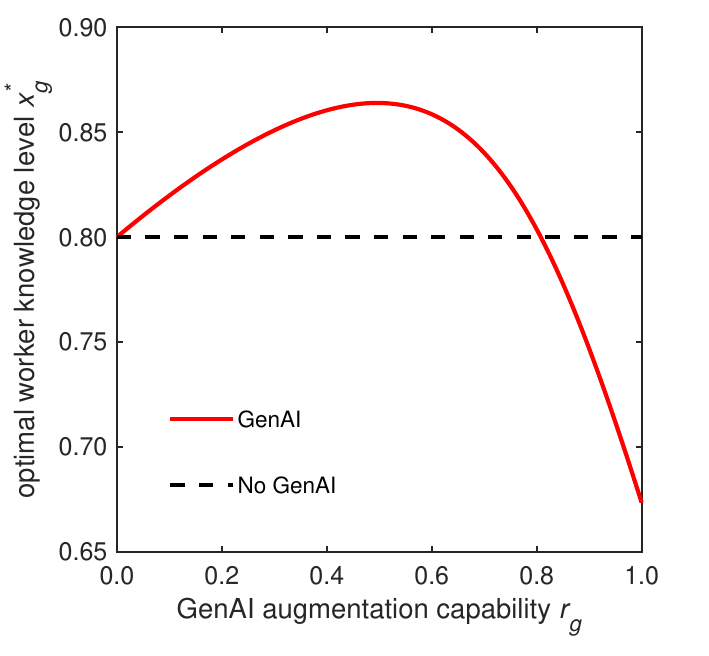}\label{fig-apppro-x-r2}
    }
    \subfloat[$A=0.8$]{
    \includegraphics[width=0.325\textwidth]{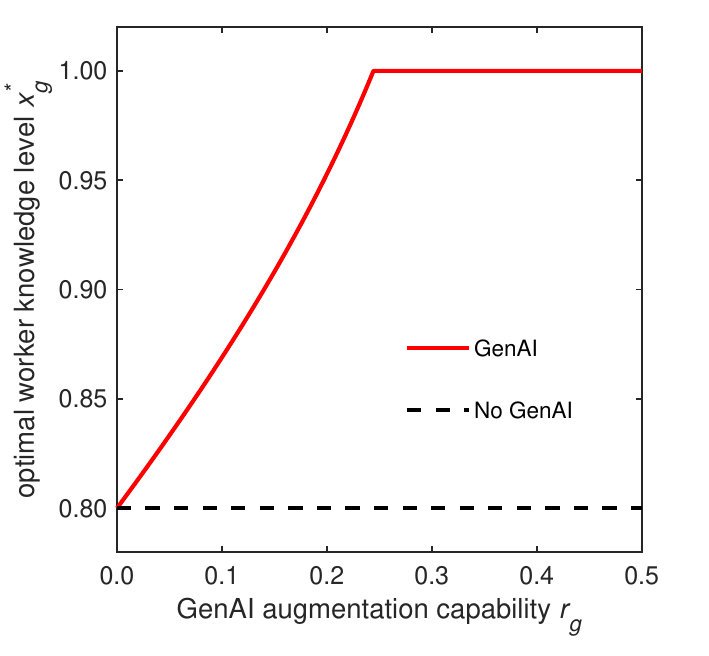}\label{fig-apppro-x-r3}
    }~\\[2mm]
    \caption{Impacts of GenAI capability enhancement on worker knowledge level under productivity improvement extension ($k=0.4$, $w=0.2$, $t_c=0.8$, $t_v=0.2$, $h=0.1$)}
    \label{fig-apppro-x-r}
\end{figure}

Figure \ref{fig-apppro-x-r} shows how introducing productivity improvements alters the comparative statics of GenAI capability under worker-level augmentation. When productivity improvements are modest (Figure \ref{fig-apppro-x-r1}), higher augmentation capability $r_g$ primarily strengthens the substitution effect of GenAI: workers can rely more on AI assistance rather than intrinsic expertise. As a result, the firm optimally deskills: $x_g^*$ decreases monotonically and remains below the no-GenAI benchmark. At an intermediate productivity gain intensity (Figure \ref{fig-apppro-x-r2}), capability improvements operate through two opposing channels. A higher $r_g$ still relaxes skill requirements for workers by expanding the augmented boundary, yet it also raises the return to processing tasks within the worker's knowledge range by increasing productivity. Consequently, the firm initially upskills to exploit the productivity gain, but once the boundary-expansion effect dominates, $x_g^*$ declines, generating an inverted-U pattern. When productivity gains are strong (Figure \ref{fig-apppro-x-r3}), the efficiency benefit from bringing tasks into the worker's intrinsic domain dominates throughout, because the firm can both leverage the higher processing rate and mitigate losses from hallucinations and validation frictions. Accordingly, $x_g^*$ increases sharply in $r_g$ and quickly reaches the upper bound $x_g^*=1$, implying pronounced upskilling. In this regime, the firm effectively operates with a fully knowledgeable frontline workforce to maximize efficiency while leveraging GenAI as a high-throughput copilot.

\begin{figure}[h]
    \centering
    \subfloat[$A=0.2$]{
    \includegraphics[width=0.32\textwidth]{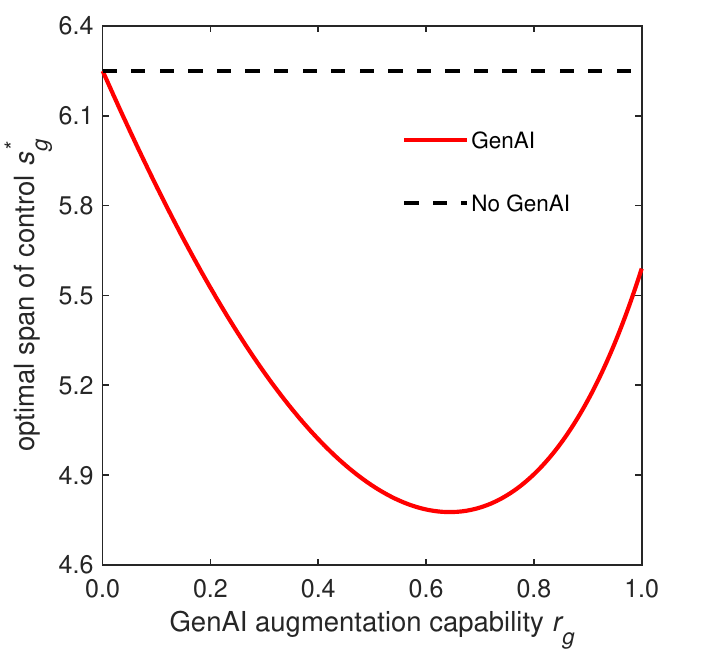}\label{fig-apppro-s-r1}
    }
    \subfloat[$A=0.5$]{
    \includegraphics[width=0.32\textwidth]{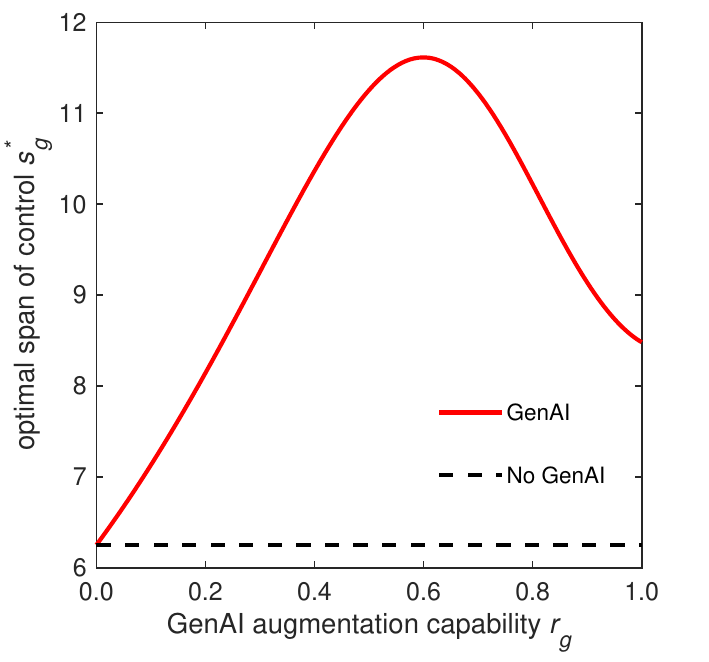}\label{fig-apppro-s-r2}
    }
    \subfloat[$A=0.8$]{
    \includegraphics[width=0.325\textwidth]{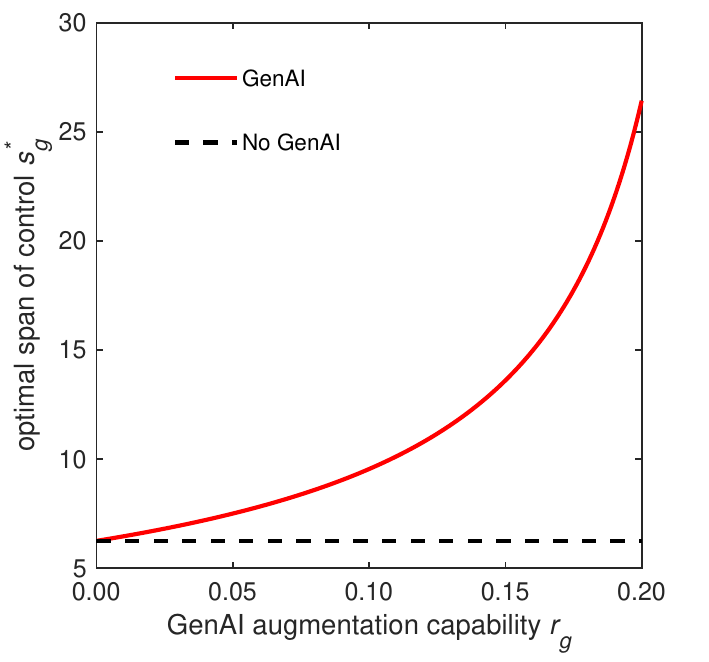}\label{fig-apppro-s-r3}
    }~\\[2mm]
    \caption{Impacts of GenAI capability enhancement on span of control under productivity improvement extension ($k=0.4$, $w=0.2$, $t_c=0.8$, $t_v=0.2$, $h=0.1$)}
    \label{fig-apppro-s-r}
\end{figure}

Figure \ref{fig-apppro-s-r} reveals that productivity improvements can change how the optimal span of control responds to augmentation capability. When productivity improvements are modest (Figure \ref{fig-apppro-s-r1}), the span of control preserves the U-shaped pattern shown in the main text; we thus refrain from a repeated discussion. At an intermediate productivity gain intensity (Figure \ref{fig-apppro-s-r2}), the span of control exhibits a non-monotonic, inverted-U pattern as augmentation capability $r_g$ increases. For relatively small $r_g$, the firm optimally increases worker knowledge level to exploit GenAI's ability to accelerate task execution within the worker's intrinsic expertise domain $[0, x]$; the resulting reduction in expert intervention expands the span of control. Once $r_g$ exceeds a certain threshold, the firm instead shifts toward deskilling, lowering $x_g^*$ to leverage GenAI to expand workers' effective capability boundary. In this regime, the decline in  $x_g^*$ increases the demand for expert oversight, which narrows the span of control and generates the inverted-U trajectory. When productivity gains are strong (Figure \ref{fig-apppro-s-r3}), the efficiency advantage dominates throughout: as $r_g$ increases, the firm continuously upskills its workforce, which diminishes the necessity for expert intervention and allows the span of control to increase monotonically. Ultimately, the hierarchy collapses to a single layer of fully knowledgeable workers who use GenAI as a copilot across all tasks, thereby maximizing processing efficiency.

\begin{figure}[h]
    \centering
    \subfloat[Worker knowledge level ($A=0.2$)]{
    \includegraphics[width=0.4\textwidth]{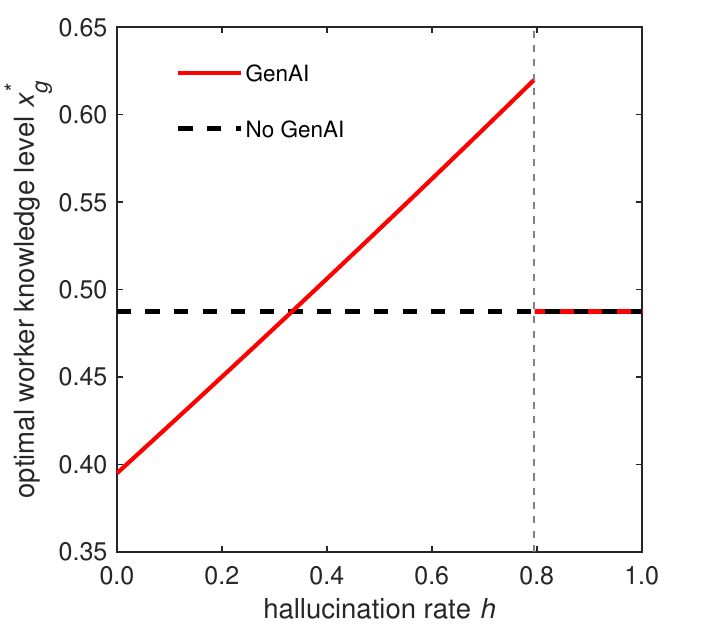}{\label{fig-apppro-x-h1}}
    }
    \quad\quad
    \subfloat[Worker knowledge level ($A=0.8$)]{
    \includegraphics[width=0.4\textwidth]{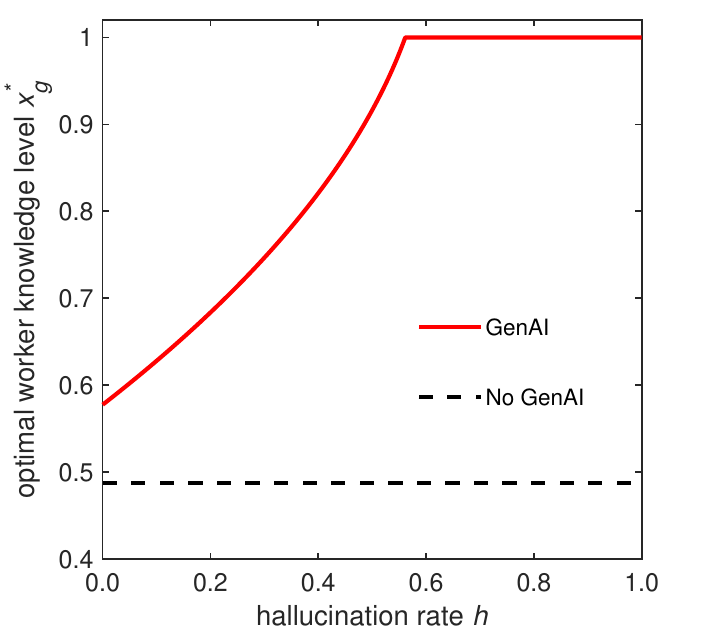}{\label{fig-apppro-x-h2}}
    }
    \quad\quad\quad
    \subfloat[Span of control ($A=0.2$)]{
    \includegraphics[width=0.4\textwidth]{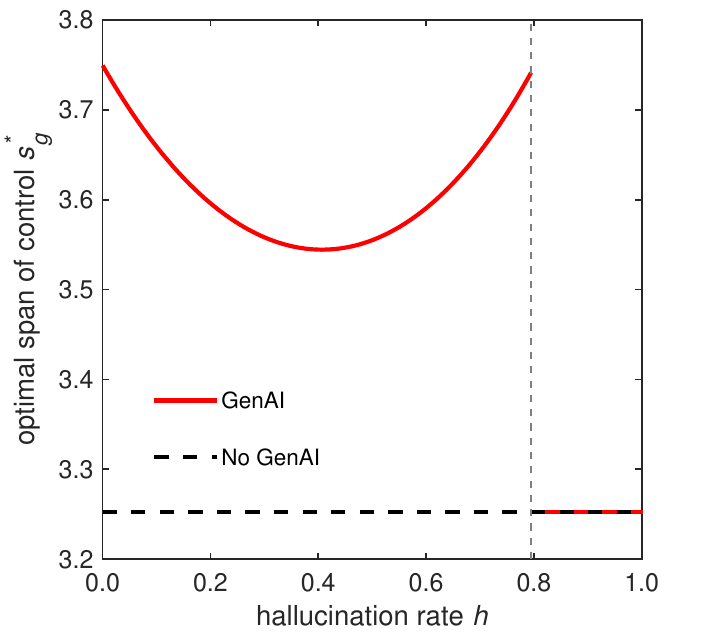}{\label{fig-apppro-s-h1}}
    }
    \quad\quad
    \subfloat[Span of control ($A=0.8$)]{
    \includegraphics[width=0.4\textwidth]{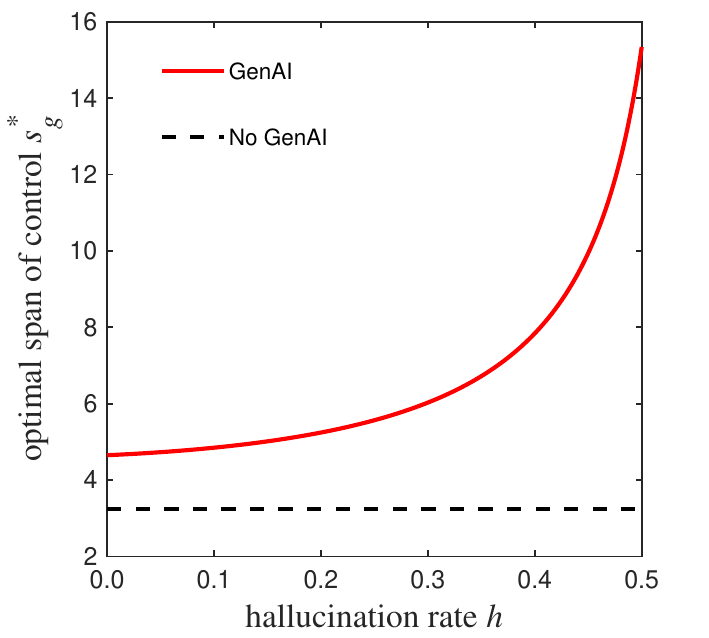}{\label{fig-apppro-s-h2}}
    }
    \caption{Impact of hallucination reduction under productivity improvement extension \\  ($k=0.8$, $w=0.25$, $t_c=0.6$, $t_v=0.25$, $r_g=0.5$)}
    \label{fig-apppro-h}
\end{figure}

Figure \ref{fig-apppro-h} illustrates how the productivity improvement driven by GenAI shapes the effect of hallucination reduction on both optimal worker knowledge levels and the span of control. When the productivity gain intensity is small (Figures \ref{fig-apppro-x-h1} and \ref{fig-apppro-s-h1}), the observed patterns mirror those discussed in \S\ref{section-w-aug}: the firm adopts GenAI only when the hallucination rate is relatively low, and conditional on adoption, further reductions in $h$ induce deskilling ($x_g^*$ declines) while the span of control follows the familiar U-shaped response. Notably, at the adoption threshold where the firm switches to GenAI, there is a discrete upward jump in the worker knowledge level. This occurs because, with productivity gains, the firm finds it is optimal to adopt GenAI even at relatively high hallucination rates to leverage productivity improvements. Choosing a higher worker knowledge level expands the mass of tasks that benefit from accelerated execution and simultaneously limits the share of tasks exposed to hallucination risk. This discrete increase in $x_g^*$ at adoption is accompanied by a corresponding jump in the span of control.

When the productivity gain intensity is high (Figures \ref{fig-apppro-x-h2} and \ref{fig-apppro-s-h2}), a decline in $h$ continues to strengthen the firm's incentive to use GenAI for workers' capability expansion, allowing it to lower $x_g^*$ while sustaining performance. The key difference is that the firm adopts GenAI even at extremely high hallucination rates and, in that region, optimally employs fully knowledgeable workers ($x_g^*=1$), using GenAI primarily as a copilot that enhances execution efficiency rather than as a capability amplifier. Although $x_g^*$ decreases as $h$ falls, it remains above the no-GenAI benchmark, implying lower expert demand and thus a larger span of control relative to no adoption. As $h$ declines further and $x_g^*$ falls, a larger mass of tasks requires expert intervention, thereby generating the monotonic decline in the span of control as shown in Figure \ref{fig-apppro-s-h2}.

\begin{figure}[h]
    \centering
    \subfloat[Worker knowledge level]{
    \includegraphics[width=0.4\textwidth]{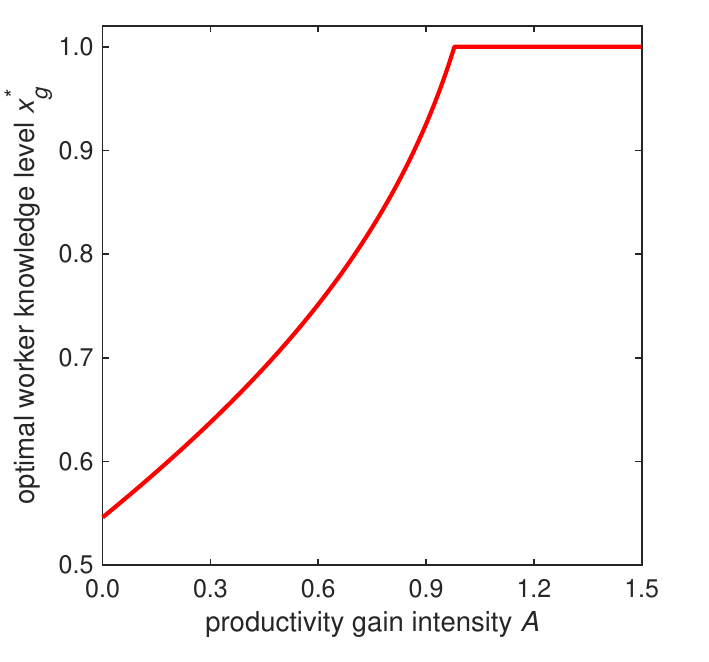}{\label{fig-apppro-x-A}}
    }
    \quad\quad
    \subfloat[Span of control]{
    \includegraphics[width=0.4\textwidth]{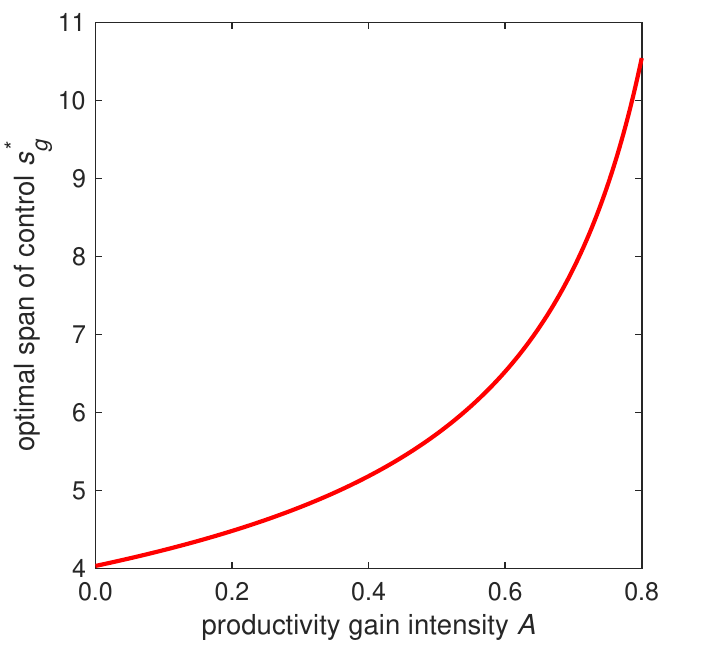}{\label{fig-apppro-s-A}}
    }
    \caption{Impact of increasing productivity gain intensity under productivity improvement extension \\  ($k=0.4$, $w=0.2$, $t_c=0.6$, $t_v=0.3$, $r_g=0.3$, $h=0.2$)}
    \label{fig-apppro-A}
\end{figure}

Finally, we examine how the productivity gain intensity $A$ affects the firm's optimal skill choice and hierarchy. A higher $A$ can be interpreted as tighter GenAI integration into frontline workflows, greater worker proficiency in utilizing GenAI, or technological advances that deliver larger throughput gains for a given augmentation level. Figure \ref{fig-apppro-x-A} shows that $x_g^*$ increases monotonically in $A$ and eventually reaches the upper bound of one. Intuitively, a larger $A$ raises the return from tasks handled within the worker's intrinsic knowledge domain, because these tasks are executed more quickly under human-AI collaboration. The firm therefore finds it optimal to hire more knowledgeable workers so that a larger share of tasks can be processed within $[0,x]$ and benefit from the productivity gain. Consistent with this mechanism, Figure \ref{fig-apppro-s-A} shows that the optimal span of control also increases monotonically in $A$: as the firm continuously increases $x_g^*$, fewer tasks require expert involvement, enabling each expert to supervise a larger frontline workforce. In the extreme case where $x_g^*=1$, the conventional two-tier hierarchy collapses to a single layer of fully knowledgeable workers who use GenAI as a copilot across all tasks, eliminating the need for senior expert consultation while maximizing processing efficiency.

\clearpage

\section{GenAI Capability-Reliability Coupling}\label{app-extension-rh}
Our baseline analysis treats the hallucination rate and GenAI capability as independent parameters. This modeling choice is not a claim about engineering trajectories; rather, it is an analytical strategy that isolates two organizationally distinct channels through which GenAI affects organizational structure. Capability determines the scope of tasks subject to AI processing, reshaping the boundaries of task allocation across the organizational hierarchy. Reliability determines the per-task oversight burden, governing how much validation and rework each AI-processed task requires. These channels have fundamentally different organizational implications, as scope affects the extensive margin of task allocation, while reliability affects the intensive margin of oversight costs. Conflating them would obscure the mechanisms driving our key results.

The empirical relationship between capability and reliability is itself context-dependent. While engineering advances such as reinforcement learning from human feedback (RLHF), retrieval-augmented generation (RAG), and chain-of-thought verification have reduced average hallucination rates as models scale, domain-specific evidence reveals a more nuanced picture.\footnote{\href{https://openai.com/index/openai-o1-system-card/?utm_source}{https://openai.com/index/openai-o1-system-card/?utm\_source}} \cite{zhou2024larger} document that more capable language models can exhibit equal or higher hallucination rates on certain task categories, and \cite{magesh2024hallucination} find that state-of-the-art legal AI tools hallucinate at rates exceeding 17\% despite representing the most capable available systems. More fundamentally, \cite{kalai2024calibrated} establish a theoretical impossibility result: calibrated language models must hallucinate at non-trivial rates due to inherent statistical properties of the generation process, implying a floor on hallucinations that does not vanish with capability improvements. 
Our baseline independence assumption is therefore best understood as capturing environments in which one dimension of GenAI performance may improve faster than another, so that capability and reliability need not advance proportionally for the specific task domain in question, a pattern consistent with the ``jagged frontier'' documented by \cite{dell2023navigating}.

In this appendix, we show that our main results are robust to an inverse relationship between $h$ and $r$, confirming that our organizational predictions are driven by structural features of GenAI (intrinsic fallibility and deployment flexibility) rather than by a specific assumption about the capability-reliability independence.
Specifically, we assume a linear relationship: $h(r_i)=b(1-r_i)$, where $r_i \in \{r_t, r_g, r_e, r_u\}$ represents the capability parameter specific to each deployment mode as defined in Sections \ref{section-w-auto}--\ref{section-expert-adoption}, and $b \in (0,1)$ is the baseline hallucination coefficient. Under this specification, $r_i$ serves as a comprehensive proxy for the quality of the GenAI technology. A higher $r_i$ implies a superior system that not only automates or augments a wider range of tasks but also does so with a lower intrinsic probability of error. In the following subsections, we revisit the four deployment modes under this interdependent specification to examine the robustness of our main insights. For ease of exposition, we refer the reader to the proofs of statements provided in Appendix \ref{app-proof} for related technical details.
\subsection{Worker-Level Automation}
\begin{proposition}\label{app-prop-wauto-rh}
\begin{enumerate}[$(a)$]
    \item The firm should adopt GenAI if and only if the hallucination coefficient $b<\hat{b}_t$.

    \item The optimal worker knowledge level under worker-level automation is 
    $$\hat{x}_t^*=\max\{r_t, \min\{1, \hat{x}_t\}\},$$
where $\hat{x}_t=\frac{(k+2w)t_c}{2k[1-bt_rr_t^2-(1-t_v-bt_r)r_t]}$. Furthermore, $\hat{x}_t^*$ increases in GenAI's automation capability $r_t$ and $\hat{x}_t^*>x_0^*$. 
\end{enumerate}    
\end{proposition}

Proposition \ref{app-prop-wauto-rh} confirms that the upskilling effect remains robust when GenAI capability and reliability are structurally linked. When the automation capability $r_t$ serves as a comprehensive proxy for GenAI quality, implying that a broader scope naturally correlates with a lower hallucination rate, GenAI adoption still drives the firm to raise worker knowledge requirements and this upskilling phenomenon intensifies as GenAI quality improves (i.e., as $r_t$ increases). Furthermore, when GenAI quality becomes sufficiently high, the organization undergoes a fundamental structural transformation (Corollary \ref{app-coro-wauto-rh}). In this regime, the combination of extensive automation capability and high reliability incentivizes the firm to employ fully knowledgeable workers ($\hat{x}_t^*=1$) who can independently validate AI-generated solutions and execute the few remaining tasks. Consequently, the traditional two-layer hierarchy collapses into a single-layer structure, rendering the specialized expert layer redundant (Figure \ref{fig-corre-wauto-x}).

\begin{corollary}\label{app-coro-wauto-rh}
After the adoption of worker-level automation, the organizational hierarchy collapses into a single-layer structure comprised solely of workers who hold complete knowledge, i.e., $\hat{x}_t^*=1$, when $r_t\ge\hat{r}_1$.    
\end{corollary}

Due to analytical intractability, we employ numerical experiments to demonstrate how advancements in GenAI quality ($r_t$) reshape organizational structure. Figure \ref{fig-corre-wauto-s} shows that the span of control initially decreases and then increases as GenAI capability improves. This finding is consistent with our main results regarding how advancements in GenAI technology influence organizational structure under an exogenous hallucination rate (Proposition \ref{prop-workerauto-s}).

\begin{figure}[h]
    \centering
    \subfloat[Worker knowledge level]{
    \includegraphics[width=0.4\textwidth]{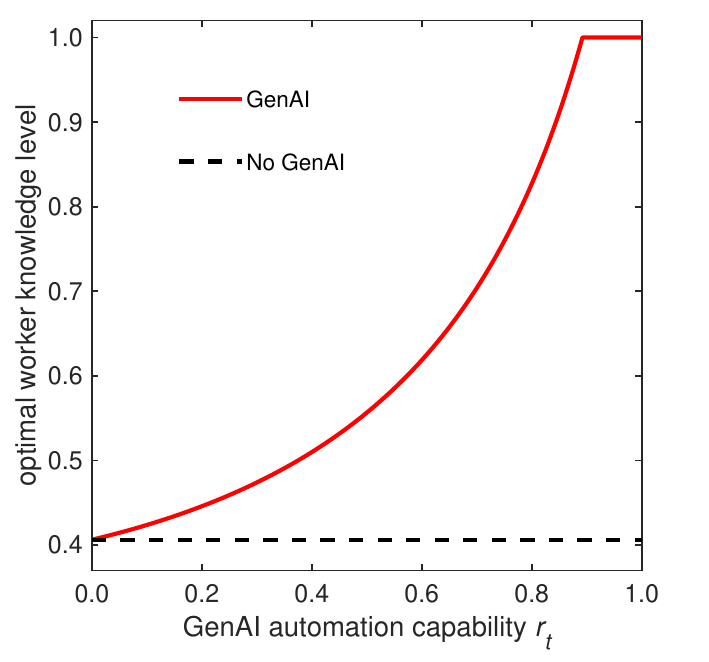}{\label{fig-corre-wauto-x}}
    }
    \quad\quad
    \subfloat[Span of control]{
    \includegraphics[width=0.4\textwidth]{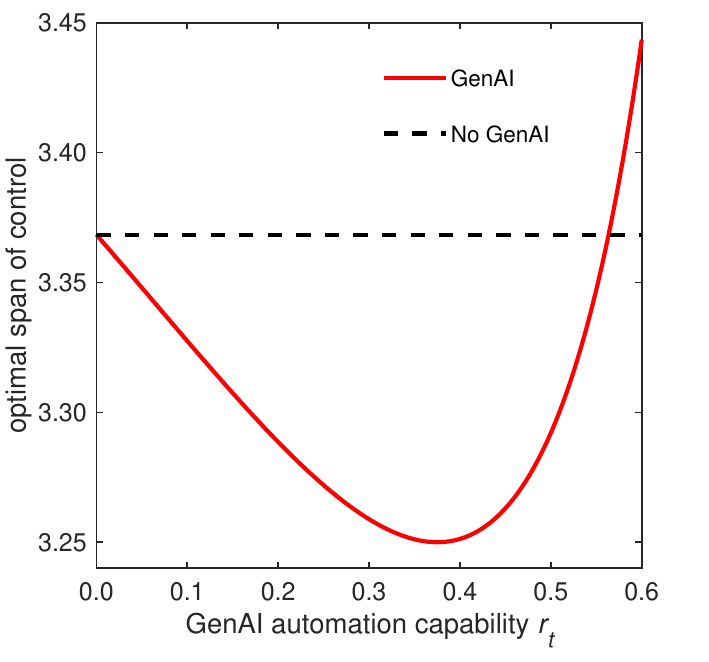}{\label{fig-corre-wauto-s}}
    }
    \caption{Impact of GenAI advancements on under worker-level automation \\
    ($k=0.8$, $w=0.25$, $t_c=0.5$, $b=0.4$, $t_v=0.3$, $t_r=0.8$)}
    \label{fig-corre-wauto}
\end{figure}

\subsection{Worker-Level Augmentation}
\begin{proposition}\label{app-prop-waug-rh-x}
\begin{enumerate}[$(a)$]
    \item The firm should adopt GenAI if and only if $r_g>\hat{r}_g$, where $\hat{r}_g=1-\frac{t_c-t_v}{bt_c}$.

    \item The optimal worker knowledge level under worker-level augmentation is 
    $$\hat{x}_g^*=\frac{(k+2w)[r_gt_v+(1-r_g)t_c+br_g(1-r_g)t_c]}{2k}.$$
    $\hat{x}_g^*$ decreases in GenAI's augmentation capability $r_g$ and $\hat{x}_g^*<x_0^*$.
\end{enumerate}    
\end{proposition}

Proposition \ref{app-prop-waug-rh-x} characterizes the adoption conditions and optimal worker knowledge level under worker-level augmentation. GenAI adoption is profitable only when the augmentation capability $r_g$ exceeds a certain threshold (i.e., $r_g>\hat{r}_g$). Given the inverse relationship between hallucination rates and capability, a high $r_g$ implicitly guarantees a low hallucination rate, a result consistent with the reliability constraint established in Lemma \ref{lemma-w-aug}. Furthermore, Proposition \ref{app-prop-waug-rh-x}(b) confirms that the deskilling effect persists in this extended setting, and as GenAI quality improves (i.e., as $r_g$ increases), it optimal for the firm to further relax worker knowledge requirements, thereby intensifying the deskilling phenomenon (see Figure \ref{fig-corre-waug-x}).

\begin{proposition}\label{app-prop-waug-rh-s}
After the adoption of worker-level augmentation, 
\begin{enumerate}[$(a)$]
    \item For $t_c>\frac{k}{k+2w}$, the span of control $\hat{s}_g^*$ decreases in GenAI's augmentation capability $r_g$ on $(\hat{r}_g, \hat{r}_2)$ and increases in $r_g$ on $(\hat{r}_2, 1)$. 

    \item For $t_c\le\frac{k}{k+2w}$,  the span of control $\hat{s}_g^*$ increases in GenAI's augmentation capability $r_g$ on $(\hat{r}_g, 1)$. 
\end{enumerate}
\end{proposition}

Proposition \ref{app-prop-waug-rh-s} and Figure \ref{fig-corre-waug-s} confirm that the non-monotonic, U-shaped trajectory of the span of control remains robust even when the hallucination rate is inversely correlated with GenAI capability. Consistent with the baseline model, the initial contraction is driven by a surge in expert referrals due to aggressive deskilling, while the subsequent expansion results from the dominance of direct augmentation gains as GenAI quality improves.

\begin{figure}[h]
    \centering
    \subfloat[Worker knowledge level]{
    \includegraphics[width=0.4\textwidth]{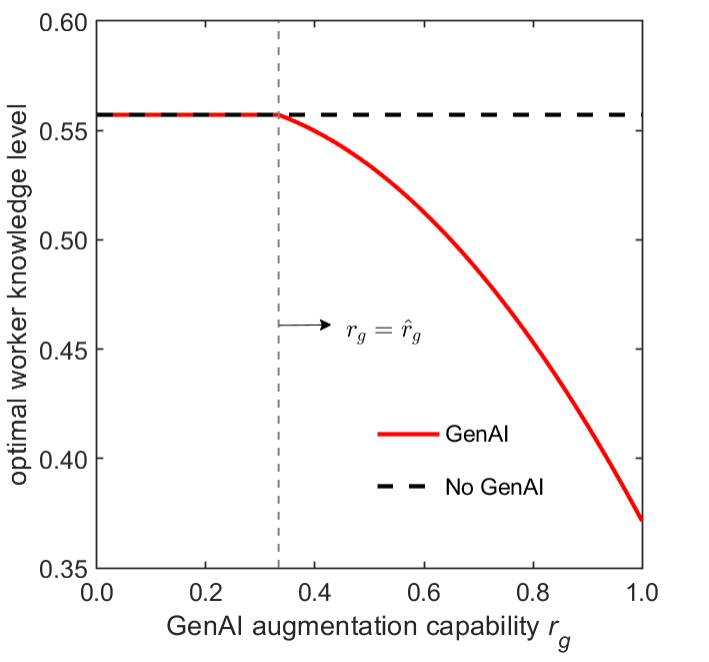}{\label{fig-corre-waug-x}}
    }
    \quad\quad
    \subfloat[Span of control]{
    \includegraphics[width=0.4\textwidth]{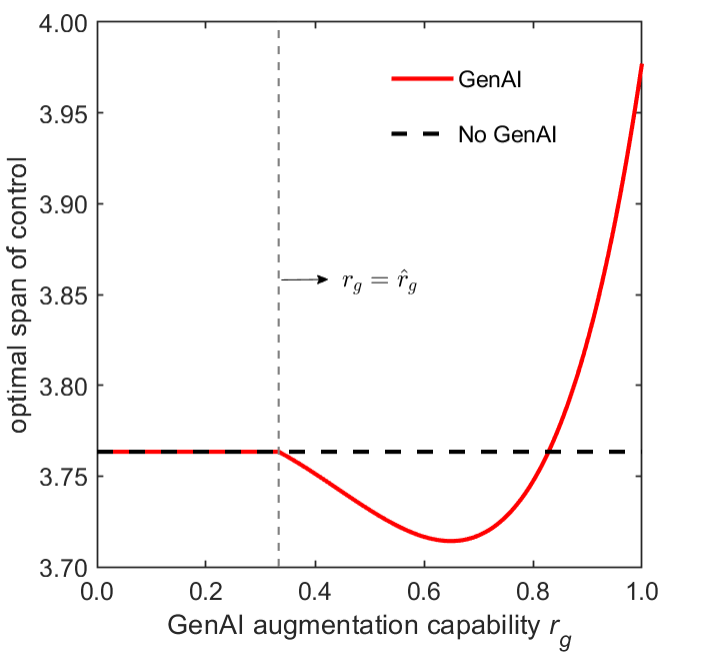}{\label{fig-corre-waug-s}}
    }
    \caption{Impact of GenAI advancements on under worker-level augmentation \\
    ($k=0.7$, $w=0.3$, $t_c=0.6$, $b=0.5$, $t_v=0.4$)}
    \label{fig-corre-waug}
\end{figure}

\subsection{Expert-Level Automation}
\begin{proposition}\label{app-prop-eauto-rh-x}
\begin{enumerate}[$(a)$]
    \item The firm should adopt GenAI if and only if $r_e>\hat{r}_e$.

    \item The optimal worker knowledge level under expert-level automation is 
    $\hat{x}_e^*=\frac{(k+2w)[t_v+b(1-r_e)t_c]}{2k}$.
    $\hat{x}_e^*$ decreases in GenAI's augmentation capability $r_e$ and $\hat{x}_e^*<x_0^*$.
\end{enumerate}        
\end{proposition}

Proposition \ref{app-prop-eauto-rh-x} characterizes the optimal adoption condition and worker knowledge level under expert-level automation. Under the interdependent specification, a high capability level implicitly guarantees a low hallucination rate; thus, substantial GenAI capability ($r_e>\hat{r}_e$) renders adoption profitable by simultaneously offloading a large portion of expert work and ensuring high reliability. A critical distinction from the main text, where capability and reliability are independent, is that the optimal worker knowledge level is no longer invariant to GenAI capability. Instead, after GenAI adoption, $\hat{x}_e^*$ strictly decreases as capability improves (Figure \ref{fig-corre-eauto-x}). This result arises because the resulting decline in hallucination rates lowers the expected time human experts spend correcting AI-guided outputs, thereby reducing the marginal benefit of maintaining high worker knowledge and incentivizing the firm to further relax worker skill requirements to economize on wages.

\begin{proposition}\label{app-prop-eauto-rh-s}
After the adoption of expert-level automation, the optimal span of control $\hat{s}_e^*$ decreases in GenAI capability $r_e$ on $(\hat{r}_e, \hat{r}_3)$ and increases in $r_e$ on $(\hat{r}_3, 1)$.     
\end{proposition}

\begin{figure}[h]
    \centering
    \subfloat[Worker knowledge level]{
    \includegraphics[width=0.4\textwidth]{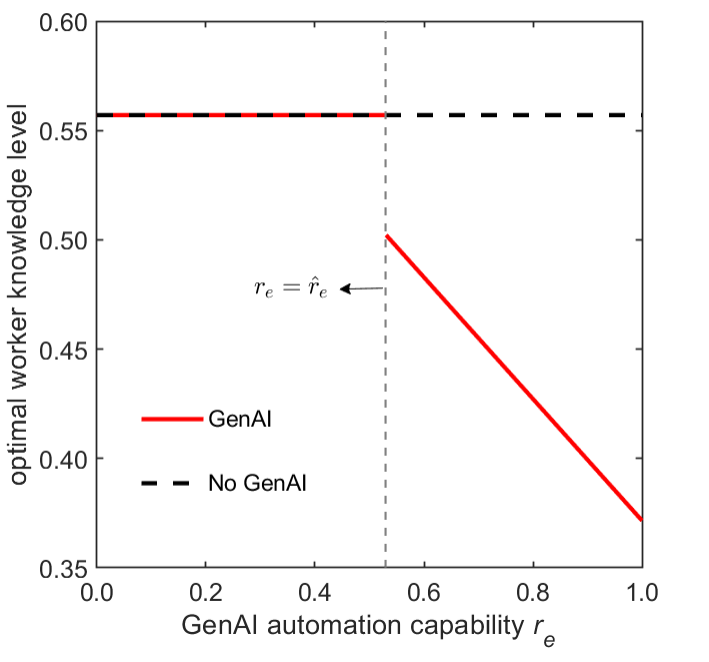}{\label{fig-corre-eauto-x}}
    }
    \quad\quad
    \subfloat[Span of control]{
    \includegraphics[width=0.4\textwidth]{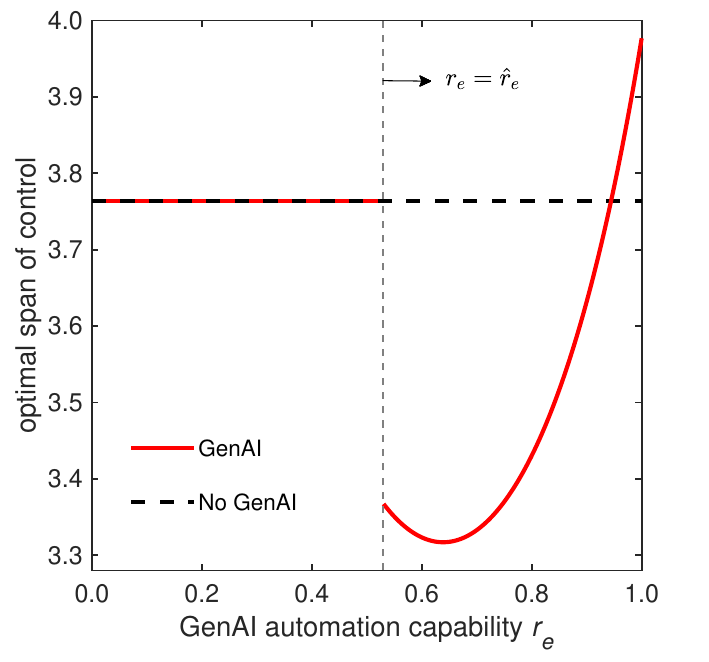}{\label{fig-corre-eauto-s}}
    }
    \caption{Impact of GenAI advancements on under expert-level automation \\
    ($k=0.7$, $w=0.3$, $t_c=0.6$, $b=0.5$, $t_v=0.4$)}
    \label{fig-corre-eauto}
\end{figure}

Proposition \ref{app-prop-eauto-rh-s} further reveals that, consistent with the baseline scenario where capability and reliability are independent, the initial adoption of expert-level automation triggers a structural discontinuity: the immediate reduction in worker knowledge level generates a surge in the validation and correction workload for experts, causing a sharp initial contraction in the span of control. However, a critical divergence from the baseline model, where the span of control monotonically expands post-adoption, is the emergence of a U-shaped trajectory as GenAI capability advances (see Figure \ref{fig-corre-eauto-s}). This non-monotonic pattern arises because the optimal worker knowledge level is no longer fixed; instead, as GenAI capability improves, the firm finds it optimal to further lower worker skill requirements. In the intermediate regime ($\hat{r}_e<r_e<\hat{r}_3$), this aggressive deskilling amplifies the volume of AI-guided outputs requiring human expert validation and correction, increasing the demand for experts and narrowing the span of control. Eventually, as GenAI capability reaches sufficiently high levels ($r_e>\hat{r}_3$), the direct efficiency gains from broad-scope automation and minimal hallucination rates dominate the deskilling-induced burden, reducing the net expert workload and allowing the span of control to expand.

Comparing the results of this extension with the baseline model clarifies the distinct mechanisms driving organizational change. In the baseline analysis, we established that pure capability expansion, when isolated from reliability, leaves the optimal worker knowledge level invariant and leads to a strictly monotonic expansion of the span of control. Consequently, the novel dynamics observed in this extension, specifically, the progressive deskilling and the initial contraction of the span of control, are not driven by the expanded automation capability itself. Instead, they are structurally attributable to the implicit reliability gains (i.e., the reduction in $h$) accompanying higher capability.

\subsection{Expert-Level Augmentation}
\begin{proposition}\label{app-prop-eaug-rh}
\begin{enumerate}[$(a)$]
    \item The firm should adopt GenAI if and only if $r_u > \hat{r}_u$, where $\hat{r}_u=\frac{b}{1+b}$.
    
    \item After GenAI adoption, the optimal worker knowledge level is 
    $\hat{x}_u^*=\frac{(k+2w)(1+b)(1-r_u)t_c}{2k}$, which decreases in GenAI's augmentation capability $r_u$ and satisfies $\hat{x}_u^*<x_0^*$.
    
    \item After GenAI adoption, the optimal span of control 
    $\hat{s}_u^*$ decreases in GenAI augmentation capability $r_u$ on $(\hat{r}_u, \hat{r}_4)$ and increases in $r_u$ on $(\hat{r}_4, 1)$.
\end{enumerate}    
\end{proposition}

Proposition \ref{app-prop-eaug-rh} confirms that the fundamental economic dynamics of expert-level augmentation remain consistent with the main text. The firm adopts GenAI only when the augmentation capability exceeds a certain threshold ($r_u > \hat{r}_u$). Under the interdependent specification, a high $r_u$ implicitly guarantees the low hallucination rate required for profitability. Post-adoption, the deskilling effect persists and as GenAI quality improves, it intensifies. Consequently, the span of control follows a U-shaped pattern with respect to GenAI advancement: initially narrowing due to the surge in referrals caused by deskilling, before eventually expanding as the direct efficiency gains from faster consultation dominate the workload.

In summary, this extension establishes the robustness of our core insights when the hallucination rate is inversely related to GenAI capability. Although this interdependence introduces additional nuances (most notably under expert-level automation), the fundamental conclusion regarding skill formation remains intact. Worker-level automation continues to generate upskilling incentives, whereas the remaining deployment modes consistently induce deskilling. Moreover, the qualitative evolution of organizational structure is preserved: across all modes, the span of control exhibits a first-decreasing-then-increasing pattern, reflecting the shifting dominance between direct efficiency gains and indirect skill adjustments.

\clearpage
\section{GenAI Deployment Cost}\label{app-cost}
In the baseline model, we focus on the organizational mechanism through which GenAI reshapes hierarchical structure. Accordingly, the main text explicitly models the organizational costs that arise when GenAI is integrated into workflows, including validation, rework, and consultation burdens. These costs are endogenous consequences of deployment mode and deployment location, and they directly determine when GenAI adoption is profitable within the baseline framework. At the same time, firms may also incur direct implementation and technology-acquisition costs, such as procurement, licensing, system integration, fine-tuning, or other investments needed to improve GenAI performance. Because such costs are often highly firm- and context-specific and are not central to the main organizational mechanism studied in the baseline model, we abstract from them in the main text and treat them separately in this appendix.

In this appendix, we extend the baseline framework by explicitly incorporating these implementation costs through endogenous technology choice. Specifically, we allow firms to strategically invest in GenAI's capability level and reliability. We primarily examine the scenario where the firm endogenously determines the capability level while the hallucination rate remains exogenously fixed. This aligns with the observation that AI models with more advanced capabilities do not necessarily exhibit lower hallucination rates \citep{zhou2024larger}. Building upon this foundational setup, we further explore a generalized dual-investment setting where both the capability level and the hallucination rate are endogenous decision variables. This reflects a common deployment process: after selecting a base AI capability, firms can make further investments in reliability, such as through fine-tuning or retrieval-augmented generation (RAG), to reduce the hallucination rate. Crucially, to faithfully capture the intrinsic fallibility of GenAI, we employ structurally distinct cost functions for capability enhancement and hallucination reduction. In particular, the cost function for lowering the hallucination rate incorporates a fundamental technological constraint: regardless of the magnitude of investment, GenAI hallucinations can never be completely eliminated, so the marginal cost of further hallucination reduction becomes arbitrarily large as the hallucination rate approaches zero.
\subsection{Worker-Level Automation}
We first consider the case where the firm strategically invests to endogenously determine the GenAI's automation capability level, $r_t$, while keeping the hallucination rate, $h$, fixed. The firm's profit function is thus given by
\begin{equation*}
    \Pi_t(x, r_t)=1-\left(w+\frac{1}{2}kx^2\right)[(1-r_t)+(t_v+ht_r)r_t]-\left(w+\frac{1}{2}k\right)(1-x)t_c-\frac{1}{2}c_t^rr_t^2. 
\end{equation*}
We capture the cost of GenAI deployment using an increasing and strictly convex cost function, $\frac{1}{2}c_t^r r_t^2$. The parameter $c_t^r > 0$ serves as the cost coefficient for capability investment. A smaller value of $c_t^r$ indicates that the firm can enhance GenAI automation capability at a lower marginal cost, reflecting a more advanced underlying GenAI technology. Consequently, the firm's decision problem is to jointly optimize the AI capability $r_t$ and its labor structure, represented by the optimal worker knowledge level $x_t$.

\begin{figure}[h]
    \centering
    \subfloat[GenAI capability and worker knowledge]{
    \includegraphics[width=0.4\textwidth]{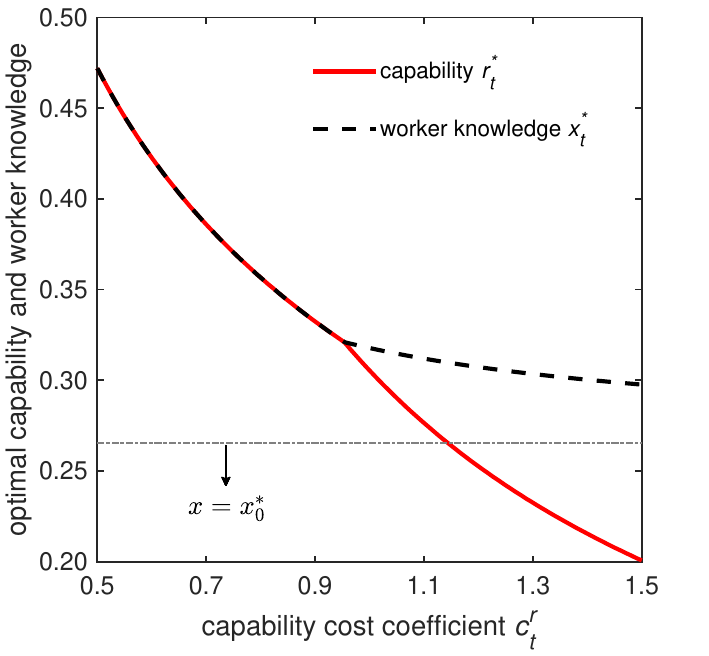}{\label{fig-cost-wauto-r-rx}}
    }
    \quad\quad
    \subfloat[Span of control]{
    \includegraphics[width=0.4\textwidth]{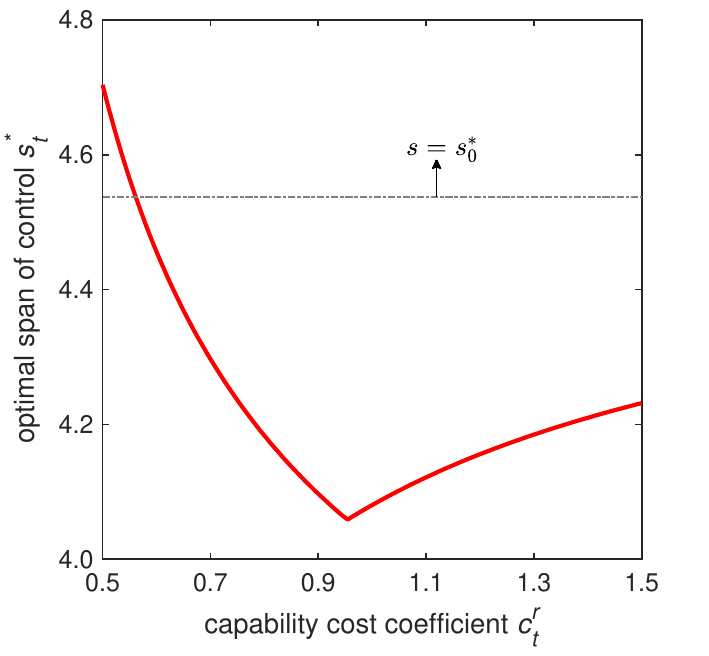}{\label{fig-cost-wauto-r-s}}
    }
    \caption{Impacts of capability cost coefficient with endogenous capability and fixed hallucination rate under worker-level automation \\ ($k=1.3$, $w=0.5$, $t_c=0.3$, $t_v=0.3$, $t_r=0.8$, $h=0.2$)}
    \label{fig-cost-wauto-r}
\end{figure}

Due to analytical intractability, we resort to numerical experiments to examine how the capability cost parameter, $c_t^r$, shapes the firm's investment in GenAI and the resulting organizational design. We interpret a decline in $c_t^r$ as technological advancement that lowers the marginal cost of capability. As $c_t^r$ falls, the firm optimally increases its investment in GenAI capability, thereby expanding the scope of tasks automated by GenAI. Crucially, Figure \ref{fig-cost-wauto-r-rx} reveals a robust upskilling effect: the optimal worker knowledge level (the dashed line) consistently exceeds the no-GenAI benchmark (the dash-dotted line), confirming that worker-level automation incentivizes the firm to hire more knowledgeable workers. Furthermore, the co-movement between worker knowledge and capability exhibits a clear regime shift. When $c_t^r$ is relatively high, the firm selects only modest capability. In this region, workers could validate the limited AI scope with lower knowledge, yet the firm sets worker knowledge strictly above the AI automation boundary to reduce costly upward referrals to experts. As $c_t^r$ decreases further, the optimal capability becomes large and validation needs become pervasive; the firm then chooses worker knowledge that aligns exactly with the AI automation scope, achieving reliable oversight of extensive AI outputs without paying additional wage premia beyond what validation requires.

Figure \ref{fig-cost-wauto-r-s} reports the corresponding implications for span of control. When $c_t^r$ is high, marginal reductions in $c_t^r$ mainly expand automation while inducing only gradual upskilling on entry-level workers. The direct automation effect therefore displaces frontline workers, whereas expert demand remains largely unchanged, so the span of control contracts. Once $c_t^r$ falls below a threshold, the expanded automation scope forces a sharp increase in worker knowledge to sustain reliable validation. This stronger upskilling reduces reliance on expert consultation, compresses expert demand, and ultimately widens the span of control.

Having established the organizational dynamics under a fixed hallucination rate, we now extend our framework to a generalized setting where the firm can simultaneously invest in expanding the GenAI's automation capability, $r_t$, and enhancing its reliability by lowering the hallucination rate, $h$. The firm's updated profit function is given by:
\begin{equation*}
    \Pi_t(x, r_t, h)=1-\left(w+\frac{1}{2}kx^2\right)[(1-r_t)+(t_v+ht_r)r_t]-\left(w+\frac{1}{2}k\right)(1-x)t_c-\frac{1}{2}c_t^rr_t^2-\frac{c_t^h}{10}\left(\frac{\overline{h}_0}{h}-1\right). 
\end{equation*}
The term $\frac{c_t^h}{10}\left(\frac{\overline{h}_0}{h}-1\right)$ captures the cost of hallucination reduction, where $\overline{h}_0$ is the baseline hallucination rate without investment and $c_t^h > 0$ is the corresponding cost coefficient. We adopt this specific functional form to reflect GenAI's intrinsic fallibility. The marginal cost of reliability investment strictly increases as $h$ decreases. More importantly, as $h$ approaches zero, the required investment cost asymptotically approaches infinity. This property formally captures the fundamental technological constraint that GenAI hallucinations can never be completely eliminated, regardless of the investment magnitude.

\begin{figure}[h]
    \centering
    \subfloat[Hallucination rate]{
    \includegraphics[width=0.33\textwidth]{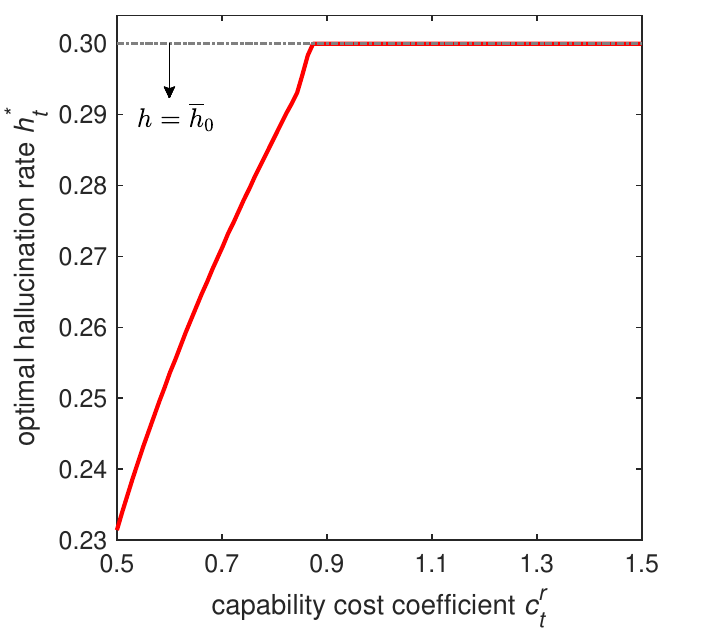}\label{fig-cost-wauto-rh-rh}
    }
    \subfloat[Capability and worker knowledge]{
    \includegraphics[width=0.32\textwidth]{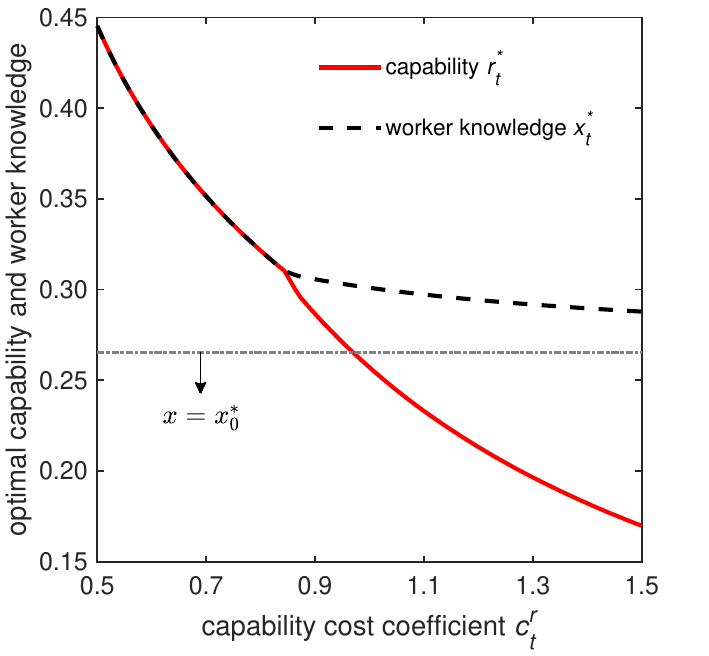}\label{fig-cost-wauto-rh-rrx}
    }
    \subfloat[Span of control]{
    \includegraphics[width=0.32\textwidth]{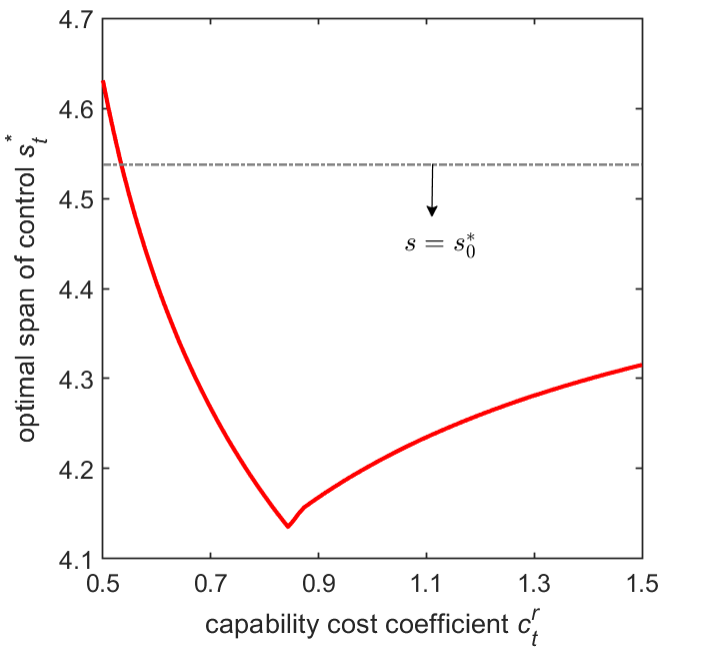}\label{fig-cost-wauto-rh-rs}
    }~\\[2mm]
    \caption{Impacts of capability cost coefficient with endogenous capability and hallucination rate under worker-level automation \\  ($k=1.3$, $w=0.5$, $t_c=0.3$, $t_v=0.3$, $t_r=0.8$, $c_t^h=0.4$, $\overline{h}_0=0.3$)}
    \label{fig-cost-wauto-rh-r}
\end{figure}

Figure \ref{fig-cost-wauto-rh-r} presents how variations in the capability cost coefficient, $c_t^r$, affect the firm's dual investment strategy and the resulting organizational implications under endogenous $r_t$ and $h$. The optimal GenAI capability, the worker knowledge level, and the span of control exhibit patterns that are qualitatively similar to those observed under the fixed-reliability case (see Figures \ref{fig-cost-wauto-rh-rrx} and \ref{fig-cost-wauto-rh-rs}). Furthermore, Figure \ref{fig-cost-wauto-rh-rh} reveals a powerful complementary effect: once $c_{t}^{r}$ drops below a critical threshold, any further decrease in the cost of capability expansion also stimulates the firm to invest more aggressively in hallucination reduction. Thus, the deployed GenAI system achieves a broader automation scope and stronger reliability simultaneously.

\begin{figure}[h]
    \centering
    \subfloat[Hallucination rate]{
    \includegraphics[width=0.33\textwidth]{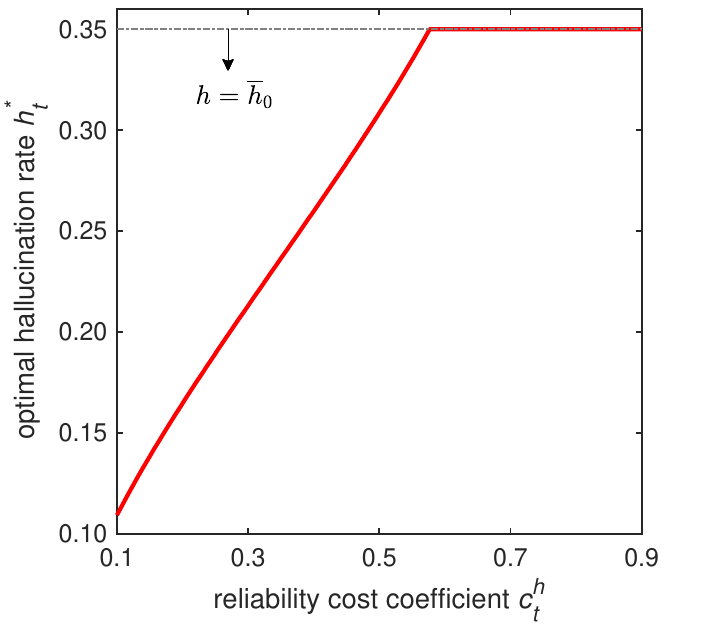}\label{fig-cost-wauto-rh-hh}
    }
    \subfloat[Capability and worker knowledge]{
    \includegraphics[width=0.32\textwidth]{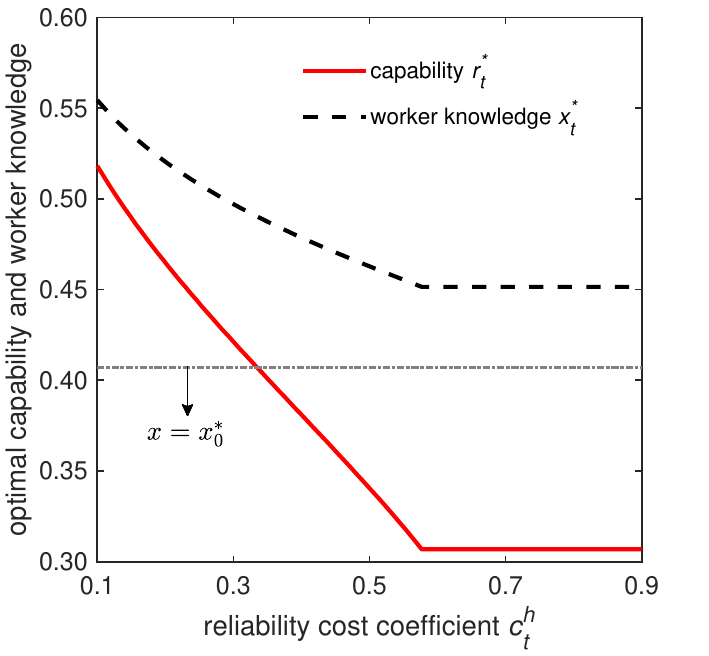}\label{fig-cost-wauto-rh-hrx}
    }
    \subfloat[Span of control]{
    \includegraphics[width=0.325\textwidth]{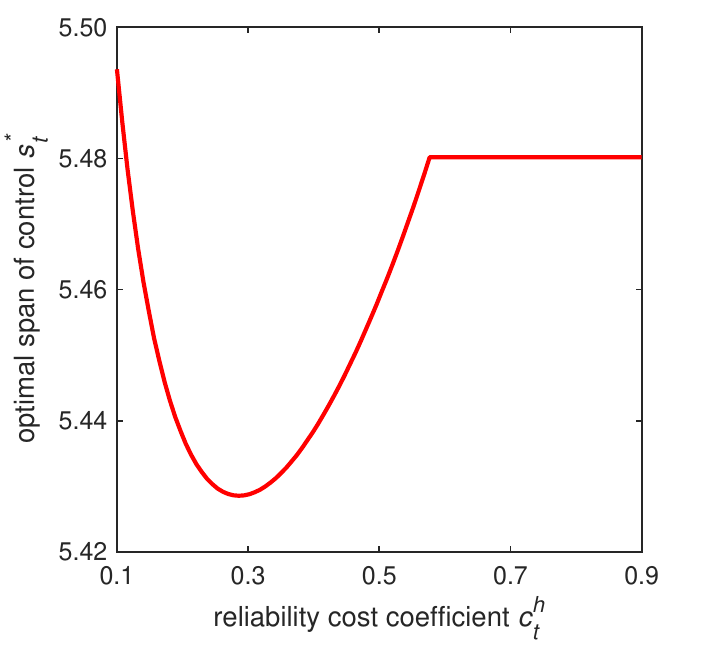}\label{fig-cost-wauto-rh-hs}
    }~\\[2mm]
    \caption{Impacts of reliability cost coefficient with endogenous capability and hallucination rate under worker-level automation \\  ($k=0.7$, $w=0.6$, $t_c=0.3$, $t_v=0.4$, $t_r=0.8$, $c_t^r=0.7$, $\overline{h}_0=0.35$)}
    \label{fig-cost-wauto-rh-h}
\end{figure}

Figure \ref{fig-cost-wauto-rh-h} illustrates how changes in the reliability cost coefficient, $c_t^h$, shape the firm's dual investment strategy in GenAI capability and reliability, as well as the subsequent impact on organizational structure. As shown in Figure \ref{fig-cost-wauto-rh-hh}, the firm initiates investment to reduce hallucinations only when $c_t^h$ falls below a critical threshold. After that, a decrease in $c_t^h$ incentivizes greater investment in reliability, effectively driving down the GenAI hallucination rate. Figure \ref{fig-cost-wauto-rh-hrx} reveals a similar complementarity in the opposite direction: a lower $c_{t}^{h}$ (making reliability cheaper) incentivizes the firm to simultaneously expand the GenAI's automation scope. These dual improvements, lower hallucination rates and higher capability, necessitate an upskilling effect on worker knowledge levels. Finally, Figure \ref{fig-cost-wauto-rh-hs} shows that as $c_t^h$ decreases, the optimal span of control follows a non-monotonic, U-shaped trajectory. This reflects a shifting dominance between the direct reduction in the workers' correction burden brought about by lower hallucination rates and the indirect upskilling effect that reduces their reliance on experts. We omit a detailed exposition here, as this mechanism closely mirrors the dynamics analyzed in \S\ref{section-wauto-s}.
\subsection{Worker-Level Augmentation}
We next incorporate endogenous adoption costs under worker-level augmentation. Mirroring the analytical sequence in the previous subsection, we begin with the case in which the firm strategically invests in the GenAI's augmentation capability $r_g$, while holding the hallucination rate $h$, fixed. We model capability investment using a strictly convex cost function $\frac{1}{2}c_g^r r_g^2$, where the parameter $c_g^r > 0$ is the mode-specific cost coefficient for capability expansion. The firm's profit function is therefore
\begin{equation*}
    \Pi_g(x, r_g)=1-\left(w+\frac{1}{2}kx^2\right)-\left(w+\frac{1}{2}k\right)\big[r_g(1-x)(t_v+ht_c)+(1-r_g)(1-x)t_c\big]-\frac{1}{2}c_g^r r_g^2.
\end{equation*}
Proposition \ref{app-cost-waug} characterizes the firm's optimal choices under worker-level augmentation and the associated organizational implications. To guarantee strict concavity of profits in $r_g$ and the existence of an interior solution, we assume $c_g^r>\underline{c}_g^r$. The technical details are provided in the proofs in Appendix \ref{app-proof}.

\begin{proposition}\label{app-cost-waug}
\begin{enumerate}[$(a)$]
    \item The firm should adopt worker-level augmentation if and only if $h<\overline{h}_g=1-\frac{t_v}{t_c}$.

    \item After GenAI adoption, the optimal augmentation capability and worker knowledge level are 
    $$r_g^*=\frac{(k+2w)[2k-(k+2w)t_c][t_c-ht_c-t_v]}{4kc_g^r-(k+2w)^2(t_c-t_v-ht_c)^2}, ~~ x_g^*=\frac{(k+2w)[r_g^*t_v+(1+r_g^*h-r_g^*)t_c]}{2k}.$$
    Furthermore, $r_g^*$ strictly increases while $x_g^*$ strictly decreases as $c_g^r$ decreases, and $x_g^*$ satisfies $x_g^*<x_0^*$.

    \item After GenAI adoption, the optimal span of control is $s_g^*=\frac{1}{(1-x_g^*)[r_g^*(t_v+ht_c)+(1-r_g^*)t_c]}$.
    \begin{enumerate}[$(1)$]
        \item If $t_c>\frac{k}{k+2w}$, $s_g^*$ decreases in $c_g^r$ for  $\underline{c}_g^r<c_g^r<c_{g1}^r$ and increases in $c_g^r$ for $c_g^r>\max\{\underline{c}_g^r, c_{g1}^r\}$. 

    \item If $t_c\le\frac{k}{k+2w}$, $s_g^*$ decreases in $c_g^r$. 
    \end{enumerate}
\end{enumerate}
    
\end{proposition}

\begin{figure}[h]
    \centering
    \subfloat[Capability]{
    \includegraphics[width=0.32\textwidth]{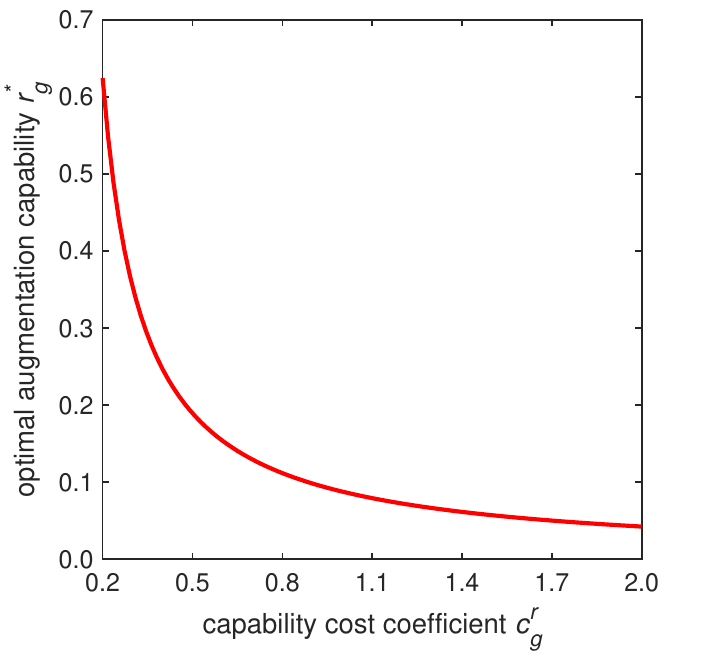}\label{fig-cost-waug-r-r}
    }
    \subfloat[Worker knowledge]{
    \includegraphics[width=0.32\textwidth]{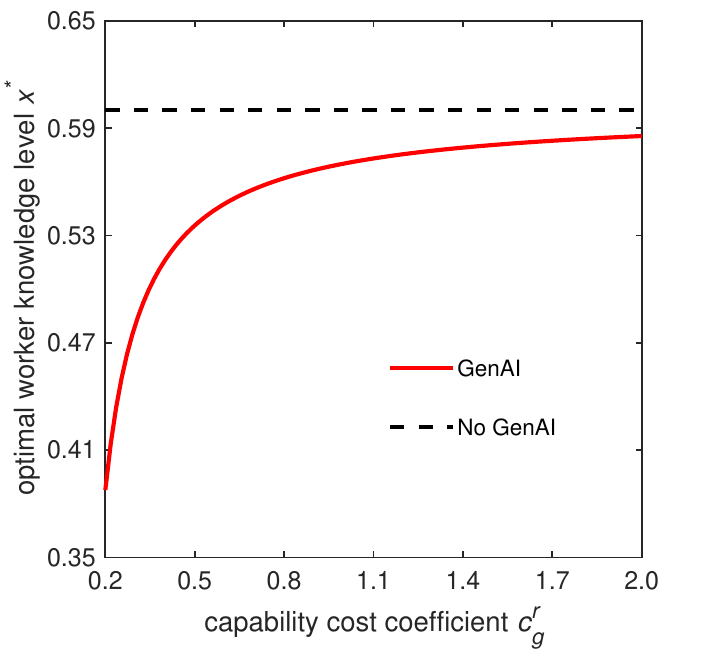}\label{fig-cost-waug-r-x}
    }
    \subfloat[Span of control]{
    \includegraphics[width=0.33\textwidth]{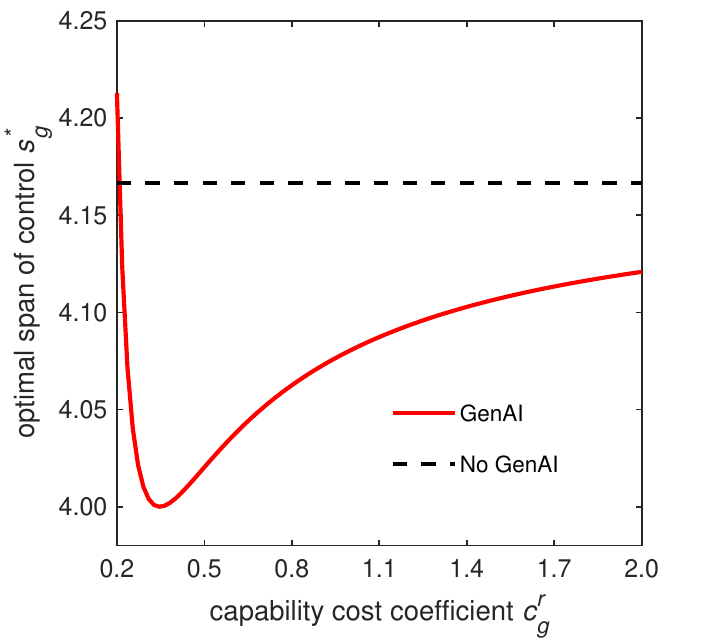}\label{fig-cost-waug-r-s}
    }~\\[2mm]
    \caption{Impacts of capability cost coefficient with endogenous capability and fixed hallucination rate  under worker-level augmentation \\  ($k=0.6$, $w=0.3$, $t_c=0.6$, $t_v=0.2$, $h=0.1$)}
    \label{fig-cost-waug-r}
\end{figure}

Proposition \ref{app-cost-waug} and Figure \ref{fig-cost-waug-r} characterize the firm's joint choice of GenAI augmentation capability and worker knowledge when the hallucination rate is fixed. As in the baseline model without adoption costs, worker-level augmentation generates a clear deskilling effect: upon adoption, the optimal worker knowledge level falls strictly below the no-GenAI benchmark. As GenAI technology advances, captured by a decline in the capability cost coefficient $c_g^r$, the firm optimally invests more in augmentation capability (Figure \ref{fig-cost-waug-r-r}). Higher capability strengthens GenAI's role as a capability amplifier, allowing the firm to further relax entry-level knowledge requirements, reduce hiring costs, and sustain performance. Proposition \ref{app-cost-waug}(c) further demonstrates the resulting organizational dynamics. When inter-tier communication is costly ($t_c>\frac{k}{k+2w}$), the span of control evolves non-monotonically as $c_g^r$ decreases, first contracting and later expanding. In particular, for ($c_g^r>\max\{\underline{c}_g^r,,c_{g1}^r\}$), declining worker knowledge increases reliance on expert validation and intervention, raising expert demand and narrowing the span of control. This mechanism implies that during early stages of GenAI improvement, organizations can face sustained, and even rising, demand for senior expertise before hierarchies ultimately flatten.

Building on the preceding analysis, we now extend the worker-level augmentation framework to allow the firm to simultaneously invest in GenAI capability and reliability. The firm's profit is therefore given by
\begin{align*}
    \Pi_g(x, r_g, h)=&1-\left(w+\frac{1}{2}kx^2\right)-\left(w+\frac{1}{2}k\right)\big[r_g(1-x)(t_v+ht_c)+(1-r_g)(1-x)t_c\big]\\
    &-\frac{1}{2}c_g^r r_g^2-\frac{c_g^h}{10}\left(\frac{\overline{h}_0}{h}-1\right).
\end{align*}
Because the economic rationale for this dual investment mirrors our discussion in the worker-level automation setting, we omit the redundant model setup details. Due to the analytical intractability of this joint optimization problem, we resort to numerical experiments. 

\begin{figure}[h]
    \centering
    \subfloat[Hallucination rate]{
    \includegraphics[width=0.4\textwidth]{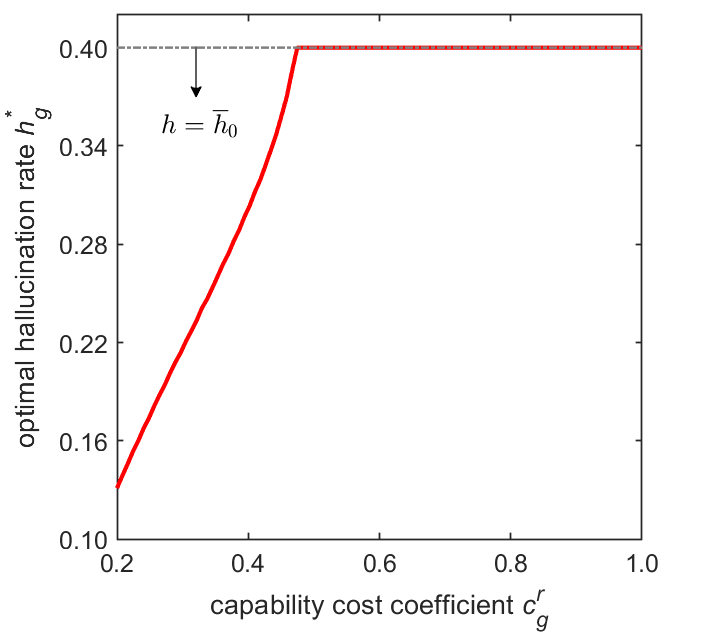}{\label{fig-cost-waug-rh-rh}}
    }
    \quad\quad
    \subfloat[Augmentation capability]{
    \includegraphics[width=0.4\textwidth]{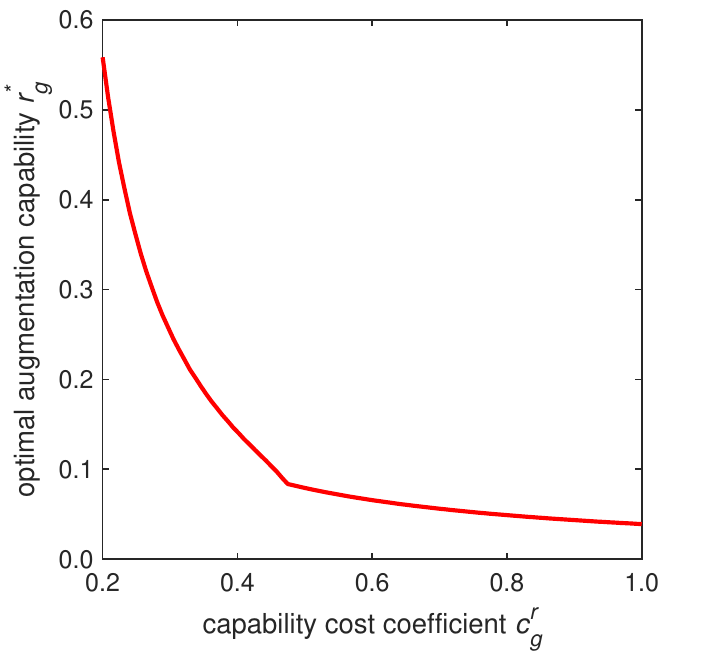}{\label{fig-cost-waug-rh-rr}}
    }
    \quad\quad\quad
    \subfloat[Worker knowledge level]{
    \includegraphics[width=0.4\textwidth]{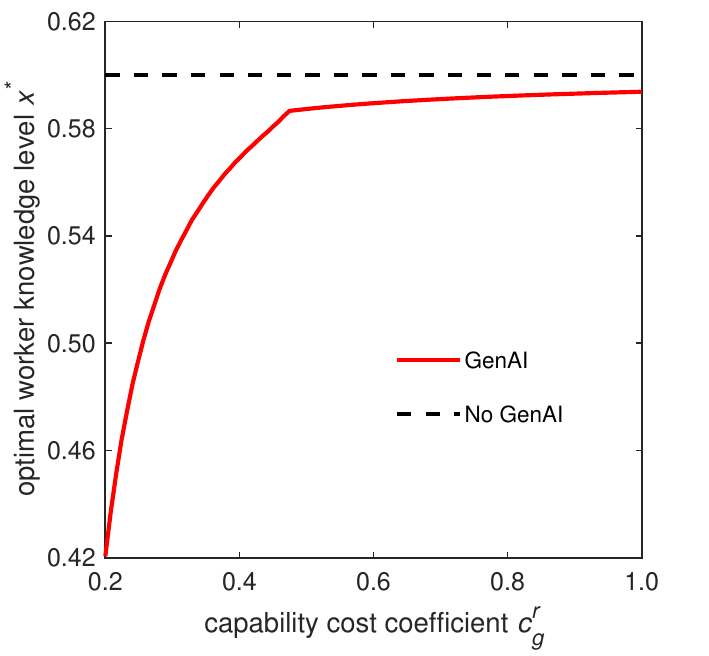}{\label{fig-cost-waug-rh-rx}}
    }
    \quad\quad
    \subfloat[Span of control]{
    \includegraphics[width=0.4\textwidth]{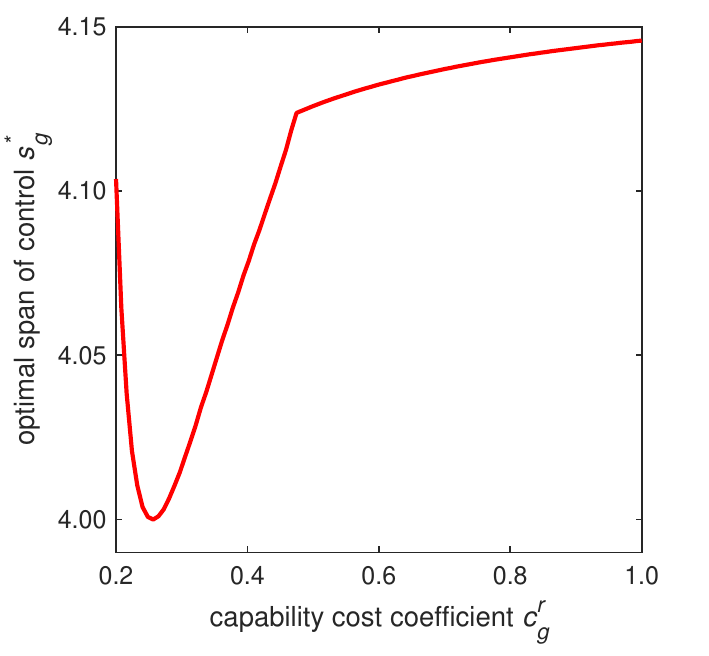}{\label{fig-cost-waug-rh-rs}}
    }
    \caption{Impacts of capability cost coefficient with endogenous capability and hallucination rate under worker-level augmentation \\  ($k=0.6$, $w=0.3$, $t_c=0.6$, $t_v=0.2$, $c_g^h=0.05$, $\overline{h}_0=0.4$)}
    \label{fig-cost-waug-rh-r}
\end{figure}

Figure \ref{fig-cost-waug-rh-r} illustrates the firm's optimal investment choices and the resulting organizational design as the capability cost coefficient, $c_g^r$, varies. Overall, the optimal GenAI augmentation capability, worker knowledge level, and span of control follow trajectories that are qualitatively similar to those under a fixed hallucination rate. The key difference is that allowing dual investments generates an amplification effect. Once $c_g^r$ drops below a specific threshold, the decline in capability costs makes it optimal for the firm to invest actively in reducing hallucinations (Figure \ref{fig-cost-waug-rh-rh}). The resulting decrease in the hallucination rate strengthens GenAI's effectiveness as a knowledge amplifier, thereby intensifying deskilling at the entry level (Figure \ref{fig-cost-waug-rh-rx}). As frontline knowledge drops more sharply, tasks increasingly require expert oversight and intervention, so the optimal span of control contracts much more steeply as $c_g^r$ continues to fall (Figure \ref{fig-cost-waug-rh-rs}).

\begin{figure}[h]
    \centering
    \subfloat[Hallucination rate]{
    \includegraphics[width=0.4\textwidth]{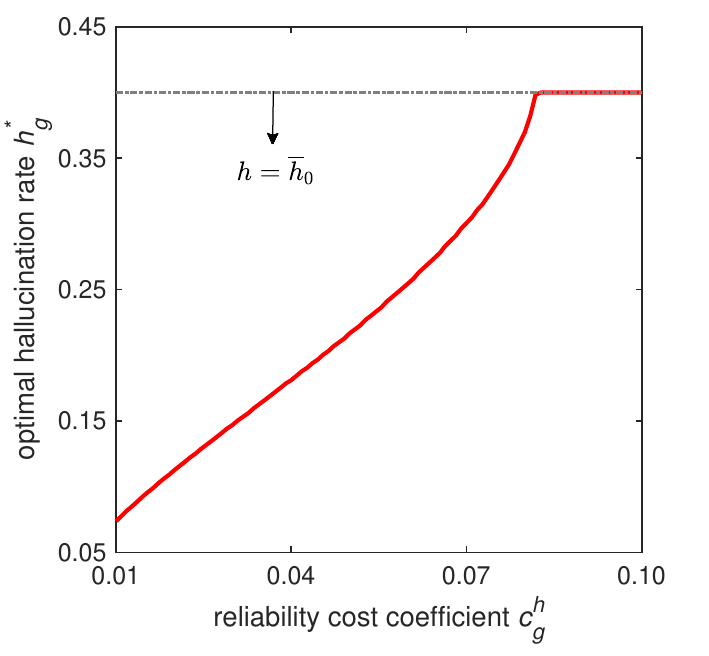}{\label{fig-cost-waug-rh-hh}}
    }
    \quad\quad
    \subfloat[Augmentation capability]{
    \includegraphics[width=0.4\textwidth]{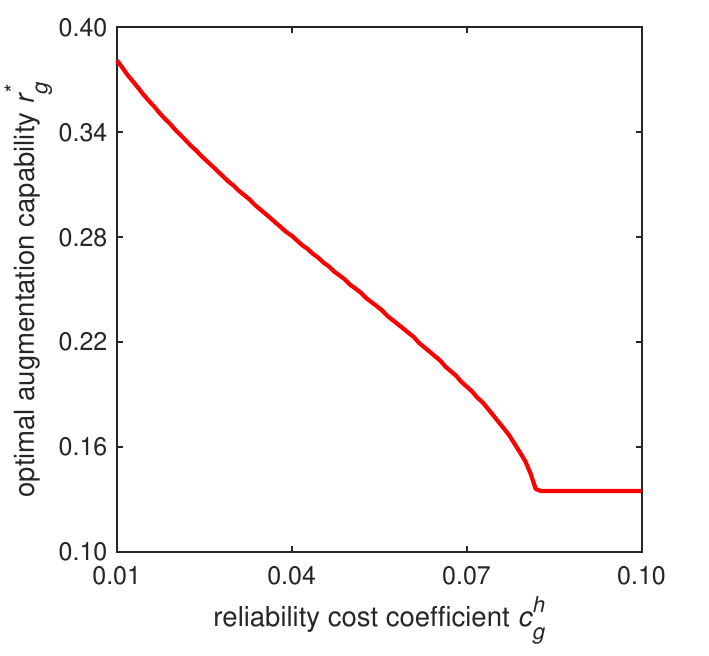}{\label{fig-cost-waug-rh-hr}}
    }
    \quad\quad\quad
    \subfloat[Worker knowledge level]{
    \includegraphics[width=0.4\textwidth]{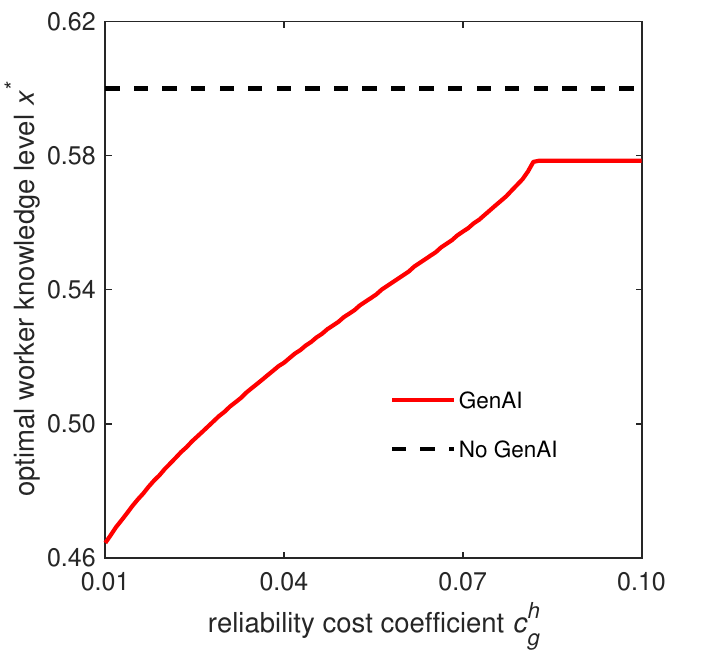}{\label{fig-cost-waug-rh-hx}}
    }
    \quad\quad
    \subfloat[Span of control]{
    \includegraphics[width=0.4\textwidth]{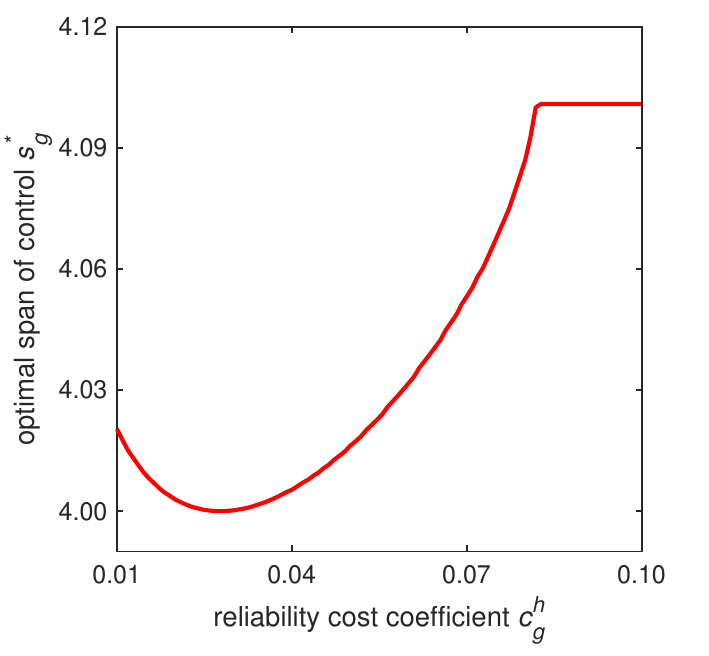}{\label{fig-cost-waug-rh-hs}}
    }
    \caption{Impacts of reliability cost coefficient with endogenous capability and hallucination rate under worker-level augmentation \\  ($k=0.6$, $w=0.3$, $t_c=0.6$, $t_v=0.2$, $c_g^r=0.3$, $\overline{h}_0=0.4$)}
    \label{fig-cost-waug-rh-h}
\end{figure}

Figure \ref{fig-cost-waug-rh-h} illustrates how the hallucination-reduction cost $c_g^h$ affects the firm's optimal investments and the resulting organizational design. Figures \ref{fig-cost-waug-rh-hh} and \ref{fig-cost-waug-rh-hr} show that the firm invests in reducing hallucinations only once $c_g^h$ falls below a threshold, and that this reliability investment is accompanied by higher investment in augmentation capability, echoing the complementary pattern under worker-level automation. Figure \ref{fig-cost-waug-rh-hx} further shows that lowering the hallucination rate reduces the optimal worker knowledge requirement, intensifying deskilling. Finally, as $c_g^h$ decreases, the optimal span of control follows a U-shaped path, contracting initially and expanding thereafter (Figure \ref{fig-cost-waug-rh-hs}). Overall, these results reinforce the robustness of the core insights in \S\ref{subsection-waug} on how GenAI improvements reshape hierarchies under worker-level augmentation.

\subsection{Expert-Level GenAI Adoption}
We now examine endogenous adoption costs under expert-level GenAI adoption. In this subsection, we focus on the scenario where the firm strategically invests in GenAI capability while keeping the hallucination rate fixed. As demonstrated in our worker-level analysis, endogenizing reliability alongside capability generates mechanisms that are qualitatively similar to those under a fixed hallucination rate. We therefore omit the analysis of endogenous hallucination reduction to maintain analytical brevity and concentrate on the organizational implications of capability investment.

We first consider the expert-level automation scenario. If the firm chooses to adopt GenAI, it must optimally set the worker knowledge level strictly below the AI's automation capability, ensuring $x < r_e$. Otherwise, GenAI would provide no solutions for problems that exceed the workers' existing capabilities, yet it would still incur strictly positive implementation expenses. Therefore, conditional on $x < r_e$, the firm's profit function is formulated as follows:
\begin{equation*}
   \Pi_e(x,r_e)=1-\left(w+\frac{1}{2}kx^2\right)-\left(w+\frac{1}{2}k\right)[(1-r_e)t_c+(r_e-x)(t_v+ht_c)]-\frac{1}{2}c_e^rr_e^2.
\end{equation*}
Consistent with our preceding analysis, we employ a strictly convex function, $\frac{1}{2}c_e^r r_e^2$, to capture the cost of GenAI deployment. Here, the parameter $c_e^r > 0$ serves as the capability cost coefficient specific to this expert-level automation configuration. Furthermore, to guarantee that the firm's optimal automation capability level remains strictly less than one, we assume that $c_e^r>\underline{c}_e^r$.

\begin{proposition}\label{app-prop-cost-eauto}
\begin{enumerate}[$(a)$]
    \item The firm should adopt expert-level automation if and only if $h<\overline{h}_e$ and $c_e^r<\overline{c}_e^r$.

    \item After GenAI adoption, the optimal GenAI automation capability and worker knowledge level are 
    $$r_e^*=\frac{(k+2w)(t_c-t_v-ht_c)}{2c_e^r}, ~~ x_e^*=\frac{(k+2w)(ht_c+t_v)}{2k}.$$
    Furthermore, $r_e^*$ strictly increases as $c_e^r$ decreases, while $x_e^*$ is independent of $c_e^r$, and $x_e^*$ satisfies $x_e^*<x_0^*$.

    \item After GenAI adoption, the optimal span of control is $s_e^*=\frac{1}{(1-r_e^*)t_c+(r_e^*-x_e^*)(t_v+ht_c)}$, which strictly increases as $c_e^r$ decreases. 
\end{enumerate}    
\end{proposition}

Proposition \ref{app-prop-cost-eauto} analytically characterizes the firm's optimal decisions under expert-level automation, validating the robustness of the core insights derived from our main model. We interpret a decline in the capability cost coefficient $c_e^r$ as GenAI advancement that lowers the marginal cost of capability. Accordingly, the firm optimally invests more, and the chosen automation capability $r_e^*$ increases as $c_e^r$ falls. A central implication in Proposition \ref{app-prop-cost-eauto}(b) is the capability insensitivity of the optimal worker knowledge level after adoption. Once $c_e^r$ drops below the adoption threshold, the optimal worker knowledge $x_e^*$ experiences a sudden drop. Following this initial deskilling effect, $x_e^*$ remains completely independent of any further reductions in $c_e^r$. Proposition \ref{app-prop-cost-eauto}(c) then describes the implied span-of-control dynamics. Unlike the baseline model with an exogenously fixed capability, endogenizing the capability choice ensures that the optimal span of control remains continuous at the exact adoption cutoff. This continuity emerges even though the adoption decision remains threshold-based, with both $r_e^*$ and $x_e^*$ exhibiting discrete jumps at $c_e^r=\overline{c}_e^r$. In the post-adoption regime, the span of control increases strictly and monotonically as the GenAI capability becomes cheaper to expand.


Finally, we turn to the expert-level augmentation scenario. Following our established analytical framework, we model the cost of capability investment using a strictly convex function $\frac{1}{2}c_u^r r_u^2$, where the parameter $c_u^r > 0$ represents the capability cost coefficient specific to this augmentation mode. Accordingly, the firm's profit function is given by
\begin{equation*}
   \Pi_u(x,r_u)=1-\left(w+\frac{1}{2}kx^2\right)-\left(w+\frac{1}{2}k\right)(1-x)[(1-r_u)t_c+ht_c]-\frac{1}{2}c_u^rr_u^2.
\end{equation*}
Consistent with our preceding analysis, we assume $c_u^r>\underline{c}_u^r$ to guarantee strict concavity of the profit function in $r_u$ and the existence of an interior solution. 

\begin{proposition}\label{app-prop-cost-eaug}
\begin{enumerate}[$(a)$]
    \item The firm should adopt expert-level augmentation if and only if $h<\overline{h}_u'$, and $\overline{h}_u'$ strictly increases as $c_u^r$ decreases.

    \item After GenAI adoption, the optimal GenAI augmentation capability and worker knowledge level are 
    $$r_u^*=\frac{(k+2w)[2k-(k+2w)(1+h)t_c]t_c}{4kc_u^r-(k+2w)^2t_c^2}, ~~ x_u^*=\frac{(k+2w)(1-r_u^*+h)t_c}{2k}.$$
    Furthermore, $r_u^*$ strictly increases while $x_u^*$ strictly decreases as $c_u^r$ decreases, and $x_u^*$ satisfies $x_u^*<x_0^*$.

    \item After GenAI adoption, the optimal span of control is $s_u^*=\frac{1}{(1-x_u^*)[(1-r_u^*)t_c+ht_c]}$.
    \begin{enumerate}[$(1)$]
        \item If $t_c>\frac{k}{(k+2w)(1+h)}$, $s_u^*$ decreases in $c_u^r$ for  $\underline{c}_u^r<c_u^r<c_{u1}^r$ and increases in $c_u^r$ for $c_u^r>\max\{\underline{c}_u^r, c_{u1}^r\}$. 

    \item If $t_c\le\frac{k}{(k+2w)(1+h)}$, $s_u^*$ decreases in $c_u^r$. 
    \end{enumerate}
\end{enumerate}
 
\end{proposition}

Proposition \ref{app-prop-cost-eaug} characterizes the firm's optimal choices under expert-level augmentation and confirms that our baseline insights are robust to endogenous adoption costs. As indicated in Proposition \ref{app-prop-cost-eaug}(a), although the endogenous adoption cost alters the hallucination cutoff, it preserves the fundamental pattern that as GenAI advances (i.e., the capability cost coefficient $c_u^r$ decreases), the adoption requirement for hallucination rates becomes less stringent. Proposition \ref{app-prop-cost-eaug}(b) shows that lower $c_u^r$ induces greater investment in augmentation capability, and that empowering experts with GenAI deskills the frontline, so the optimal worker knowledge level declines as $c_u^r$ decreases. Proposition \ref{app-prop-cost-eaug}(c) then links this to hierarchy: when communication is costly, the span of control is non-monotonic as $c_u^r$ falls, contracting initially as deskilling raises upward referrals, and expanding later as improvements in expert efficiency dominate.


In summary, this appendix extends our baseline framework by formally incorporating the endogenous costs of GenAI adoption and allowing the firm to strategically invest in both AI capability and reliability across different deployment modes. Most importantly, the analytical and numerical results presented throughout this section consistently confirm that our main insights remain highly robust. In particular, the fundamental skill asymmetries resulting from different deployment strategies are fully preserved. Furthermore, the non-monotonic structural dynamics of the span of control continue to manifest in most cases.
Therefore, explicitly accounting for implementation costs does not overturn our main organizational predictions; it instead reinforces the core economic forces through which GenAI reshapes hierarchy.

\clearpage

\section{Endogenous Expert Knowledge Level}\label{app-expert}
In the main text, we assume that experts possess complete knowledge over the entire difficulty spectrum $[0, 1]$. In this appendix, we relax this assumption by allowing the firm to endogenously determine the optimal expert knowledge level, denoted by $y$. This extension reflects the practical reality that acquiring comprehensive expertise is costly, and organizations may rationally choose to leave a certain fraction of highly complex tasks unresolved. We first formulate the firm's profit maximization problem under the no-GenAI benchmark:
\begin{equation*}
\begin{aligned}
    \mathop{\max}\limits_{x, y}&\quad \Pi_0=y-\left(w+\frac{1}{2}kx^2\right)-\left(w+\frac{1}{2}ky^2\right)(1-x)t_c,\\[2mm]
    \text{s.t.} & \quad x\le y\le 1. 
\end{aligned}
\end{equation*}
This objective function differs from the baseline in two respects. First, the firm can resolve problems only up to complexity $y$, reducing the total revenue from one to $y$. Second, the wage paid to each expert becomes $w + \frac{1}{2}ky^2$ rather than the baseline $w + \frac{1}{2}k$, because the expert knowledge level is now the endogenous decision variable $y$. Notably, expert staffing remains unchanged compared to the baseline. Since entry-level workers escalate all tasks beyond their own knowledge boundary $x$, every task in $(x,1]$ generates a communication cost of $t_c$, regardless of whether it can ultimately be resolved by experts \citep{ide2025artificial}. Accordingly, the total expert time, and hence the required expert headcount, remains $(1-x)t_c$.

\begin{proposition}\label{app-experty-ben}
For $k<\overline{k}$, the optimal expert knowledge level in the no-GenAI benchmark is $y_0^*=1$, where $\overline{k}=\frac{1}{t_c}$.    
\end{proposition}

Proposition \ref{app-experty-ben} establishes that under a relatively small knowledge premium ($k<\overline{k}$), it remains optimal for the firm to hire experts with complete knowledge in the no-GenAI benchmark ($y_0^*=1$). Consequently, the optimal worker knowledge level and span of control perfectly match the results in Lemma \ref{lemma-noGenAI}. This confirms that our analysis in the main text effectively characterizes environments with a moderately low knowledge premium. Furthermore, supplementary numerical analysis reveals that when the knowledge premium $k$ is sufficiently high, a further increase in $k$ results in a decrease in the optimal expert knowledge level.

\subsection{Worker-Level Automation}
We first consider the scenario of worker-level automation. Applying the endogenous expert knowledge framework, the firm's profit maximization problem is formulated as follows:
\begin{equation*}
\begin{aligned}
    \mathop{\max}\limits_{x, y}&\quad \Pi_t=y-\left(w+\frac{1}{2}kx^2\right)[(1-r_t)+(t_v+ht_r)r_t]-\left(w+\frac{1}{2}ky^2\right)(1-x)t_c,\\[2mm]
    \text{s.t.} & \quad r_t\le x\le y\le 1. 
\end{aligned}
\end{equation*}
Because this automation applies exclusively to entry-level labor, the mechanism through which endogenous expert knowledge shapes the profit function remains structurally identical to the benchmark without GenAI. Specifically, the expert capability $y$ simultaneously dictates the total realized revenue and determines the individual expert compensation as $w+\frac{1}{2}ky^2$. The required expert headcount remains $(1-x)t_c$ driven by the communication volume within the difficulty interval $(x, 1]$.

\begin{figure}[h]
    \centering
    \subfloat[Impact of knowledge premium]{
    \includegraphics[width=0.4\textwidth]{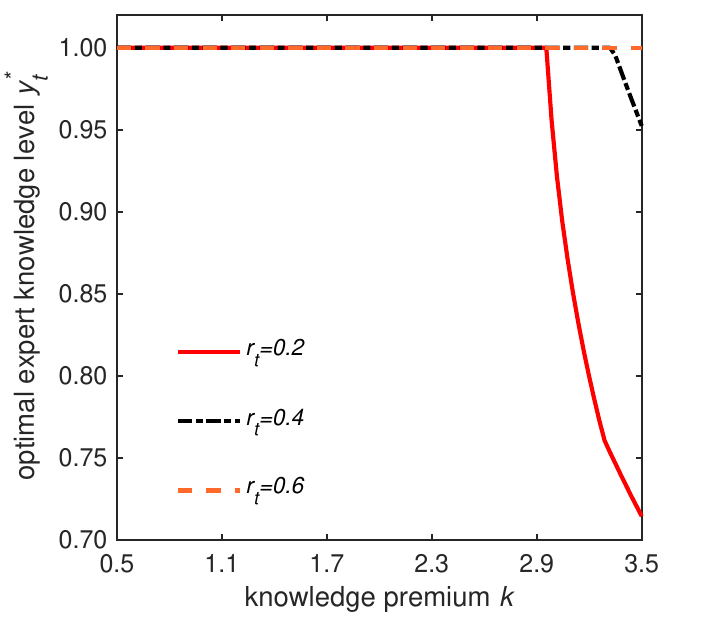}{\label{fig-y-wauto-k}}
    }
    \quad\quad
    \subfloat[Impact of capability]{
    \includegraphics[width=0.4\textwidth]{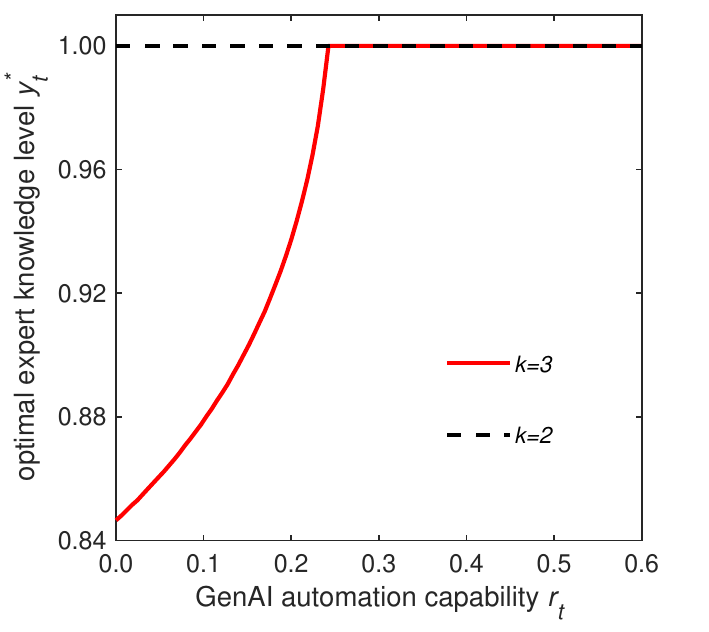}{\label{fig-y-wauto-r}}
    }
    \caption{Optimal expert knowledge level under worker-level automation  \\ ($w=0.2$, $t_c=0.5$, $t_v=0.3$, $t_r=0.8$, $h=0.1$)}
    \label{fig-y-wauto}
\end{figure}

Figure \ref{fig-y-wauto} illustrates the numerical results for worker-level automation. Figure \ref{fig-y-wauto-k} demonstrates that it is optimal for the firm to maintain complete expert knowledge ($y_t^*=1$) unless the knowledge premium $k$ is exceptionally large. Furthermore, Figure \ref{fig-y-wauto-r} reveals that even under a prohibitively high $k$ (i.e., $k=3$), the optimal expert knowledge level increases with the GenAI automation capability $r_t$ and returns to 1 once $r_t$ is not excessively small. Overall, these patterns confirm that our baseline assumption of full expert knowledge remains highly robust. 

\subsection{Worker-Level Augmentation}
Next, we examine the worker-level augmentation scenario and the firm's optimization problem with the endogenous expert knowledge level becomes:
\begin{equation*}
\begin{aligned}
    \mathop{\max}\limits_{x, y}&\quad \Pi_g=y-\left(w+\frac{1}{2}kx^2\right)-\left(w+\frac{1}{2}ky^2\right)[r_g(1-x)(t_v+ht_c)+r_g(1-x)t_c],\\[2mm]
    \text{s.t.} & \quad x+r_g(1-x)\le y\le 1. 
\end{aligned}
\end{equation*}
Consistent with the preceding cases, endogenizing the expert knowledge level affects the profit function primarily by scaling the firm's expected revenue to $y$ and adjusting the per-expert wage to $w + \frac{1}{2}ky^2$. The constraint $x+r_g(1-x)\le y$ ensures that the expert knowledge level weakly exceeds the augmented worker capability boundary, allowing experts to credibly verify AI-assisted outputs.

\begin{figure}[h]
    \centering
    \subfloat[Impact of knowledge premium]{
    \includegraphics[width=0.4\textwidth]{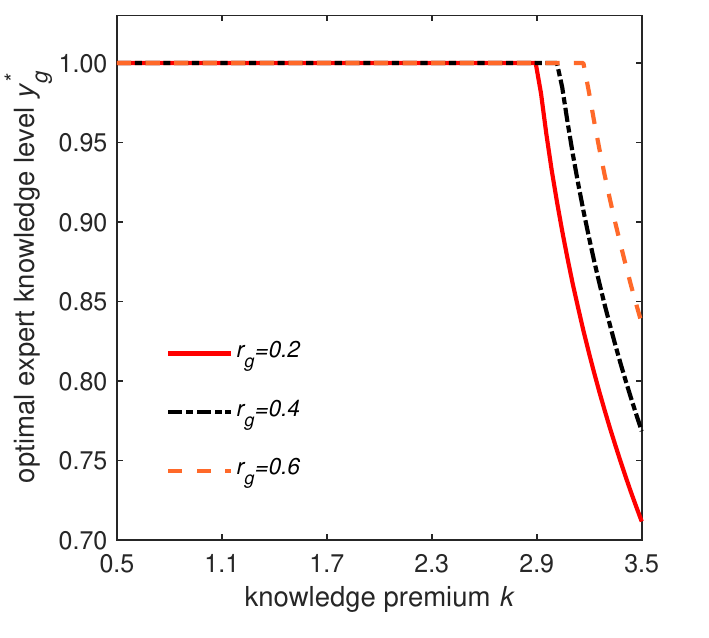}{\label{fig-y-waug-k}}
    }
    \quad\quad
    \subfloat[Impact of capability]{
    \includegraphics[width=0.4\textwidth]{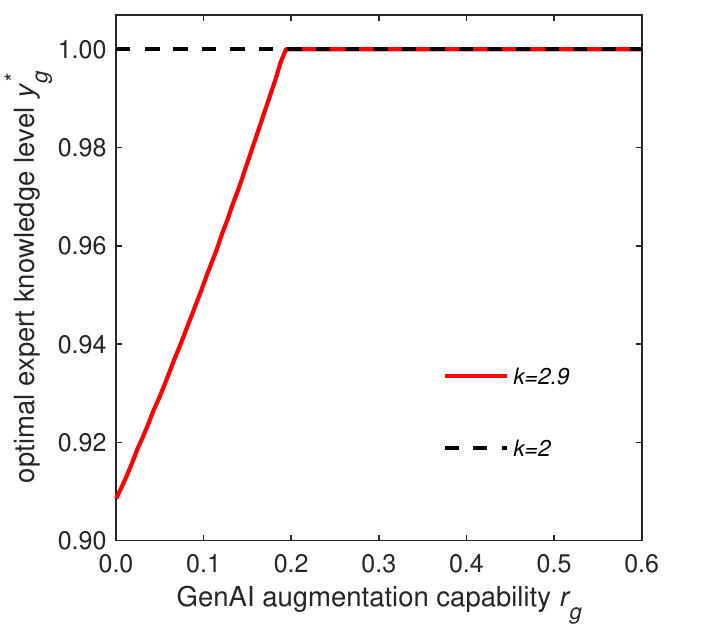}{\label{fig-y-waug-r}}
    }
    \caption{Optimal expert knowledge level under worker-level augmentation  \\ ($w=0.2$, $t_c=0.5$, $t_v=0.3$, $h=0.1$)}
    \label{fig-y-waug}
\end{figure}

Turning to worker-level augmentation, Figure \ref{fig-y-waug} shows patterns similar to those under worker-level automation. Figure \ref{fig-y-waug-k} indicates that departures from full expert knowledge ($y_g^*<1$) arise only when the knowledge premium $k$ is extremely large. Figure \ref{fig-y-waug-r} further shows that higher GenAI capability $r_g$ pushes the optimal expert knowledge back to its upper bound, even when $k$ is significantly high. Therefore, these findings reinforce the robustness of the baseline assumption of full expert knowledge across most parameter settings.

\subsection{Expert-Level GenAI Adoption}
We now turn to expert-level GenAI adoption. Under expert-level automation, GenAI partially substitutes for experts by autonomously handling a subset of escalations from frontline workers. For the technology to deliver any value, its capability must strictly exceed the workers' intrinsic knowledge boundary, necessitating the condition $x \le r_e$. Otherwise, GenAI cannot address any of the escalated problems. Incorporating the endogenous expert knowledge framework, the firm's profit maximization problem is formulated as follows, where the constraint $x\le r_e\le y$ enforces the capability ordering among the worker, GenAI, and experts: GenAI must cover problems beyond the worker's knowledge boundary, while expert knowledge must weakly exceed GenAI capability so that experts can accurately validate and, when necessary, assist workers in correcting AI-guided outputs.
\begin{equation*}
\begin{aligned}
    \mathop{\max}\limits_{x, y}&\quad \Pi_e=y-\left(w+\frac{1}{2}kx^2\right)-\left(w+\frac{1}{2}ky^2\right)[(r_e-x)(t_v+ht_c)+(1-r_e)t_c],\\[2mm]
    \text{s.t.} & \quad x\le r_e\le y\le 1. 
\end{aligned}
\end{equation*}

Regarding expert-level augmentation, the firm's optimization problem within the endogenous expert knowledge framework is presented below. Under this mode, the enhanced expert consultation efficiency facilitated by GenAI compresses the standard communication cost from $t_c$ to the augmented time multiplier $(1-r_u)t_c+ht_c$. 
\begin{equation*}
\begin{aligned}
    \mathop{\max}\limits_{x, y}&\quad \Pi_u=y-\left(w+\frac{1}{2}kx^2\right)-\left(w+\frac{1}{2}ky^2\right)(1-x)[(1-r_u)t_c+ht_c],\\[2mm]
    \text{s.t.} & \quad x\le y\le 1. 
\end{aligned}
\end{equation*}

\begin{figure}[h]
    \centering
    \subfloat[Expert-level automation]{
    \includegraphics[width=0.4\textwidth]{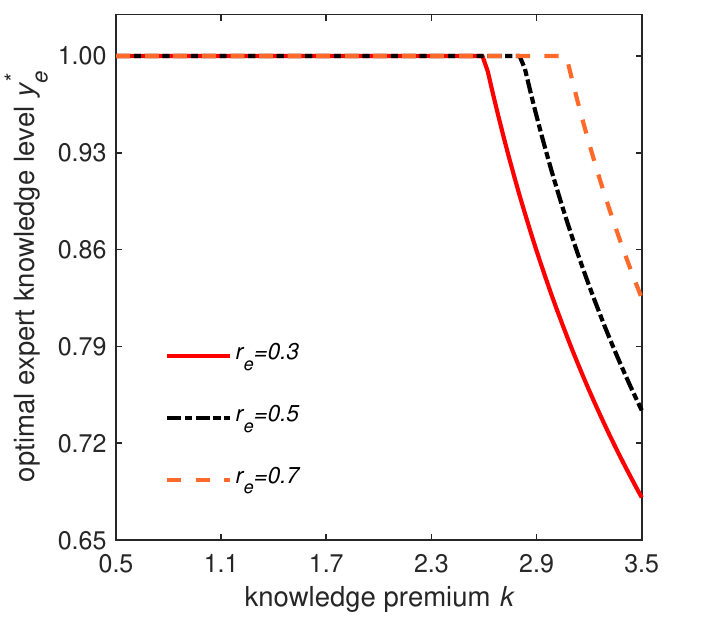}{\label{fig-y-eauto-k}}
    }
    \quad\quad
    \subfloat[Expert-level augmentation]{
    \includegraphics[width=0.4\textwidth]{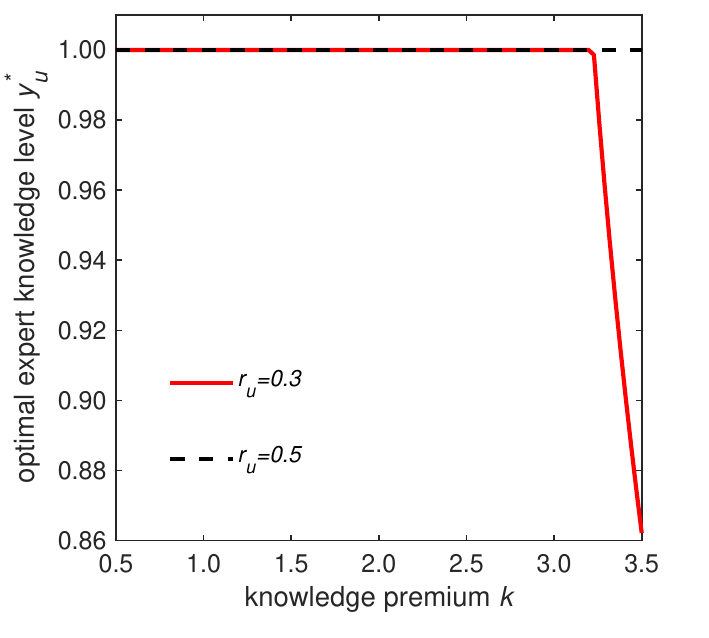}{\label{fig-y-eaug-k}}
    }
    \caption{Optimal expert knowledge level under expert-level GenAI adoption \\ ($w=0.2$, $t_c=0.5$, $t_v=0.3$, $h=0.1$)}
    \label{fig-y-expert}
\end{figure}

Finally, Figure \ref{fig-y-expert} illustrates the numerical outcomes under expert-level GenAI adoption. Figure \ref{fig-y-eauto-k} (automation) and Figure \ref{fig-y-eaug-k} (augmentation) exhibit patterns similar to the worker-level cases. In both modes, the optimal expert knowledge level remains at $y^*=1$ whenever the knowledge premium $k$ is below a threshold; reductions in expert capability arise only when the knowledge acquisition cost is exceptionally high.

In summary, this extension endogenizes the expert knowledge boundary and examines its role in the no-GenAI benchmark and all four GenAI deployment architectures. The results show that, except under extreme knowledge acquisition costs, the profit-maximizing firm maintains a full expert knowledge base. Thus, normalizing the expert boundary to its maximum value in the main analysis is not merely a tractability device, but a robust implication of the optimal design. As a result, the central mechanisms and organizational predictions developed in the main text remain intact.

\newpage
\clearpage
\section{Proofs of Statements}\label{app-proof}
We present all the proofs of Lemmas, Propositions, and Corollaries in this appendix.\\[3mm]
\noindent\textbf{Proof of Lemma \ref{lemma-noGenAI}}:\\
In the absence of GenAI, the firm's optimization problem is 
\begin{equation*}
    \mathop{\max}_x\;\Pi_0(x)=1-\left(w+\frac{1}{2}kx^2\right)-\left(w+\frac{1}{2}k\right)(1-x)t_c.
\end{equation*}
Since $\frac{\partial^2\Pi_0}{\partial x^2}=-k<0$, $\Pi_0(x)$ is strictly concave in $x$. From the first-order condition, we can derive the optimal worker knowledge level $$x_0^*=\frac{(k+2w)t_c}{2k}.$$ 
Under the assumption $0<t_c<\overline{t}_c$, we have $0<x_0^*<1$. Plugging $x_0^*$ into $n_0$, $s_0$ and $\Pi_0$, we have that the optimal number of experts is $$n_0^*=(1-x_0^*)t_c=\frac{[2k-(k+2w)t_c]t_c}{2k},$$ 
the optimal span of control is 
$$s_0^*=\frac{1}{(1-x_0^*)t_c}=\frac{2k}{[2k-(k+2w)t_c]t_c},$$ 
and the firm's optimal profit is $$\Pi_0^*=\frac{[2k-(k+2w)t_c]^2+4k(2-k-2w)}{8k}.$$ 
This proves the result in Lemma \ref{lemma-noGenAI}.     \Halmos\\[3mm]
\noindent\textbf{Proof of Lemma \ref{lemma-adoption-wauto}}:\\
If the firm adopts worker-level automation, the firm's optimization problem is 
\begin{equation*}
\begin{aligned}
\max_{x} \quad & \Pi_t(x)
= 1-\left(w+\frac{1}{2}kx^2\right)[(1-r_t)+(t_v+ht_r)r_t]-\left(w+\frac{1}{2}k\right)(1-x)t_c, \\
\text{s.t.} \quad & r_t \le x \le 1.
\end{aligned}
\end{equation*}
By comparing the expressions for $\Pi_0(x)$ and $\Pi_t(x)$, we have that if $h\ge\frac{1-t_v}{t_r}$, worker-level automation directly increases the firm's demand for workers, implying $\Pi_0(x)\ge\Pi_t(x)$. Hence, we only need to focus on the case $h<\frac{1-t_v}{t_r}$ in the subsequent analysis. 

Since $\frac{\partial^2\Pi_t}{\partial x^2}=-k(hr_tt_r+r_tt_v+1-r_t)<0$, $\Pi_t(x)$ is strictly concave in $x$. From the first-order condition, the unconstrained optimal worker knowledge level can be solved as 
$$x_t=\frac{(k+2w)t_c}{2k[(1-r_t)+(t_v+ht_r)r_t]}=\frac{x_0^*}{(1-r_t)+(t_v+ht_r)r_t}.$$
Consider the case of $h<\frac{1-t_v}{t_r}$ and $r_t\le x_0^*$. Under such conditions, $x_t>x_0^*\ge r_t$ and thus $\Pi_t|_{x=\min\{1, x_t\}}>\Pi_t|_{x=x_0^*}>\Pi_0^*$. Therefore, the firm should adopt GenAI and the optimal worker knowledge level is $x_t^*=\min\{1, x_t\}$.

Consider the case of $h<\frac{1-t_v}{t_r}$ and $r_t>x_0^*$. It can be verified that $x_t\ge r_t$ is equivalent to $h\le h_{1a}$, where 
$$h_{1a}=\frac{2kr_t^2(1-t_v)+(k+2w)t_c-2kr_t}{2kt_rr_t^2},$$
and $h_{1a}<\frac{1-t_v}{t_r}$ for $r_t>x_0^*$. Therefore, when $h\le h_{1a}$ and $r_t>x_0^*$, we have $x_t\ge r_t$ and thus $\Pi_t|_{x=\min\{1, x_t\}}>\Pi_t|_{x=x_0^*}>\Pi_0^*$. When $h_{1a}<h<\frac{1-t_v}{t_r}$ and $r_t>x_0^*$, the firm's optimal choice is $x_t^*=r_t$ under worker-level automation. It can be verified that $\Pi_t|_{x=r_t}>\Pi_0^*$ is equivalent to $h<h_{2a}$, where 
$$h_{2a}=\frac{1}{4kr_tt_r(kr_t^2+2w)}\{-(k+2w)^2t_c^2+4kr_t(k+2w)t_c-4kr_t[kr_t-(1-t_v)(kr_t^2+2w)]\},$$
and $h_{1a}<h_{2a}<\frac{1-t_v}{t_r}$ for $r_t>x_0^*$. 

Combining the discussions above, we have that the firm should adopt worker-level automation if and only if $h<\overline{h}_t$, where 
\begin{equation*}
     \overline{h}_t
    =\left \{ 
    \begin{aligned}
    &\; \frac{1-t_v}{t_r}, \quad\quad\quad && \text{if \;} r_t\le x_0^*; \\
    &\; h_{2a}, \quad\quad\quad && \text{otherwise. \;}      \quad\quad\quad  \\
    \end{aligned}
    \right.
\end{equation*}
This proves the result in Lemma \ref{lemma-adoption-wauto}.   \Halmos\\[3mm]
\noindent\textbf{Proof of Proposition \ref{prop-workerauto-x}}:\\
Based on the proof of Lemma \ref{lemma-adoption-wauto}, we have that under the condition $h<\overline{h}_t$, the firm's optimal worker knowledge level is 
$$x_t^*=\max\{r_t, \min\{1, x_t\}\}.$$ 
$x_t^*\ge \min\{1, x_t\}>x_0^*$, and $x_t^*$ is continuous in both $r_t$ and $h$. Since 
$$\frac{\partial x_t}{\partial r_t}=\frac{(k+2w)(1-t_v-ht_r)t_c}{2k[1-r_t+r_t(t_v+ht_r)]^2}>0,$$ 
$x_t^*$ (weakly) increases in $r_t$. Furthermore, the expression for $x_t^*$ can be rewritten as 
\begin{equation*}
     x_t^*
    =\left \{ 
    \begin{aligned}
    &\; r_t, \quad\quad\quad && \text{if \;} h\ge h_{1a}; \\
    &\; x_t, \quad\quad\quad && \text{if \;} h_{3a}<h<h_{1a}; \\
    &\; 1, \quad\quad\quad && \text{if \;} h\le h_{3a}.
    \end{aligned}
    \right.
\end{equation*}
where $$h_{3a}=\frac{2k(1-t_v)r_t+(k+2w)t_c-2k}{2kt_rr_t}.$$ 
Since $\frac{\partial x_t}{\partial h}=-\frac{(k+2w)t_ct_rr_t}{2k[1-r_t+r_t(t_v+ht_r)]^2}<0$, $x_t^*$ (weakly) decreases in $h$.    \Halmos\\[3mm]
\noindent\textbf{Proof of Proposition \ref{prop-workerauto-s}}:\\
It can be verified that $x_t\ge 1$ for $r_t\ge r_0$ and $x_t<1$ for $r_t<r_0$, where $r_0=\frac{2k-(k+2w)t_c}{2k(1-t_v-ht_r)}$. Since $t_c<\frac{2k}{k+2w}$, $r_0>0$ and $r_0<1$ if and only if $h<h_0$, where $h_0=\frac{(k+2w)t_c-2kt_v}{2kt_r}$. Therefore, after worker-level automation adoption, $x_t^*=1$ for $h<h_0$ and $r\ge r_0$ and the hierarchy collapses
into a single-layer structure composed solely of workers with complete knowledge. 

In the subsequent analysis, we focus on the case where $h\ge h_0$ or $r_t<r_0$. The optimal span of control is given by 
$$s_t^*=\frac{1-r_t+(t_v+ht_r)r_t}{(1-x_t^*)t_c},$$
where $x_t^*=\max\{r_t, x_t\}$. For $r_t>x_t$ and $x_t^*=r_t$, the span of control is 
$$s_{t1}=\frac{1-r_t+(t_v+ht_r)r_t}{(1-r_t)t_c},$$
and $\frac{\partial s_{t1}}{\partial r_t}=\frac{ht_r+t_v}{(1-r_t)^2t_c}>0$.
For $r_t\le x_t$ and $x_t^*=x_t$, the span of control is 
$$s_{t2}=\frac{1-r_t+(t_v+ht_r)r_t}{(1-x_t)t_c}=\frac{2k[1-r_t+r_t(t_v+ht_r)]^2}{t_c[2k-(k+2w)t_c-2k(1-t_v-ht_r)r_t]},$$
and 
$$\frac{\partial s_{t2}}{\partial r_t}=\frac{4k(1-t_v-ht_r)[1-r_t+r_t(t_v+ht_r)][k(1-t_v-ht_r)r_t+(k+2w)t_c-k]}{t_c[2k-(k+2w)t_c-2k(1-t_v-ht_r)r_t]^2}.$$
$\frac{\partial s_{t2}}{\partial r_t}<0$ for $r_t<r_{1a}$ and $\frac{\partial s_{t2}}{\partial r_t}>0$ for $r_t>r_{1a}$, where $r_{1a}=\frac{k-(k+2w)t_c}{k(1-t_v-ht_r)}$ and $r_{1a}<r_0$ for $h<\overline{h}_t$. Since $r_{1a}>0$ if and only if $t_c<\frac{k}{k+2w}$, we have that $\frac{\partial s_t^*}{\partial r_t}>0$ consistently holds for $t_c\ge\frac{k}{k+2w}$.

Under the condition of $t_c<\frac{k}{k+2w}$, we need to compare $r_t$ and $x_t$ to determine how an increase in $r_t$ affects $s_t^*$. 
$$x_t-r_t=\frac{F_1(r_t)}{2k[1-r_t+r_t(t_v+ht_r)]},$$
where $F_1(r_t)=2k(1-t_v-ht_r)r_t^2-2kr_t+(k+2w)t_c$. For $h\le h_{1b}$, the discriminant of the quadratic function $\Delta_{F1}\le 0$, where 
$$h_{1b}=\frac{2t_c(1-t_v)(k+2w)-k}{2t_ct_r(k+2w)},$$
and $h_{1b}<h_0$.
Therefore, when $h\le h_{1b}$, $x_t\ge r_t$ consistently holds. Consequently, $\frac{\partial s_t^*}{\partial r_t}<0$ for $0<r_t<r_{1a}$, and $\frac{\partial s_t^*}{\partial r_t}>0$ otherwise. For $h\ge h_0$, $F_1|_{r_t=1}\le 0$ and thus $x_t\ge r_t$ for $0<r_t\le r_{2a}$ and $x_t<r_t$ for $r_{2a}<r_t<1$, where 
$$r_{2a}=\frac{1-\sqrt{1-4(1-t_v-ht_r)x_0^*}}{2(1-t_v-ht_r)},$$
and $r_t=r_{2a}$ is the smaller root of the equation $F_1(r_t)=0$. 
As a result, $\frac{\partial s_t^*}{\partial r_t}<0$ for $0<r_t<\min\{r_{1a}, r_{2a}\}$, and $\frac{\partial s_t^*}{\partial r_t}>0$ otherwise. For $h_{1b}<h<h_0$, $F_1|_{r_t=1}>0$ and the quadratic function $F_1(r_t)$ has two roots in the interval $(0, 1)$. It can be verified that $r_{1a}<r_{3a}$, where $r_t=r_{3a}$ is the larger root of the equation $F_1(r_t)=0$, where 
$$r_{3a}=\frac{1+\sqrt{1-4(1-t_v-ht_r)x_0^*}}{2(1-t_v-ht_r)}.$$
Consequently, $\frac{\partial s_t^*}{\partial r_t}<0$ for $0<r_t<\min\{r_{1a}, r_{2a}\}$, and $\frac{\partial s_t^*}{\partial r_t}>0$ otherwise.
Combining the discussions above, we have that for $h\ge h_0$ or $r_t<r_0$, the optimal span of control $s_t^*$ strictly decreases in $r_t$ on $(0, r_1)$ and strictly increases in $r_t$ otherwise, where 
\begin{equation*}
     r_1
    =\left \{ 
    \begin{aligned}
    &\; r_{1a}, \quad\quad\quad && \text{if \;} h\le h_{1b}; \\
    &\; \min\{r_{1a}, r_{2a}\}, \quad\quad\quad && \text{otherwise. \;}      \quad\quad\quad  \\
    \end{aligned}
    \right.
\end{equation*}

Regarding the impact of $h$ on $s_t^*$, we have that for $r_t>x_t$ and $x_t^*=r_t$, $s_t^*=s_{t1}$ and 
$$\frac{\partial s_{t1}}{\partial h}=\frac{t_rr_t}{(1-r_t)t_c}>0.$$
For $r_t\le x_t$ and $x_t^*=x_t$, $s_t^*=s_{t2}$ and 
$$\frac{\partial s_{t2}}{\partial h}=\frac{4kt_rr_t[1-r_t+r_t(t_v+ht_r)][t_rr_tkh+k-(k+2w)t_c-r_tk(1-t_v)]}{t_c[2k-(k+2w)t_c-2k(1-t_v-ht_r)r_t]^2}.$$
Therefore, $\frac{\partial s_{t2}}{\partial h}<0$ for $h<h_{2b}$ and $\frac{\partial s_{t2}}{\partial h}>0$ for $h>h_{2b}$, where 
$$h_{2b}=\frac{(k+2w)t_c-k+r_tk(1-t_v)}{kt_rr_t}.$$
As discussed in the proof of Lemma \ref{lemma-adoption-wauto}, for $r_t\le x_0^*$, $x_t>r_t$ consistently holds. Consequently, for $r_t\le x_0^*$, $\frac{\partial s_t^*}{\partial h}<0$ for $0<h<h_{2b}$, and $\frac{\partial s_t^*}{\partial h}>0$ otherwise. For $r_t>x_0^*$, $x_t\ge r_t$ for $h\le h_{1a}$ and $x_t<r_t$ for $h>h_{1a}$. Therefore, $\frac{\partial s_t^*}{\partial h}<0$ for $0<h<\min\{h_{2b}, h_{1a}\}$, and $\frac{\partial s_t^*}{\partial h}>0$ otherwise. Combining the discussions above, we have that for $h\ge h_0$ or $r_t<r_0$, the optimal span of control $s_t^*$ strictly decreases in $h$ on $(0, h_1)$ and strictly increases in $h$ otherwise, where 
\begin{equation*}
     h_1
    =\left \{ 
    \begin{aligned}
    &\; h_{2b}, \quad\quad\quad && \text{if \;} r_t\le x_0^*; \\
    &\; \min\{h_{2b}, h_{1a}\}, \quad\quad\quad && \text{otherwise. \;}      \quad\quad\quad  \\
    \end{aligned}
    \right.
\end{equation*}
This proves the results in Proposition \ref{prop-workerauto-s}.  \Halmos\\[3mm]
\noindent\textbf{Proof of Lemma \ref{lemma-w-aug}}:\\
If the firm adopts worker-level augmentation, the firm's optimization problem is 
\begin{equation*}
    \mathop{\max}_x\;\Pi_g(x)=1-\left(w+\frac{1}{2}kx^2\right)-\left(w+\frac{1}{2}k\right)[r_g(1-x)(t_v+ht_c)+(1-r_g)(1-x)t_c].
\end{equation*}
Since $\frac{\partial^2\Pi_g}{\partial x^2}=-k<0$, $\Pi_g(x)$ is strictly concave in $x$. Therefore, from the first-order condition, the unconstrained optimal knowledge level of workers can be solved as
$$x_g=\frac{(k+2w)[r_gt_v+(1+r_gh-r_g)t_c]}{2k}.$$
One can check that $x_g<1$ for $h<h_{1c}$ and $x_g\ge 1$ for $h\ge h_{1c}$, where 
$$h_{1c}=\frac{2k-(k+2w)[(1-r_g)t_c+r_gt_v]}{(k+2w)r_gt_c}.$$
Therefore, if the firm adopts worker-level augmentation, the optimal worker knowledge level is 
\begin{equation*}
     x_g^*
    =\left \{ 
    \begin{aligned}
    &\; 1, \quad\quad\quad && \text{if \;} h\ge h_{1c}; \\
    &\; x_g, \quad\quad\quad && \text{otherwise. \;}      \quad\quad\quad  \\
    \end{aligned}
    \right.
\end{equation*}
When $h\ge h_{1c}$, it is straightforward to see $\Pi_g^*=\Pi_g|_{x=1}<\Pi_0^*$, and thus the firm should not adopt GenAI. When $h<h_{1c}$, by plugging $x_g^*=x_g$ into $\Pi_g$, we have 
\begin{equation*}
    \Pi_g^*-\Pi_0^*=\frac{r_g(2w+k)(t_c-t_v-ht_c)[4k-(k+2w)(2t_c-r_gt_c+r_gt_v)-t_c(k+2w)r_gh]}{8k}.
\end{equation*}
One can check that $\Pi_g^*>\Pi_0^*$ for $h<\overline{h}_g$ or $h>h_{2c}$ and $\Pi_g^*\le\Pi_0^*$ for $\overline{h}_g\le h\le h_{2c}$, where 
$$\overline{h}_g=1-\frac{t_v}{t_c}, ~~ h_{2c}=\frac{4k-(k+2w)[(2-r_g)t_c+r_gt_v]}{(k+2w)r_gt_c}.$$
We have $\overline{h}_g<h_{1c}<h_{2c}$ under the assumption $0<t_c<\frac{2k}{k+2w}$. Therefore, the firm should adopt worker-level augmentation (i.e., $\Pi_g^*>\Pi_0^*$) if and only if $h<\overline{h}_g$, and the optimal worker knowledge level is $x_g^*=x_g$.       \Halmos\\[3mm]
\noindent\textbf{Proof of Proposition \ref{prop-w-aug-equ}}:\\
Based on the proof of Lemma \ref{lemma-w-aug}, we have that under the condition $h<\overline{h}_g$, the optimal worker knowledge level is 
$$x_g^*=\frac{(k+2w)[r_gt_v+(1+r_gh-r_g)t_c]}{2k}.$$
Then, it can be verified that 
$$x_g^*-x_0^*=-\frac{r_g(k+2w)(t_c-t_v-ht_c)}{2k}<0,$$
$$\frac{\partial x_g^*}{\partial r_g}=-\frac{(k+2w)(t_c-t_v-ht_c)}{2k}<0,$$  $$\frac{\partial x_g^*}{\partial h}=\frac{(k+2w)r_gt_c}{2k}>0.$$

\par This gives the results in Proposition \ref{prop-w-aug-equ}.    \Halmos\\[3mm] 
\noindent\textbf{Proof of Proposition \ref{prop-waug-span}}:\\
When $h<\overline{h}_g$, the firm adopts workerk-level GenAI augmentation and the optimal span of control is 
\begin{align*}
    s_g^*=&\frac{1}{r_g(1-x_g^*)(t_v+ht_c)+(1-r_g)(1-x_g^*)t_c}  \\[2mm]
         =&\frac{2k}{[t_c-r_g(t_c-t_v-ht_c)][(2w+k)(t_c-t_v-ht_c)r_g+2k-(k+2w)t_c]}.
\end{align*}
We have 
$$\frac{\partial s_g^*}{\partial r_g}=\frac{4k(t_c-t_v-ht_c)[k-(k+2w)t_c+(k+2w)(t_c-t_v-ht_c)r_g]}{[t_c-r_g(t_c-t_v-ht_c)]^2[(2w+k)(t_c-t_v-ht_c)r_g+2k-(k+2w)t_c]^2}.$$
$\frac{\partial s_g^*}{\partial r_g}<0$ for $r_g<r_2$ and $\frac{\partial s_g^*}{\partial r_g}>0$ for $r_g>r_2$, where 
$$r_2=\frac{(k+2w)t_c-k}{(k+2w)(t_c-t_v-ht_c)}.$$ 
Therefore, there exists a threshold $r_2$ such that $s_g^*$ strictly decreases in $r_g$ on $(0, r_2)$ and strictly increases in $r_g$ on $(r_2, 1)$.

Regarding the impact of changes in hallucination rates, we have 
$$\frac{\partial s_g^*}{\partial h}=\frac{4kr_gt_c[r_gt_c(k+2w)h+(k+2w)t_c-k-r_g(k+2w)(t_c-t_v)]}{[t_c-r_g(t_c-t_v-ht_c)]^2[(2w+k)(t_c-t_v-ht_c)r_g+2k-(k+2w)t_c]^2}.$$
$\frac{\partial s_g^*}{\partial h}<0$ for $h<h_2$ and $\frac{\partial s_g^*}{\partial h}>0$ for $h>h_2$, where 
$$h_2=\frac{r_g(k+2w)(t_c-t_v)-(k+2w)t_c+k}{r_g(k+2w)t_c}.$$ 
Therefore, there exists a threshold $h_2$ such that $s_g^*$ strictly decreases in $h$ on $(0, h_2)$ and strictly increases in $h$ on $(h_2, \overline{h}_g)$.    \Halmos\\[3mm]
\noindent\textbf{Proof of Lemma \ref{lemma-adoption-eauto}}:\\
We first derive the firm's optimal worker knowledge level $x_e^*$ under expert-level automation and then compare it with the benchmark case without GenAI to obtain the condition under which the firm should adopt GenAI.
Under expert-level automation, the firm's profit function is given by:
\begin{equation*}
     \Pi_e(x) 
    =\left \{ 
    \begin{aligned}
    &1-\left(w+\frac{1}{2}kx^2\right)-\left(w+\frac{1}{2}k\right)[(1-r_e)t_c+(r_e-x)(t_v+ht_c)], && \text{if $x\le r_e$}; \\
    &1-\left(w+\frac{1}{2}kx^2\right)-\left(w+\frac{1}{2}k\right)(1-x)t_c, && \text{if $x\ge r_e$}.\\
    \end{aligned}
    \right.
\end{equation*}
If the firm chooses $x\le r_e$, $\frac{\partial^2\Pi_e}{\partial x^2}=-k<0$ and thus $\Pi_e(x)$ is strictly concave in $x$. From the first-order condition, the unconstrained optimal knowledge level of workers can be solved as
$$x_e=\frac{(k+2w)(t_v+ht_c)}{2k}.$$
Therefore, in the case $x\le r_e$, the firm chooses $x=x_e$ for $r_e\ge x_e$ and $x=r_e$ for $r_e\le x_e$. 

If the firm chooses $x\ge r_e$, the adoption of GenAI plays no role and $\Pi_e(x)=\Pi_0(x)$. The unconstrained optimal worker knowledge level is $x_0^*=\frac{(k+2w)t_c}{2k}$. Therefore, in the case $x\ge r_e$, the firm chooses $x=x_0^*$ for $r_e\le x_0^*$ and $x=r_e$ for $r_e\ge x_0^*$. It can be verified that $x_e<x_0^*$ for $h<\overline{h}_e$ and $x_e\ge x_0^*$ for $h\ge\overline{h}_e$, where $\overline{h}_e=1-\frac{t_v}{t_c}$. 

Consider the case of $h<\overline{h}_e$. It can be verified that the firm's globally optimal choice is $x_e^*=x_0^*$ for $r_e<x_e$ and $x_e^*=x_e$ for $r_e>x_0^*$. For $x_e\le r_e\le x_0^*$, we need to compare $\Pi_e|_{x=x_e}$ and $\Pi_e|_{x=x_0^*}$.
$$\Pi_e|_{x=x_e}-\Pi_e|_{x=x_0^*}=\frac{(k+2w)(t_c-ht_c-t_v)F_2(r_e)}{8k},$$
where 
$$F_2(r_e)=4kr_e-(k+2w)(ht_c+t_c+t_v).$$ 
Therefore, $\Pi_e|_{x=x_e}>\Pi_e|_{x=x_0^*}$ for $r_e>\overline{r}_e$ and $\Pi_e|_{x=x_e}\le\Pi_e|_{x=x_0^*}$ for $r_e\le\overline{r}_e$, where $\overline{r}_e=\frac{(k+2w)(ht_c+t_c+t_v)}{4k}$ and $x_e<\overline{r}_e<x_0^*$. Consequently, under expert-level automation, the optimal worker knowledge level is $x_e^*=x_0^*$ for $r_e\le\overline{r}_e$ and $x_e^*=x_e$ for $r_e>\overline{r}_e$. When $r_e\le\overline{r}_e$, $\Pi_e^*=\Pi_e|_{x=x_0^*}=\Pi_0^*$ and thus the firm should not adopt GenAI; when $r_e>\overline{r}_e$, $\Pi_e^*=\Pi_e|_{x=x_e^*}>\Pi_e|_{x=x_0^*}=\Pi_0^*$, and the firm should adopt GenAI. 

Consider the case of $h\ge\overline{h}_e$. It can be verified that under expert-level automation, the optimal worker knowledge level is $x_e^*=x_0^*$ for $r_e<x_0^*$, $x_e^*=r_e$ for $x_0^*\le r_e\le x_e$, and $x_e^*=x_e$ for $r_e>x_e$, and that $\Pi_e^*\le\Pi_0^*$ consistently holds. Therefore, the firm should not adopt GenAI. 

Combining the discussions above, we have that the firm should adopt expert-level GenAI adoption if and only if $h<\overline{h}_e$ and $r_e>\overline{r}_e$.  \Halmos\\[3mm]   
\noindent\textbf{Proof of Proposition \ref{prop-eauto-x}}:\\
Based on the proof of Lemma \ref{lemma-adoption-eauto}, under the condition of $h<\overline{h}_e$ and $r_e>\overline{r}_e$, the optimal worker knowledge level is $x_e^*=\frac{(k+2w)(t_v+ht_c)}{2k}$. 
$$x_e^*-x_0^*=-\frac{(k+2w)(t_c-ht_c-t_v)}{2k}<0.$$
It is straightforward to see that $x_e^*$ is independent of $r_e$ and strictly increasing in $h$.   \Halmos\\[3mm]
\noindent\textbf{Proof of Proposition \ref{prop-eauto-span}}:\\
After expert-level automation adoption ($h<\overline{h}_e$ and $r_e>\overline{r}_e$), the optimal span of control is 
$$s_e^*=\frac{1}{(1-r_e)t_c+(r_e-x_e^*)(t_v+ht_c)}.$$
We can have 
$$\frac{\partial s_e^*}{\partial r_e}=\frac{t_c-ht_c-t_v}{[(1-r_e)t_c+(r_e-x_e^*)(t_v+ht_c)]^2}>0.$$
By plugging $r_e=\overline{r}_e$ into the expression of $s_e^*$, we have 
$$s_e^*|_{r_e=\overline{r}_e}-s_0^*=-\frac{(ht_c+t_c+t_v)(t_c-ht_c-t_v)(k+2w)}{2[(1-\overline{r}_e)t_c+(\overline{r}_e-x_e^*)(t_v+ht_c)][2k-(k+2w)t_c]t_c}<0.$$
Therefore, under the condition $h<\overline{h}_e$, once $r_e$ exceeds the threshold $\overline{r}_e$, the optimal span of control drops discretely and thereafter strictly increases in $r_e$. 

$$\frac{\partial s_e^*}{\partial h}=\frac{t_c[(k+2w)t_ch-kr_e+(k+2w)t_v]}{k[(1-r_e)t_c+(r_e-x_e^*)(t_v+ht_c)]^2}.$$
Therefore, $\frac{\partial s_e^*}{\partial h}<0$ for $0<h<h_3$ and $\frac{\partial s_e^*}{\partial h}>0$ for $h_3<h<\overline{h}_e$, where $h_3=\frac{kr_e-(k+2w)t_v}{(k+2w)t_c}$.  \Halmos\\[3mm]
\noindent\textbf{Proof of Lemma \ref{lemma-adoption-eaug}}:\\
Under expert-level augmentation, the firm's profit function is given by
$$\Pi_u(x)=1-\left(w+\frac{1}{2}kx^2\right)-\left(w+\frac{1}{2}k\right)(1-x)[(1-r_u)t_c+ht_c].$$
By comparing with the profit function in the pre-GenAI benchmark, we have that the firm should adopt GenAI if and only if $(1-r_u)t_c+ht_c<t_c$, i.e., $h<r_u$.    \Halmos\\[3mm]
\noindent\textbf{Proof of Proposition \ref{prop-e-aug}}:\\
Since $\frac{\partial^2\Pi_u}{\partial x^2}=-k<0$, from the first-order condition, the optimal worker knowledge level can be solved as 
$$x_u^*=\frac{(k+2w)(1-r_u+h)t_c}{2k}.$$
Furthermore, $x_u^*<x_0^*$ for $h<r_u$, and 
$$\frac{\partial x_u^*}{\partial r_u}=-\frac{(k+2w)t_c}{2k}<0,$$ 
$$\frac{\partial x_u^*}{\partial h}=\frac{(k+2w)t_c}{2k}>0.$$
Consequently, $x_u^*$ strictly decreases in $r_u$ and strictly increases in $h$. 

After the adoption of expert-level augmentation (i.e., $h<r_u$), the optimal span of control is 
$$s_u^*=\frac{1}{(1-x_u^*)[(1-r_u)t_c+ht_c]}.$$
We can have 
$$\frac{\partial s_u^*}{\partial r_u}=\frac{t_c[t_c(k+2w)r_u+k-t_c(k+2w)(1+h)]}{k(1-x_u^*)^2[(1-r_u)t_c+ht_c]^2}.$$
Therefore, $\frac{\partial s_u^*}{\partial r_u}<0$ for $0<r_u<r_4$ and $\frac{\partial s_u^*}{\partial r_u}>0$ for $r_4<r_u<1$, where $r_4=\frac{t_c(k+2w)(1+h)-k}{t_c(k+2w)}$.
$$\frac{\partial s_u^*}{\partial h}=\frac{t_c[t_c(k+2w)h+t_c(1-r_u)(k+2w)-k]}{k(1-x_u^*)^2[(1-r_u)t_c+ht_c]^2}.$$
Therefore, $\frac{\partial s_u^*}{\partial h}<0$ for $0<h<h_4$ and $\frac{\partial s_u^*}{\partial h}>0$ for $h_4<h<r_u$, where $h_4=\frac{k-t_c(k+2w)(1-r_u)}{t_c(k+2w)}$.  \Halmos\\[3mm]
\noindent\textbf{Proof of Proposition \ref{app-prop-wauto-rh}}:\\
Substituting $h=b(1-r_t)$ into the proof of Lemma \ref{lemma-adoption-wauto} and Proposition \ref{prop-workerauto-x} yields the GenAI adoption condition and the optimal worker knowledge level characterized in Proposition \ref{app-prop-wauto-rh}, where 
\begin{equation*}
     \hat{b}_t
    =\left \{ 
    \begin{aligned}
    &\; \frac{1-t_v}{(1-r_t)t_r}, \quad\quad\quad && \text{if \;} r_t\le x_0^*; \\
    &\; \frac{1-t_v}{(1-r_t)t_r}-\frac{(2kr_t-kt_c-2wt_c)^2}{4kr_tt_r(1-r_t)(kr_t^2+2w)}, \quad\quad\quad && \text{otherwise. \;}      \quad\quad\quad  \\
    \end{aligned}
    \right.
\end{equation*}
Under the condition $b<\hat{b}_t$, it can be verified that $\hat{x}_t^*\ge \min\{\hat{x}_t, 1\}>x_0^*$ and 
$$\frac{\partial \hat{x}_t}{\partial r_t}=\frac{(k+2w)t_c[t_r(2r_t-1)b+1-t_v]}{2k[1-bt_rr_t^2-(1-t_v-bt_r)r_t]^2}>0.$$
This proves the results in Proposition \ref{app-prop-wauto-rh}.    \Halmos\\[3mm]
\noindent\textbf{Proof of Corollary \ref{app-coro-wauto-rh}}:\\
The optimal worker knowledge level $\hat{x}_t^*=1$ if and only if $\hat{x}_t\ge 1$. It can be verified that $\hat{x}_t\ge 1$ is equivalent to $F_3(r_t)\ge 0$, where 
$$F_3(r_t)=2bkt_rr_t^2+2k(1-t_v-bt_r)r_t+(k+2w)t_c-2k.$$
Based on the properties of quadratic functions, we have that $F_3(r_t)\ge 0$ for $\hat{r}_1\le r_t<1$ and $F_3(r_t)<0$ for $0<r_t<\hat{r}_1$, where $r_t=\hat{r}_1$ is the larger root of the equation $F_3(r_t)=0$ and 
$$\hat{r}_1=\frac{1}{2bkt_r}\left\{-k(1-t_v-bt_r)+\sqrt{k^2t_r^2b^2-2kt_r[(k+2w)t_c-k(1+t_v)]b+k^2(1-t_v)^2}\right\}.$$
This proves the results in Corollary \ref{app-coro-wauto-rh}.   \Halmos\\[3mm]
\noindent\textbf{Proof of Proposition \ref{app-prop-waug-rh-x}}:\\
Substituting $h=b(1-r_g)$ into the proof of Lemma \ref{lemma-w-aug}  yields the GenAI adoption condition and the optimal worker knowledge level characterized in Proposition \ref{app-prop-waug-rh-x}. Under the condition $\hat{r}_g<r_g<1$, it can be verified that $\hat{x}_g^*<x_0^*$ and 
$$\frac{\partial \hat{x}^*_g}{\partial r_g}=\frac{(k+2w)[(1-2r_g)t_cb-t_c+t_v]}{2k}<0.$$
This proves the results in Proposition \ref{app-prop-waug-rh-x}.    \Halmos\\[3mm]
\noindent\textbf{Proof of Proposition \ref{app-prop-waug-rh-s}}:\\
After GenAI adoption ($\hat{r}_g<r_g<1$), the optimal span of control is 
$$\hat{s}_g^*=\frac{1}{(1-\hat{x}_g^*)[r_gt_v+b(1-r_g)r_gt_c+(1-r_g)t_c]}.$$ 
It can be verified that $\frac{\partial\hat{s}_g^*}{\partial r_g}<0$ is equivalent to $F_4(r_g)<0$, where 
$$F_4(r_g)=(k+2w)bt_cr_g^2+(k+2w)(t_c-bt_c-t_v)r_g+k-(k+2w)t_c.$$
Based on the properties of quadratic functions, for $t_c>\frac{k}{k+2w}$, we have $F_4(r_g)<0$ for $\hat{r}_g<r_g<\hat{r}_2$ and $F_4(r_g)>0$ for $\hat{r}_2<r_g<1$, where $r_g=\hat{r}_2$ is the larger root of the equation $F_4(r_u)=0$ and  
\begin{align*}
    \hat{r}_2=\frac{1}{2b(k+2w)t_c}\{&(k+2w)(bt_c-t_c+t_v)+[(k+2w)t_c^2b^2+2t_c(kt_c+kt_v+2t_cw+2t_vw-2k)b \\
    &+(k+2w)(t_c-t_v)^2]^\frac{1}{2}(k+2w)^{\frac{1}{2}}\}.
\end{align*}
By contrast, for $t_c\le\frac{k}{k+2w}$, $F_4(r_g)\ge 0$ holds for all $\hat{r}_g<r_g<1$. This gives the results in Proposition \ref{app-prop-waug-rh-s}.  \Halmos\\[3mm]
\noindent\textbf{Proof of Proposition \ref{app-prop-eauto-rh-x}}:\\
Substituting $h=b(1-r_e)$ into the proof of Lemma \ref{lemma-adoption-eauto} yields the GenAI adoption condition and the optimal worker knowledge level characterized in Proposition \ref{app-prop-eauto-rh-x}, where 
$$\hat{r}_e=\max\left\{1-\frac{t_c-t_v}{bt_c}, \frac{(k+2w)(bt_c+t_c+t_v)}{4k+bkt_c+2bt_cw}\right\}.$$
Under the condition $\hat{r}_e<r_e<1$, it can be verified that $\hat{x}_e^*<x_0^*$ and $\frac{\partial \hat{x}^*_e}{\partial r_e}<0$.  \Halmos\\[3mm]
\noindent\textbf{Proof of Proposition \ref{app-prop-eauto-rh-s}}:\\
After GenAI adoption ($\hat{r}_e<r_e<1$), the optimal span of control is 
$$\hat{s}_e^*=\frac{1}{(1-r_e)t_c+(r_e-\hat{x}_e^*)[t_v+b(1-r_e)t_c]}.$$ 
We have 
$\frac{\partial\hat{s}_e^*}{\partial r_e}=\frac{F_5(r_e)\hat{s}_e^{*2}}{k}$, 
where
$$F_5(r_e)=bt_c[2k+b(k+2w)t_c]r_e-t_c^2(k+2w)b^2-t_c[k+(k+2w)t_v]b+k(t_c-t_v).$$
Consequently, $\frac{\partial\hat{s}_e^*}{\partial r_e}<0$ on $(\hat{r}_e, \hat{r}_3)$ and $\frac{\partial\hat{s}_e^*}{\partial r_e}>0$ on $(\hat{r}_3, 1)$, where $r_e=\hat{r}_3$ is the root of the equation $F_5(r_e)=0$ and 
$$\hat{r}_3=\frac{t_c^2(k+2w)b^2+t_c[k+(k+2w)t_v]b-k(t_c-t_v)}{bt_c[2k+b(k+2w)t_c]}.$$ 
This gives the results in  Proposition \ref{app-prop-eauto-rh-s}. \Halmos\\[3mm]
\noindent\textbf{Proof of Proposition \ref{app-prop-eaug-rh}}:\\
Substituting $h=b(1-r_u)$ into the proof of Lemma \ref{lemma-adoption-eaug} and Proposition \ref{prop-e-aug} yields the GenAI adoption condition and the optimal worker knowledge level characterized in Proposition \ref{app-prop-eaug-rh}.
Under the condition $\hat{r}_u<r_u<1$, it can be verified that $\hat{x}_u^*<x_0^*$ and $\frac{\partial \hat{x}^*_u}{\partial r_u}<0$.

After GenAI adoption ($\hat{r}_u<r_u<1$), the optimal span of control is 
$$\hat{s}_u^*=\frac{1}{(1-\hat{x}_u^*)[(1-r_u)t_c+b(1-r_u)t_c]}.$$ 
We have 
$\frac{\partial\hat{s}_u^*}{\partial r_u}=\frac{t_c(1+b)F_6(r_u)\hat{s}_u^{*2}}{k}$, 
where
$$F_6(r_u)=(1+b)(k+2w)t_cr_u+k-(1+b)(k+2w)t_c.$$
Therefore, $\frac{\partial\hat{s}_u^*}{\partial r_u}<0$ on $(\hat{r}_u, \hat{r}_4)$ and $\frac{\partial\hat{s}_u^*}{\partial r_u}>0$ on $(\hat{r}_4, 1)$, where $r_u=\hat{r}_4$ is the root of the equation $F_6(r_u)=0$ and $\hat{r}_4=\frac{(1+b)(k+2w)t_c-k}{(1+b)(k+2w)t_c}$.   \Halmos\\[3mm]
\noindent\textbf{Proof of Proposition \ref{app-cost-waug}}:\\
Considering the cost regarding GenAI capability investment, the firm's profit function becomes
\begin{equation*}
    \Pi_g(x, r_g)=1-\left(w+\frac{1}{2}kx^2\right)-\left(w+\frac{1}{2}k\right)\big[r_g(1-x)(t_v+ht_c)+(1-r_g)(1-x)t_c\big]-\frac{1}{2}c_g^r r_g^2.
\end{equation*}
Based on the proof of Lemma \ref{lemma-w-aug}, we have that given the GenAI capability level, the unconstrained optimal knowledge level of workers can be solved as
$$x_g=\frac{(k+2w)[r_gt_v+(1+r_gh-r_g)t_c]}{2k}.$$
To guarantee strict concavity of the profit function in $r_g$ and the existence of an interior solution, we assume $c_g^r>\underline{c}_g^r$, where $\underline{c}_g^r=\max\{c_1, c_2\}$, 
$$c_1=\frac{(k+2w)^2(t_c-t_v-ht_c)^2}{4k}, ~~ c_2=\frac{(k+2w)(t_c-ht_c-t_v)[(k+2w)(ht_c+t_v)-2k]}{4k}.$$
Plugging $x=x_g$ into $\Pi_g(x, r_g)$, we have 
$$\frac{\partial^2\Pi_g}{\partial r_g^2}=-\frac{4kc_g^r-(k+2w)^2(t_c-t_v-ht_c)^2}{4k}<0.$$
From the first-order condition, the unconstrained optimal capability level is 
$$r_g^*=\frac{(k+2w)[2k-(k+2w)t_c](t_c-ht_c-t_v)}{4kc_g^r-(k+2w)^2(t_c-t_v-ht_c)^2}.$$
It can be verified that 
$$1-r_g^*=\frac{4kc_g^r-(k+2w)(t_c-ht_c-t_v)[2k-(k+2w)(ht_c+t_v)]}{4kc_g^r-(k+2w)^2(t_c-t_v-ht_c)^2}.$$
Therefore, since $r_g^*\le 0$ for $h\ge \overline{h}_g$ and $0<r_g^*<1$ for $h< \overline{h}_g$, the firm should adopt GenAI if and only if $h< \overline{h}_g$, where $\overline{h}_g=1-\frac{t_v}{t_c}$. This gives Proposition \ref{app-cost-waug}(a). 

By plugging $r_g=r_g^*$ into the expression of $x_g$, we have that after GenAI adoption, the optimal worker knowledge level is
$$x_g^*=\frac{(k+2w)[r_g^*t_v+(1+r_g^*h-r_g^*)t_c]}{2k}.$$
It can be verified that 
$$x_0^*-x_g^*=\frac{(k+2w)^2[2k-(k+2w)t_c](t_c-ht_c-t_v)^2}{2k[4kc_g^r-(k+2w)^2(t_c-t_v-ht_c)^2]}>0, $$
$$\frac{\partial x_g^*}{\partial c_g^r}=\frac{2(k+2w)^2[2k-(k+2w)t_c](t_c-ht_c-t_v)^2}{[4kc_g^r-(k+2w)^2(t_c-t_v-ht_c)^2]^2}>0.$$
This gives Proposition \ref{app-cost-waug}(b). 

After GenAI adoption, the optimal span of control is 
$$s_g^*=\frac{1}{r_g^*(1-x_g^*)(t_v+ht_c)+(1-r_g^*)(1-x_g^*)t_c}.$$
It can be verified that $\frac{\partial s_g^*}{\partial c_g^r}>0$ is equivalent to $[4(k+2w)t_c-4k]c_g^r-(k+2w)^2(t_c-ht_c-t_v)^2>0$. Therefore, for $t_c\le\frac{k}{k+2w}$, $\frac{\partial s_g^*}{\partial c_g^r}\le 0$ consistently holds and $s_g^*$ increases as $c_g^r$ decreases. For $t_c>\frac{k}{k+2w}$, $\frac{\partial s_g^*}{\partial c_g^r}<0$ for $c_g^r<c_{g1}^r$ and $\frac{\partial s_g^*}{\partial c_g^r}>0$ for $c_g^r>c_{g1}^r$, where 
$$c_{g1}^r=\frac{(k+2w)^2(t_c-ht_c-t_v)^2}{4(k+2w)t_c-4k}.$$ 
Therefore, if $t_c>\frac{k}{k+2w}$, $s_g^*$ strictly increases as $c_g^r$ decreases for $\underline{c}_g^r<c_g^r<c_{g1}^r$ and $s_g^*$ strictly decreases as $c_g^r$ decreases for $c_g^r>\max\{c_{g1}^r, \underline{c}_g^r\}$. This gives Proposition \ref{app-cost-waug}(c).   \Halmos\\[3mm] 
\noindent\textbf{Proof of Proposition \ref{app-prop-cost-eauto}}:\\
Considering the cost regarding GenAI capability investment, the firm's profit function becomes
\begin{equation*}
   \Pi_e(x,r_e)=1-\left(w+\frac{1}{2}kx^2\right)-\left(w+\frac{1}{2}k\right)[(1-r_e)t_c+(r_e-x)(t_v+ht_c)]-\frac{1}{2}c_e^rr_e^2.
\end{equation*}
Based on the proof of Lemma \ref{lemma-adoption-eauto}, we have that given the GenAI capability level, the unconstrained optimal worker knowledge level can be solved as 
$$x_e^*=\frac{(k+2w)(ht_c+t_v)}{2k}.$$
Plugging $x=x_e^*$ into $\Pi_e(x,r_e)$, we have $\frac{\partial^2\Pi_e}{\partial r_e^2}=-c_e^r<0$. Therefore, from the first-order condition, the unconstrained optimal capability level is 
$$r_e^*=\frac{(k+2w)(t_c-ht_c-t_v)}{2c_e^r}.$$
Consequently, the firm should adopt GenAI only if $h<\overline{h}_e$, where $\overline{h}_e=1-\frac{t_v}{t_c}$; otherwise, $r_e^*<0$. In the following analysis, we focus on the case of $h<\overline{h}_e$.
To guarantee that the firm's optimal automation capability level remains strictly less than one, we assume that $c_e^r>\underline{c}_e^r$, where 
$$\underline{c}_e^r=\frac{(k+2w)(t_c-ht_c-t_v)}{2}.$$
Furthermore, it can be verified that
$$r_e^*-x_e^*=\frac{(k+2w)[(t_c-ht_c-t_v)k-(ht_c+t_v)c_e^r]}{2kc_e^r}.$$
$$\Pi_e(x_e^*, r_e^*)-\Pi_0^*=\frac{(k+2w)^2(t_c-t_v-ht_c)[k(t_c-ht_c-t_v)-(ht_c+t_c+t_v)c_e^r]}{2kc_e^r}.$$
Therefore, $r_e^*>x_e^*$ if and only if $c_e^r<\frac{(t_c-ht_c-t_v)k}{ht_c+t_v}$ and $\Pi_e(x_e^*, r_e^*)>\Pi_0^*$ if and only if $c_e^r<\overline{c}_e^r$, where $\overline{c}_e^r=\frac{(t_c-ht_c-t_v)k}{ht_c+t_c+t_v}$. Under the condition $h<\overline{h}_e$, we have $\overline{c}_e^r<\frac{(t_c-ht_c-t_v)k}{ht_c+t_v}$. Therefore, the firm should adopt GenAI if and only if $h<\overline{h}_e$ and $c_e^r<\overline{c}_e^r$. The optimal GenAI capability and worker knowledge level are $r_e^*$ and $x_e^*$, respectively. It is straightforward to see that $x_e^*$ is independent of $c_e^r$, and we have $x_e^*<x_0^*$.

After GenAI adoption, the optimal span of control is 
$$s_e^*=\frac{1}{(1-r_e^*)t_c+(r_e^*-x_e^*)(t_v+ht_c)}.$$
It can be verified that $s_e^*|_{c_e^r=\overline{c}_e^r}=s_0^*$, and 
$$\frac{\partial s_e^*}{\partial c_e^r}=-\frac{2k^2(k+2w)(t_c-ht_c-t_v)^2}{4(c_e^r)^2k^2[(1-r_e^*)t_c+(r_e^*-x_e^*)(t_v+ht_c)]^2}<0.$$
Therefore, the optimal span of control remains continuous at the adoption cutoff $c_e^r=\overline{c}_e^r$ and after that, $s_e^*$ strictly increases as $c_e^r$ decreases.  \Halmos\\[3mm]
\noindent\textbf{Proof of Proposition \ref{app-prop-cost-eaug}}:\\
Considering the cost regarding GenAI capability investment, the firm's profit function becomes
\begin{equation*}
   \Pi_u(x,r_u)=1-\left(w+\frac{1}{2}kx^2\right)-\left(w+\frac{1}{2}k\right)(1-x)[(1-r_u)t_c+ht_c]-\frac{1}{2}c_u^rr_u^2.
\end{equation*}
Based on the proof of Lemma \ref{lemma-adoption-eaug}, we have that given the GenAI capability level, the unconstrained optimal worker knowledge level can be solved as 
$$x_u=\frac{(k+2w)(1-r_u+h)t_c}{2k}.$$
To guarantee strict concavity of the profit function in $r_u$ and the existence of an interior solution, we assume $c_u^r>\underline{c}_u^r$, where $\underline{c}_u^r=\max\{c_3, c_4\}$, 
$$c_3=\frac{(k+2w)^2t_c^2}{4k}, ~~ c_4=\frac{[2k^2+4kw-h(k+2w)^2t_c]t_c}{4k}.$$
Plugging $x=x_u$ into $\Pi_u(x,r_u)$, we have
$$\frac{\partial\Pi_u}{\partial r_u}=-\frac{4kc_u^r-(k+2w)^2t_c^2}{4k}<0.$$
From the first-order condition, the unconstrained optimal capability level is 
$$r_u^*=\frac{(k+2w)[2k-(k+2w)(1+h)t_c]t_c}{4kc_u^r-(k+2w)^2t_c^2}.$$
It can be verified that 
$$1-r_u^*=\frac{4kc_u^r-[2k^2+4kw-h(k+2w)^2t_c]t_c}{4kc_u^r-(k+2w)^2t_c^2}.$$
$1-r_u^*>0$ for $c_u^r>\underline{c}_u^r$. Furthermore, $r_u^*>0$ if and only if $h<\frac{2k}{(k+2w)t_c}-1$. 

Plugging $r_u=r_u^*$ into $\Pi_u(x_u,r_u)$, it can be verified that 
$$\Pi_u(x_u,r_u^*)-\Pi_0^*=\frac{(k+2w)t_cF_7(h)}{8k[4kc_u^r-(k+2w)^2t_c^2]}.$$
where 
$$F_7(h)=4k(k+2w)c_u^rt_ch^2-8c_u^rk[2k-(k+2w)t_c]h+t_c(k+2w)(2k-kt_c-2wt_c)^2.$$
It can be verified that $F_7|_{h=\frac{2k}{(k+2w)t_c}-1}<0$. Therefore, we have that the firm should adopt GenAI (i.e., $\Pi_u(x_u,r_u^*)>\Pi_0^*$ and $r_u^*>0$) if and only if $h<\overline{h}_u'$, where $h=\overline{h}_u'$ is the smaller root of the equation $F_7(h)=0$ and 
$$\overline{h}_u'=\frac{[2k-(k+2w)t_c][2kc_u^r-\sqrt{kc_u^r[4kc_u^r-(k+2w)^2t_c^2]}]}{2kc_u^r(k+2w)t_c}.$$
It can be verified that 
$$\frac{\partial\overline{h}_u'}{\partial c_u^r}=-\frac{(k+2w)[2k-(k+2w)t_c]t_c}{4c_u^r\sqrt{kc_u^r[4kc_u^r-(k+2w)^2t_c^2]}}<0.$$
Therefore, $\overline{h}_u'$ strictly increases as $c_u^r$ decreases. This gives Proposition \ref{app-prop-cost-eaug}(a).

By plugging $r_u=r_u^*$ into the expression of $x_u$, we have that after GenAI adoption, the optimal worker knowledge level is 
$$x_u^*=\frac{(k+2w)(1-r_u^*+h)t_c}{2k}.$$
It can be verified that $x_u^*<x_0^*$ and $\frac{\partial x_u^*}{\partial c_u^r}>0$. This gives Proposition \ref{app-prop-cost-eaug}(b).

After GenAI adoption, the optimal span of control is 
$$s_u^*=\frac{1}{(1-x_u^*)[(1-r_u^*)t_c+ht_c]}.$$
It can be verified that $\frac{\partial s_u^*}{\partial c_u^r}>0$ is equivalent to $[4t_c(1+h)(k+2w)-4k]c_u^r-(k+2w)^2t_c^2>0$. Therefore, for $t_c\le\frac{k}{(k+2w)(1+h)}$, $\frac{\partial s_u^*}{\partial c_u^r}\le 0$ consistently holds and $s_u^*$ increases as $c_u^r$ decreases. For $t_c>\frac{k}{(k+2w)(1+h)}$, $\frac{\partial s_u^*}{\partial c_u^r}<0$ for $c_u^r<c_{u1}^r$ and $\frac{\partial s_u^*}{\partial c_u^r}>0$ for $c_u^r>c_{u1}^r$, where 
$$c_{u1}^r=\frac{(k+2w)^2t_c^2}{4t_c(1+h)(k+2w)-4k}.$$
Therefore, if $t_c>\frac{k}{(k+2w)(1+h)}$, $s_u^*$ strictly increases as $c_u^r$ decreases for $\underline{c}_u^r<c_u^r<c_{u1}^r$ and $s_u^*$ strictly decreases as $c_u^r$ decreases for $c_u^r>\max\{c_{u1}^r, \underline{c}_u^r\}$. This gives Proposition \ref{app-prop-cost-eaug}(c).   \Halmos\\[3mm] 
\noindent\textbf{Proof of Proposition \ref{app-experty-ben}}:\\
To establish the strict global monotonicity, we directly examine the first-order partial derivative of the profit function $\Pi_0(x,y)$ with respect to $y$:
$$\frac{\partial \Pi_0}{\partial y} = 1 - kyt_c(1-x).$$
Since $0\le x\le y\le 1$ and $k<\overline{k}=\frac{1}{t_c}$, 
$$\frac{\partial \Pi_0}{\partial y} = 1 - kyt_c(1-x)> 1-1=0.$$
Therefore, for any feasible pair $(x,y)$ within the region $0 \le x \le y \le 1$, the marginal profit with respect to $y$ is strictly positive. This implies that for any given $y$, regardless of the firm's optimal response for $x$, further expanding $y$ will consistently and strictly yield a higher profit. Consequently, the global optimal solution is uniquely determined at $y_0^* = 1$.  \Halmos\\[3mm]

\end{APPENDICES}

\end{document}